\documentclass{aa}  

\usepackage[colorlinks=true, linkcolor = blue,
            urlcolor  = blue,
            citecolor = blue,
            anchorcolor = blue]{hyperref}
\usepackage{graphicx}
\usepackage{txfonts}
\usepackage{natbib}
\usepackage{enumitem} 
\usepackage{color}
\usepackage{threeparttable}
\usepackage{placeins}
\usepackage{cancel}
\bibpunct{(}{)}{;}{f}{}{,} 

\newcommand{\teffa}{$T_{\mathrm{eff}}^{\mathrm{H}\alpha}$}

\newcommand{\teff}{$T_{\mathrm{eff}}$}

\newcommand{\teffHa}{$T_{\mathrm{eff}}^{\mathrm{H}\alpha}$}

\newcommand{\logg}{$\mathrm{log}\,g$}
\newcommand{\loggiso}{$\mathrm{log}\,g_{\mathrm{iso}}$}

\newcommand{\loggmg}{$\mathrm{log}\,g_{\mathrm{Mg}}$}
\newcommand{\titan}{\textsc{Titans}}
\newcommand{\loggseis}{$\mathrm{log}\,g_{\mathrm{seis}}$}

\usepackage[normalem]{ulem}

\begin{document}

  \title{Titans metal-poor reference stars II.
  \thanks{Based on observations collected at the MERCATOR telescope installed at Roque de los Muchachos Observatory and ESO archival data.}}

  \subtitle{Red giants and CEMP stars}
    
   \author{R. E. Giribaldi\inst{1}
          \and
          S. Van Eck\inst{1}
          \and
          T. Merle\inst{1}
          \and 
          A. Jorissen\inst{1}
          \and
          P. Krynski\inst{1}
          \and
          L. Planquart\inst{1}
          \and
          M. Valentini\inst{2}
          \and
          C. Chiappini\inst{2}
          \and
        H. Van Winckel\inst{3}
          }

   \institute{ Institut d'Astronomie et d'Astrophysique, Universit\'e libre de Bruxelles, CP 226, Boulevard du Triomphe, 1050 Brussels, Belgium \\
             \email{riano.giribaldi@ulb.be, rianoesc@gmail.com}
         \and
   Leibniz-Institut f\"ur Astrophysik Potsdam (AIP), An der Sternwarte 16, 14482 Potsdam, Germany          
         \and
Institute of Astronomy, KU Leuven, Celestijnenlaan 200D, 3001 Leuven, Belgium
             }
             
    \date{Received  / Accepted }

 
  \abstract
   {
    Representative samples of F-, G-, K-type stars located out of the Solar Neighbourhood has started to be available in spectroscopic surveys. The fraction of metal-poor ([Fe/H]~$\lesssim -0.8$~dex) giants
    becomes increasingly relevant to far distances. 
   In metal-poor stars, effective temperatures (\teff)  based on LTE spectroscopy and on former colour-\teff\ relations of still wide use have been reported to be inaccurate. It is necessary to re-calibrate chemical abundances based on these \teff\ scales in the multiple available surveys to bring them to the same standard scale for their simultaneous use. For that, a complete sample of standards is required, which so far, is restricted to a few stars with quasi-direct \teff\ measurements.}
   {We aim at providing a legacy sample of metal-poor standards with proven accurate atmospheric parameters. We add 47 giants to the sample of metal-poor dwarfs of \cite{giribaldi2021A&A...650A.194G}, thereby constituting the \titan\ metal-poor reference stars.}
   {\teff\ was derived by 3D non-LTE H$\alpha$ modelling, whose accuracy was tested against interferometry and InfraRed Flux Method (IRFM).
   Surface gravity (\logg) was derived by fitting Mg~I~b triplet lines, whose accuracy was tested against asteroseismology.
   Metallicity was derived using \ion{Fe}{ii} lines, which was verified to be identical to the [Fe/H] derived from non-LTE spectral synthesis.}
   {\teff\ from 3D non-LTE H$\alpha$ is equivalent to interferometric  and IRFM temperatures within a $\pm$46~K uncertainty. 
   We achieved precision of $\sim$50~K for 34 stars with spectra with the highest S/N. For \logg, we achieved a total uncertainty of $\pm$0.15~dex. For [Fe/H], we obtained a total uncertainty of $\pm$0.09~dex.
   We find that the ionization equilibrium of Fe lines under LTE is not valid in metal-poor giants.
   LTE leads to a small but significant metallicity underestimation of $\sim$0.1~dex when derived from weak \ion{Fe}{i} lines, only provided accurate \teff\ and \logg.
   This bias
   disappears totally under non-LTE.
   }
   {
   }
   
   \keywords{stars: fundamental parameters --
             stars: atmospheres --
             stars: late-type --
             stars: low-mass --
             stars: carbon --
             stars: chemically peculiar --
             techniques: spectroscopic 
             }

   \maketitle
%

\section{Introduction}
Spectroscopic surveys are nowadays the main data sources to study the Milky Way formation and evolution via kinematic, dynamic, and chemical abundance analyses when combined with Gaia astrometric data \citep{Gai(a),gaiaDR32022arXiv220800211G}.
Gaia-ESO \citep{2012Msngr.147...25G,2013Msngr.154...47R,gilmore2022A&A...666A.120G,randich2022A&A...666A.121R}, Galactic Archaeology with HERMES \citep[GALAH,][]{2015MNRAS.449.2604D,buder2021MNRAS.506..150B}, Large Sky Area Multi-Object Fiber Spectroscopic Telescope \citep[LAMOST,][]{2012RAA....12.1197C},  Apache Point Observatory Galactic Evolution Experiment \citep[APOGEE,][]{majewski2017,jonsson2020AJ....160..120J}, and The Radial Velocity Experiment \citep[RAVE,][]{2020AJ....160...83S} are projects currently providing stellar parameters  and updating them in subsequent catalogue releases.
Supported by Gaia distances,
they have prospected the Galaxy plane from nearly the bulge center to about 20~kpc away, within about 10 kpc height off the plane.
Future projects such as 4-metre Multi-Object Spectroscopic Telescope \citep[4MOST,][]{2019Msngr.175....3D}, and WEAVE \citep{2016ASPC..507...97D,2016SPIE.9908E..1GD} are expected to  register spectra of many distant stars in such a way that studies will be performed with representative samples out of the Solar Neighbourhood.
This will naturally cut off distant dwarf stars, which are relatively faint, whereas red giants will be the main, if not the only, objects from which reliable abundances associated to certain point in the time-line of the Galaxy evolution can be obtained, provided that their ages are possible to estimate \citep[e.g.][]{valentini2019A&A...627A.173V,montalban2021NatAs...5..640M}.
Maximising the profits of the efforts made to implement spectroscopic surveys depend on our ability to retrieve accurate stellar parameters, abundances, and ages \citep[e.g.][]{jofre2019ARA&A..57..571J}.

Generally, it is assumed that main sequence dwarfs and red giants share the same abundance scale. However, it has been shown that, under local thermodynamic equilibrium (LTE), the giants' metallicity ([Fe/H]\footnote{[A/B] = $ \log{\left( \frac{N(\text{A})}{N(\text{B})} \right )_\text{Star}} - \log{\left( \frac{N(\text{A})}{N(\text{B})} \right )_\text{Sun}} $, where $N$ denotes the number abundance of a given element.}) scale is consistent with that of dwarfs only with 
meticulous adapted line lists, and when their effective temperature (\teff) and surface gravity (\logg) are not constrained by assuming excitation and ionization equilibrium, but via independent methods \citep{dutra-ferreira2016A&A...585A..75D}.
Further, it has been observed that at low metallicity ranges, diffusion and mixing phenomena may induce substantial variations in Fe and other elements across the way between the turnoff and the red giant branch  (RGB) \citep{korn2007,lind2008,nord2012,gru2014}.
Moreover, tests have shown that classical spectroscopic methods may produce highly discrepant abundance outcomes for red giants \citep[e.g.][]{Lebzelter2012A&A...547A.108L, jofre2017A&A...601A..38J, casali2020A&A...643A..12C}.
 
It is therefore reasonable that scale incompatibilities, that is to say, inaccuracies, also arise for other elements further than Fe when based on inaccurate \teff\ and \logg. For example, \cite{giribaldi2023A&A...673A..18G} found an offset of $-0.07$~dex in [Mg/Fe] ratios of metal-poor dwarfs induced by temperatures underestimated by $\sim$150~K.
Here we show that in metal-poor red giants, the [Mg/Fe] offset can be exacerbated to about $-0.1$~dex due to only 100~K underestimation, which is the typical \teff\ offset of the literature values we compiled. 

In addition to these difficulties, we must also consider that observational line profiles are more accurately reproduced by three-dimensional (3D) hydrodynamic model atmospheres that take non-LTE effects into account \citep[e.g.][]{bergemann2017ApJ...847...15B,amarsi2018,gallagher2020A&A...634A..55G,wang2021MNRAS.500.2159W,amarsi2022A&A...668A..68A}, which inherently lead to more accurate abundance determinations. 
However, 3D non-LTE abundance determination is not yet feasible for the large amount of survey spectra, thus 1D LTE abundances converted to the 3D non-LTE scale using correction grids will be the most efficient option during some years in the near future \citep[e.g.][]{amarsi2019A&A...630A.104A,amarsi2022A&A...668A..68A,wang2021MNRAS.500.2159W}.
Therefore, for proper application of 3D non-LTE corrections, 1D LTE abundances free from potential \teff, \logg, and [Fe/H] systematics are paramount.
These corrections have already shown how strong their impact is in the chemo-dynamic analysis employed to identify stellar populations of the primitive Milky Way and to analyse their evolution
\citep[e.g.][]{bergemann2017ApJ...847...16B,amarsi2019A&A...630A.104A,giribaldi2019}.

Standard stars have been historically tracked and studied to calibrate acquisition instruments and astrophysical models.
For instance, most photometric systems adopted Vega as spectro-photometric standard. Its brightness and good observability for northern telescopes supported its choice.
However, its rapid rotation and the presence of a debris disc around it devise a spectrum that requires different sets of atmospheric parameters to be reproduced in the infrared and the visible  \citep[e.g.][]{casagrande2006MNRAS.373...13C,gray2007ASPC..364..305G}. This led to small but significant corrections in the zero-points of several photometric systems  \citep{bohlin2007ASPC..364..315B,2007ASPC..364..227M},
so that widely used color-dependent relations to infer \teff\ of F-, G-, and K-type stars were shown to be substantially biased to cooler determinations \citep{Casagrande2010}. 

Solar twins\footnote{Stars with atmospheric parameters matching the solar ones within uncertainties \citep[e.g.][]{porto1997,porto2014} or within small arbitrary differences, typically 100~K, 0.1~dex, and 0.1~dex in \teff, [Fe/H], and \logg\ \citep[e.g.][]{2009A&A...508L..17R}.} have also been largely tracked as standard stars \citep[e.g.][]{cayrel1996,cayrel1989,porto1997,porto2014,melendez2007ApJ...669L..89M,giribaldi2019A&A...629A..33G,yana-galarza2021MNRAS.504.1873Y}.
Among several scientific purposes, solar twins are needed to infer the solar colours and magnitudes in different photometric systems \citep[e.g.][]{neckel1986A&A...159..175N, pasquini2008A&A...489..677P,Casagrande2010,casagrande2021MNRAS.507.2684C}, given that the Sun's proximity prevents their direct measurement. 
Those magnitudes, as well as twins' spectra are required to update our knowledge of the physics of flux-wavelength distribution, which is still largely incomplete in the ultraviolet ($\sim$2348~\AA) and near ultraviolet ($\sim$3130~\AA), where many lines remain unidentified \citep[e.g.][]{bell1994MNRAS.268..771B,giribaldi2023ExA....55..117G}, and transitions and continuum opacities need to be modeled \citep[e.g.][]{short2009ApJ...691.1634S}.

Standard F-, G-, and K-type stars have been compiled within the so-called Gaia Benchmarks \citep{jofre2014,heiter2015A&A...582A..49H,hawkins2016A&A...592A..70H} mainly for the practical purpose of calibrating Gaia's parameter database, which currently is in its third data release  \citep[DR3,][]{gaiaDR32022arXiv220800211G}.
The atmospheric parameters \teff\ and \logg\ of most of these benchmarks are highly reliable in terms of accuracy because they have been inferred quasi-directly via interferometric measurements of their angular diameter and a modified version of the Stefan-Boltzmann relation. Therefore, they are ideal objects for diagnosing the accuracy of model atmospheres and the completeness of physical models of line formation \citep[e.g.][]{amarsi2016,amarsi2018,amarsi2022A&A...668A..68A}. However, despite the collective effort made by all authors that provided parameters for the Gaia Benchmarks (see references in the papers) and the use of the most sophisticated computational tools, biases in calibrated survey stellar parameters imply that more standard stars are required.
For example, offsets present in the GALAH DR2 parameter calibrations where the Gaia Benchmarks are standards \citep[][Fig. 14]{buder2018}, and possible metallicity systematics reported in the GALAH internal DR2 data set   \cite[][Fig. 11]{wheeler2020ApJ...898...58W}.
This is a relevant problem, although not evident, for users of catalogued stellar parameters, as they are often interested in element abundances, which are unavoidably affected by the parameter biases.
In addition, parameter biases could be transferred to new spectroscopic and photometric surveys, as they start to use spectroscopic surveys previously released as references for validation \citep[e.g.][]{wheeler2020ApJ...898...58W,2020AJ....160...83S,andrae2022arXiv220606138A}.

The Gaia Benchmarks presented a  paucity in the metal-poor range, approximately for metallicity lower than $-1$~dex, which was later mitigated by \cite{hawkins2016A&A...592A..70H}, who provided parameters for eleven moderately metal-poor stars ($-1.5 <$ [Fe/H] $<-1$~dex), where only two were giants. \cite{karovicova2020A&A...640A..25K} improved parameters for two already studied Benchmarks and provided new ones for other six stars, all of them being red giants.
These studies provide eighteen metal-poor stars, from which only twelve have reliable parameters according to the authors themselves. 

Carbon-enhanced metal-poor (CEMP) stars are characterized by a carbon overabundance with [C/Fe] usually larger than 0.7 \cite[e.g.][]{2005ARA&A..43..531B}.
Their fraction increases with decreasing metallicity: they represent 10–30\% of stars with [Fe/H]$<-2$ 
but up to 80\% of stars with [Fe/H]$<-4$ 
\citep{2006ApJ...652L..37L, 2014ApJ...797...21P, 2018ApJ...861..146Y}.
Those which are also enriched in heavy elements are further separated in CEMP-r, CEMP-s or CEMP-rs stars, and bear the signature of the rapid (r-), slow (s-) or potentially intermediate (i) processes of nucleosynthesis.
Their atmospheric parameters are frequently derived by traditional methods such as the excitation and ionization equilibrium of Fe lines, and sometimes by further constraining \teff\ with color-\teff\ relations \citep[e.g.][]{hansen2018ApJ...858...92H,karinkuzhi2021A&A...645A..61K}.
One of the objectives of this work is to ascertain the correct \teff-\logg-[Fe/H] scale of CEMP stars, and in particular to use spectroscopic indicators of \logg\ free from evolutionary modeling, such as Mg~I~b triplet lines. Since the abundances of heavy elements are quite sensitive to stellar parameter uncertainties, it is important to constitute a sample of benchmark CEMP stars as well.

\begin{figure*}
    \includegraphics[width=0.49\linewidth]{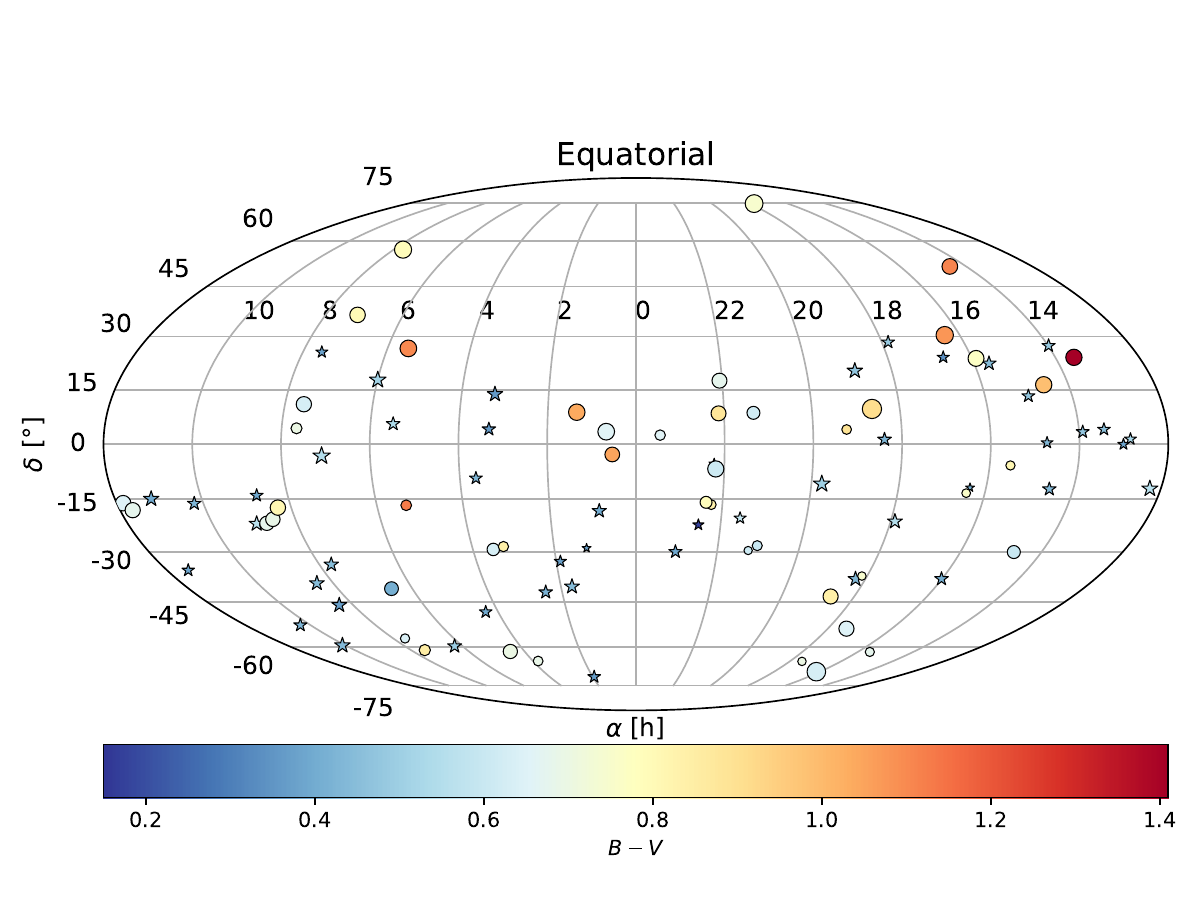}
    \includegraphics[width=0.49\linewidth]{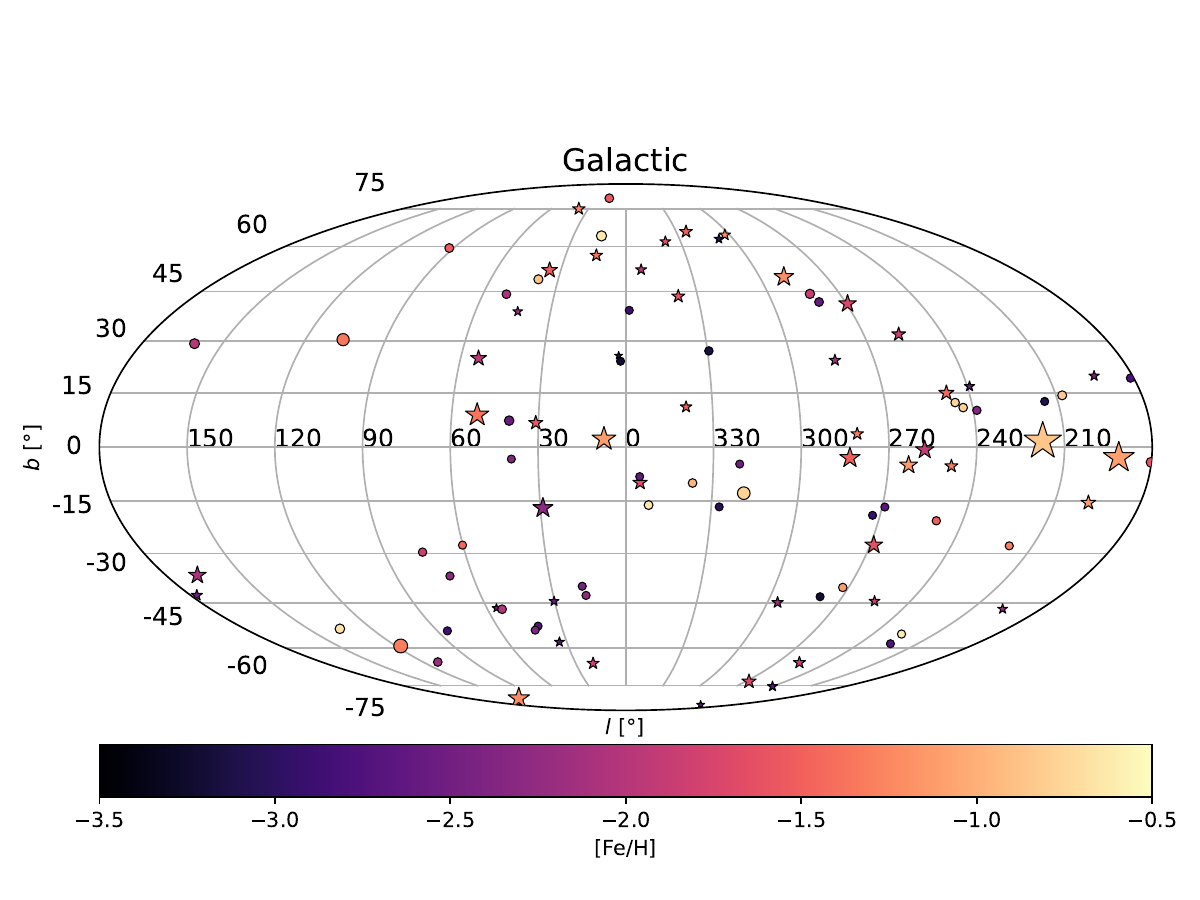}
\caption{Mollweide projections of the 95 \titan\ benchmark stars: 48 dwarfs (\titan\ I, stars) and 47 giants (\titan\ II, present work, circles). Left: in equatorial coordinates, color-coded by the index color $B-V$, the size of the star depends on the apparent $G$ magnitude: the larger the brighter. Right: in galactic coordinates, color-coded by the metallicity determined in this work (see Table~\ref{tab:results}), the size of the star scales with the Gaia parallax.}
\label{fig:skymap}
\end{figure*}

\cite{giribaldi2021A&A...650A.194G}, referred to as Paper~I henceforth, provided atmospheric parameters for 41 metal-poor dwarfs (48 dwarfs in total including seven Gaia Benchmarks) named the \titan\footnote{In Greek mythology, the \titan\ are the offsprings of Gaia.}~I metal-poor reference stars. 
They substantially fill the paucity of the Gaia Benchmarks between $-3 <$ [Fe/H] $<-1$~dex. 
Their parameters are not based on quasi-direct \teff, as those of most of the Gaia Benchmarks, but on  H$\alpha$ Balmer profiles  synthesised by 3D non-LTE models \citep{amarsi2018}. 
In Paper~I, the outcomes of both methods were proven  compatible for F-, G-, K-type stars with a wide range of metallicity, although metal-poor red giants remained to be tested.
Here we provide accurate atmospheric parameters for 47 metal-poor red giants with metallicity values between $-3.2$ and $-0.5$~dex and \logg\ between 1 and 3.5~dex;
we refer them to as \titan~II henceforth.
We employed the same 3D non-LTE H$\alpha$ models as in Paper~I, and we included tests with asteroseismic \logg\ and non-LTE [Fe/H] to scrutinise rigorously the accuracy of our atmospheric parameters.

This paper is organised as follows. Section~\ref{sec:data} describes the acquisition and selection of our observational data.
Section~\ref{sec:reduction} describes the data reduction. Section~\ref{sec:parameters} describes the determination of the atmospheric parameters. In Sect.~\ref{sec:accuracy} we show the scrutiny of our atmospheric parameters, and provide accuracy diagnostics of various widely used methods for deriving \teff, \logg, and [Fe/H].
Section~\ref{sec:abundance} describes our determination of Mg, C, N, and O abundances.
Finally, in Sect.~\ref{sec:conclusions} we list our conclusions in such a way that a reader interested on a certain accuracy test can be inmediately directed to the related section and figure.

\section{Sample selection and observational data}
\label{sec:data}
CEMP giants and their spectra were selected from the list of stars analysed on \cite{karinkuzhi2021A&A...645A..61K}. The spectra were acquired with the HERMES spectrograph \citep{hermes2011A&A...526A..69R} mounted in the 1.2~m \textit{Mercator} telescope at Roque de los Muchachos Observatory located at La Palma, Canary Islands, which covers the wavelength range 3800-9000 \AA\ at a nominal resolution $R \sim 86\,000$.  We selected stars whose H$\alpha$ line profiles show wavelength bins free of metal or molecular line contamination, as those bins are required to derive \teff\ by fitting observational with synthetic profiles; details of this procedure are provided in Sect.~\ref{sec:parameters}.

To select non-CEMP giants, we initially searched UVES \citep{dekker2000} and HARPS \citep{Mayor2003} spectra covering the H$\alpha$ line ($6562.797$~\AA) with signal-to-noise ratio (S/N) higher than 200 and resolution higher than $R = 40\,000$ in the European Southern Observatory (ESO) archive using the science portal of processed data\footnote{\url{http://archive.eso.org/scienceportal/home}}. 
We searched in the SIMBAD database to recover Gaia parallax ($\varpi$), and the $B$ and $V$ magnitudes.
We selected stars that seem to remain on the RGB in the $M_V$ vs. $B-V$ space as Fig.~\ref{fig:giants_selection} shows (reddening effects were ignored).
Once these candidates were pre-selected, we excluded the metal-rich stars
([Fe/H] $> -0.8$~dex) after cross-matching the resulting list with the PASTEL catalogue \citep{soubiran2016A&A...591A.118S}.
In the same sub-sample, we included one star with an optimal quality spectrum in the HERMES archive, HD~115444, with 
parameters and abundances 
in \cite{westin2000ApJ...530..783W,carrera2013MNRAS.434.1681C} and \cite{hansen2015A&A...583A..49H}.
Lastly, we removed stars with evident emission close to the H$\alpha$ core, since this line is used as a temperature indicator in the present work.

For validation purposes, three additional star samples were considered. First, we included a sub-sample of stars  with asteroseismologic measurements
from the Transiting Exoplanet Survey Satellite data \citep[TESS,][]{ricker2014SPIE.9143E..20R, Stassun2018AJ....156..102S}  available in \citet{Hon2021ApJ...919..131H} 
(so that asteroseismic \logg\ can be derived, see Sect.~\ref{sec:astero}).
We cross-matched the TESS Input Catalogue \citep[TIC,][]{Stassun2018AJ....156..102S} with the PASTEL database 
\citep{soubiran2016A&A...591A.118S}, requiring [Fe/H] $<-0.8$~dex.

Second, we included 
four metal-poor giants with interferometric measurements analysed by \cite{karovicova2020A&A...640A..25K}: 
HD~2665, HD~122563, HD~175305, and HD~221170,
and acquired HERMES high-resolution spectra between 
30/06/2022 and 02/07/2022. 

Third, we 
searched in the ESO archive for stars with effective temperatures  directly derived by the InfraRed Flux Method \citep[IRFM,][]{Blackwell1977,Blackwell1979,Blackwell1980}, in the catalogues of \cite{Casagrande2010}, \cite{hawkins2016A&A...592A..70H} and \cite{casagrande2021MNRAS.507.2684C}, which share the same absolute photometric calibration scale.

Fig.~\ref{fig:skymap} presents the map in equatorial and Galactic coordinates of the \titan\ benchmarks including the \titan\ I sample of metal-poor dwarfs presented in \citet{giribaldi2021A&A...650A.194G}. The distribution on the sky is quite homogeneous with a small deficiency at higher declinations. All dwarfs have  $B-V < 0.6$ while all giants have $B-V > 0.4$. The slight overlap is due to the presence of a few subgiants in the \titan\ I sample. 
The complete sample covers the $G$ magnitude range [6, 14], and the median $G$ magnitude is 9.6. In the Galactic map, the \titan\ II giants stars are all systematically further away and above the the Galactic plane compared to the \titan\ I dwarfs.  

\begin{figure*}[t]
    \centering
    \includegraphics[width=0.33\linewidth]{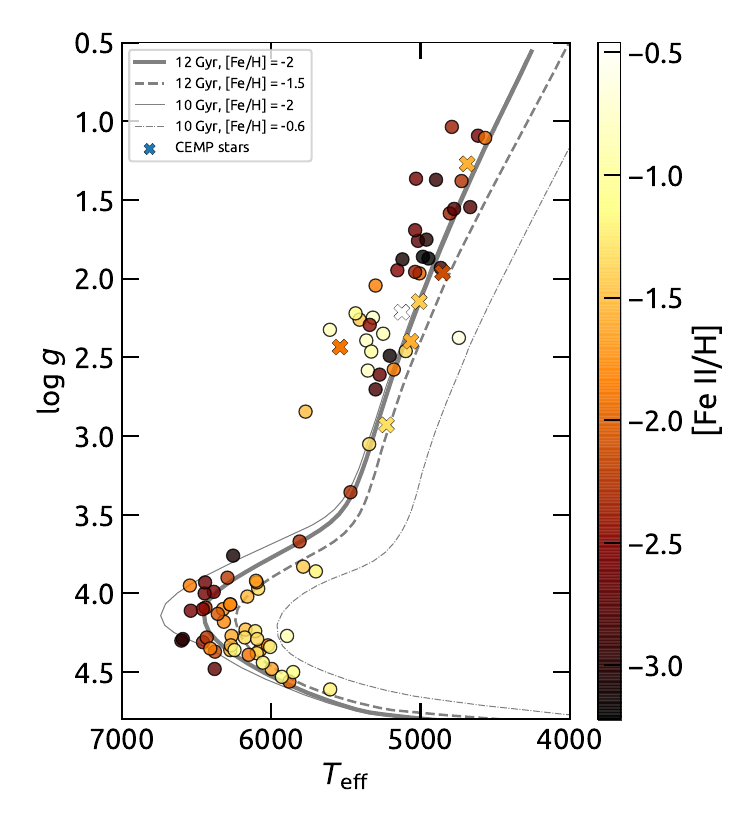}
    \includegraphics[width=0.33\linewidth]{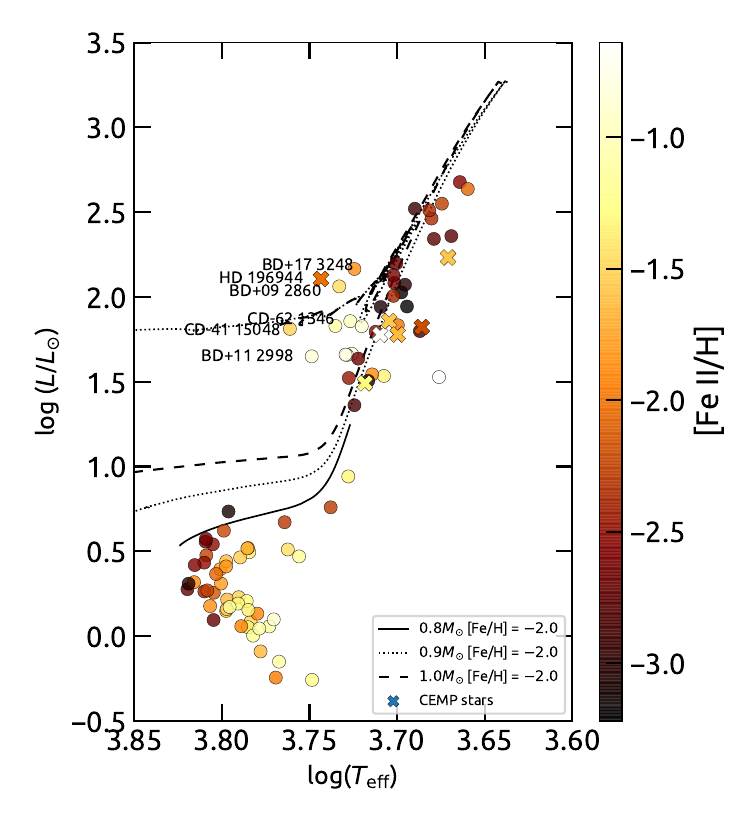}
    \includegraphics[width=0.33\linewidth]{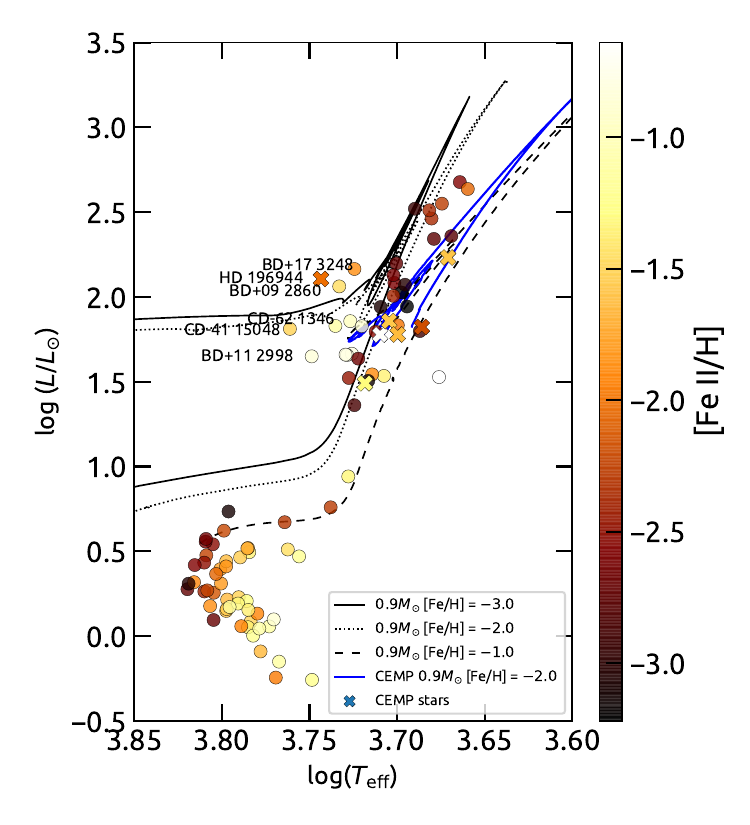}
    \caption{\tiny \titan\ in the Kiel diagram (left panel) and HR diagram (mid and right panels). 
    Surface gravities of \titan~II giants (\logg\ $< 3.5$~dex) are listed in Table~\ref{tab:results}, while those of \titan~I dwarfs come from \cite{giribaldi2021A&A...650A.194G}. CEMP stars are marked with a cross symbol.
    The Kiel diagram displays Yonsey-Yale isochrones \citep{kim2002,yi2003} as reference. HR diagrams display STAREVOL evolutionary tracks \citep{siess2008A&A...489..395S} for [Fe/H] $= -2$~dex and different masses, as labelled (mid panel), and for a mass $\mathcal{M}=0.9\mathcal{M}_\odot$ and various metallicity values, as labeled (right panel).
    The track representing CEMP stars (blue line) has the abundance ratio C/O = 2 \citep[details are given in][]{karinkuzhi2021A&A...645A..61K}.
    Tracks were constrained to maximum ages lower than the age of the universe, 13.8 Gigayears.
    Candidates to have started the Horizontal Branch are tagged by their identifiers in Table~\ref{tab:results}.}
    \label{fig:kiel}
\end{figure*}

In total we obtained a sample of 47 stars with their literature parameters listed in Table~\ref{tab:titans_giants}
where the red giants are classified according to the \teff\ or \logg\ determination method: 
interferometry and IRFM for \teff, and asteroseismology for \logg.
CEMP stars are listed separately at the end of the table. 
Among all stars, four have interferometric \teff\ determinations,
six have IRFM \teff\ determinations (one in common with interferometry),
and four have asteroseismic \logg. 
These stars are the standards required to assess the accuracy of the parameter determinations by the methods presented in this work and further applied to the stars labelled as "Other giants" or "CEMP giants" in Table~\ref{tab:titans_giants}. Seven stars are CEMP, whereas 28 stars are not carbon-enriched.
The Kiel and HR diagrams of the stellar sample are presented in Fig.~\ref{fig:kiel}, where the parameters are those derived in this work (Table~\ref{tab:results}).

\begin{table*}
\caption{Preliminary parameters from the literature. The "Ref" column indicates 
the stellar parameter reference. 
The "Inst." column specifies the 
origin of the spectra used in the present paper:
HERMES (HE), HARPS (HA), UVES (UV) or FEROS (FE). 
The $E(B-V)$ column provides reddening values extracted from \cite{Capitanio2017} when available, or alternatively from \cite{schlafly2011ApJ...737..103S}. 
CEMP-s and -rs stand for carbon-enriched metal poor stars enriched in either s, or in a mixture of s and r elements, 
following the classification in
\cite{karinkuzhi2021A&A...645A..61K}.}
\label{tab:titans_giants}
\centering
\tiny 
\begin{threeparttable}
\begin{tabular}{lccccccccc}
\hline\hline
Name & \teff\ (K) & \logg\ & [Fe/H] & $\theta$ (mas) & Ref(\teff/\logg/[Fe/H]/$\theta$) & Inst. & $E(B-V)$\\
\hline
\multicolumn{8}{c}{Giants with interferometric measurements}\\
\hline
HD~2665 & $4883 \pm 95$ & $2.21 \pm 0.03$ & $-2.10 \pm 0.10$ & $0.395 \pm 0.004$ & K20 & HE & 0.024\\
HD~122563 & $4635 \pm 34$ & $1.40 \pm 0.04$ & $-2.75 \pm 0.12$ & $0.925 \pm 0.011$ & K20 & HE & 0.027\\
HD~175305 & $4850 \pm 118$ & $2.50 \pm 0.03$ & $-1.52 \pm 0.08$ & $0.484 \pm 0.006$ & K20 & HE & 0.011\\
HD~221170 & $4248 \pm 128$ & $1.25 \pm 0.04$ & $-2.40 \pm 0.13$ & $0.596 \pm 0.005$ & K20 & HE & 0.058\\
\hline
\multicolumn{8}{c}{Giants with direct IRFM measurements}\\
\hline
HD~45282 & $5299 \pm 87$ & 3.16 & $-1.45$ & $0.270 \pm 0.009$ & Ca10 & UV & 0.002\\
HD~175305 & $5059 \pm 80$ & $2.53 \pm 0.14$ & $-1.29$ & $0.447 \pm 0.006$ & Hw16 & HE & 0.011\\
HD~218857 & $5162 \pm 80$ & $2.66 \pm 0.32$ & $-1.78$ & $0.194 \pm 0.003$ & Hw16 & UV & 0.018\\
BPS BS 17569$-049$ & 4781 & 1.20 & $-2.88$ & -- & Ca21/Ca04/Ca04/-- & UV & 0.034\\
BPS CS 22953$-003$ & 5200 & 2.30 & $-2.84$ & -- & Ca21/Ca04/Ca04/-- & UV & 0.015\\
BPS~CS~22892$-052$ & 4941 & 1.60 & $-3.03$ & -- & Ca21/Ca04/Ca04/-- & UV & 0.027\\
\hline
\multicolumn{8}{c}{Giants with asteroseismic measurements}\\
\hline
HD~3179 & $5290 \pm 105$ & 2.58  & $-1.00$ & -- & Ho21/St18/PA/-- & FE & 0.017\\
HD~13359 & $5322 \pm 106$ & 2.48 & $-0.98$ &  -- & Ho21/St18/PA/-- & FE/UV & 0.014\\
HD~17072 & $5364 \pm 107$ & 2.61 & $-0.98$ & -- & Ho21/St18/PA/-- & UV & 0.004\\
HD~221580 & $5244 \pm 104$ & 2.40 & $-1.20$ & -- & Ho21/St18/PA/-- & HA & 0.009\\
TIC 168924748 & $5148 \pm 102$ & 2.68 & $-0.61$ & -- & Ho21/Bu21/Ca21/--  & -- & 0.096\\
TIC 404605506 & $5228 \pm 104$ & 2.44 & $-0.64$ & -- & Ho21/Bu21/Ca21/--  & -- & 0.057\\
\hline
\multicolumn{8}{c}{Other giants}\\
\hline
HD~6229 & 5055 & 2.12 & $-1.21$ & -- & Ar19 & UV & 0.047\\
HD~105740 & 4718 & 2.76 & $-0.66$ & -- & Ar19 & UV & 0.022\\
HD~115444 & 4500 & 0.70 & $-3.15$ & -- & Jh12 & HE & 0.012\\
HD~186478 & 4716 & 1.57 & $-2.31$ & -- & Ar19 & UV & 0.150\\
BD$+09$ 2860 & 5369 & 2.57 & $-1.58$ & -- & Ar19 & UV & 0.025\\
BD$+09$ 2870 & 4664 & 1.47 & $-2.34$ & -- & Ar19 & UV & 0.025\\
BD+11 2998 & 5480 & 3.00 & $-1.12$ & --  & Kl12  & UV & 0.031\\
BD$+17$~3248 & 5240 & 2.44 & $-2.07$ & -- & Cr08 & UV & 0.051\\
BD$-18\,5550$ & 4750 & 1.40 & $-3.06$ & -- & Ca04 & UV & 0.172\\
CD$-30$ 298 & 5223 & 2.92 & $-3.50$ & -- & Be17 & UV & 0.020\\
CD$-41$ 15048 & 5726 & 2.04 & $-2.24$ & -- & Be17 & UV & 0.013 \\
CD$-62$~1346 & 5163 & 1.96 & $-1.54$ & -- & Ar19 & UV & 0.032\\
BPS CS 22186$-025$  & 4900 & 1.25 & $-3.00$ & -- & Ca04 & UV & 0.025\\
BPS CS 22189$-009$  & 4900 & 1.70 & $-3.49$ & -- & Ca04 & UV & 0.022\\
BPS CS 22891$-209$  & 4700 & 1.00 & $-3.29$ & -- & Ca04 & UV & 0.066\\
BPS CS 22896$-154$  & 5250 & 2.70 & $-2.69$ & -- & Ca04  & UV & 0.047\\
BPS CS 22897$-008$  & 4900 & 1.70 & $-3.41$ & -- & Ca04  & UV & 0.014\\
BPS CS 22949$-048$  & 4800 & 1.50 & $-3.25$ & -- & Sp11  & UV & 0.031\\
BPS CS 22956$-050$  & 4900 & 1.70 & $-3.33$ & -- & Ca04  & UV & 0.029\\
BPS CS 22966$-057$  & 5300 & 2.20 & $-2.62$ & -- & Ca04  & UV & 0.016\\
BPS CS 29491$-053$  & 4700 & 1.30 & $-3.04$ & -- & Ca04  & UV & 0.017\\
BPS CS 29502$-042$  & 5100 & 2.50 & $-3.19$ & -- & Ca04  & UV & 0.056\\
BPS CS 29518$-051$  & 5200 & 2.60 & $-2.69$ & -- & Ca04  & UV & 0.013\\
BPS CS 30325$-094$  & 4950 & 2.00 & $-3.30$ & -- & Ca04  & UV & 0.036\\
BPS~CS~31082$-001$ & 4876 & 1.80 & $-2.81$  & -- & Ha18 & UV & 0.014\\
\hline
\multicolumn{8}{c}{CEMP giants}\\
\hline
HD~26 (CEMP-s) & 5169 & $2.46$ & $-0.98$ & --  & Kk21 & HE & 0.044\\
HD~5223 (CEMP-rs) & 4650 & 1.03 & $-2.00$ & -- & Kk21 & HE & 0.037\\
HD~76396 (CEMP-rs) & 4750 & 2.00 & $-2.27$ & -- & Kk21 & HE & 0.025\\
HD~196944 (CEMP-rs) & 5168 & 1.28 & $-2.50$ & -- & Kk21 & HE & 0.025\\
HD~201626 (CEMP-s) & 5084 & 2.18 & $-1.75$ & -- &  Kk21 & HE & 0.056\\
HD~224959 (CEMP-rs) & 4969 & 1.26 & $-2.36$ & -- & Kk21 & HE & 0.029\\ 
HE~0507-1653 (CEMP-s) & 5035 & 2.39 & $-1.35$ & -- & Kk21 & HE & 0.048\\
\hline
\end{tabular}
\begin{tablenotes}
\item{} \textbf{Notes.} {
HD~175305 is repeated twice because it has both interferometric and IRFM data.
}
\item{} \textbf{References.} {K20: \cite{karovicova2020A&A...640A..25K}; C10: \cite{Casagrande2010}; Hw16: \cite{hawkins2016A&A...592A..70H}; Ca21: \cite{casagrande2021MNRAS.507.2684C}; Ca04: \cite{cayrel2004A&A...416.1117C}; 
Sp11: \cite{spite2011A&A...528A...9S}; Bu21: \cite{buder2021MNRAS.506..150B}; Ho21: \cite{Hon2021ApJ...919..131H}; St18: \cite{Stassun2018AJ....156..102S}; Bu21: \cite{buder2021MNRAS.506..150B}; PA: the PASTEL catalogue \cite{soubiran2016A&A...591A.118S}; Kk21 \cite{karinkuzhi2021A&A...645A..61K}; Ar19: \cite{arentsen2019A&A...627A.138A}; Jh12: \cite{johnson2002ApJS..139..219J}; Kl12: \cite{koleva2012A&A...538A.143K}; Cr08: \cite{carney2008AJ....135..196C}; Ha18: \cite{hansen2018ApJ...858...92H}; Be17: \cite{beers2017ApJ...835...81B}.
} 
\end{tablenotes}
\end{threeparttable}
\end{table*}

\FloatBarrier
\begin{table*}
\caption{Determined atmospheric parameters of the giant Titans.}
\label{tab:results}
\centering
\tiny 
\begin{threeparttable}
\begin{tabular}{lccccccrc}
\hline\hline
Name & \teffa~[K] & other \teff~[K] & \logg & [\ion{Fe}{i}/H]$_{\mathrm{LTE}}$ & [\ion{Fe}{ii}/H]  & log~($L/L_\odot$) & $R\; [R_{\odot}]$ & A(Mg)$_{\mathrm{NLTE}}$\\
\hline
\multicolumn{9}{c}{Giants with interferometric measurements}\\
\hline
HD~2665 & $5007 \pm 65$ & $4990\pm66\,^{\mathrm{Ph}}$ & $1.97\pm0.15$ &  $-2.07 \pm 0.07$ & $-1.99 \pm 0.08$   & $1.833 \pm 0.004$ & $10.96 \pm 0.29$ & $5.81\pm0.09$\\
HD~122563 & $4615 \pm 69$ & $4635\pm34\,^{\mathrm{IntK20}}$ & $1.09\pm0.15$ &  $-2.81 \pm 0.07$ & $-2.71 \pm 0.03$  & $2.677 \pm 0.010$ & $34.10 \pm 1.08$ &$5.34\pm0.08$\\
HD~175305 & $5099 \pm 59$ & $5059\pm80\,^{\mathrm{Hk}}$  & $2.46\pm0.15$ & $-1.35 \pm 0.07$ & $-1.35 \pm 0.03$  & $1.534 \pm 0.004$ & $7.49 \pm 0.18$ & $6.52\pm0.08$\\
HD~$221170$ & $4567 \pm 45$ & -- & $1.11\pm0.15$  & $-2.09 \pm 0.06$ & $-2.07 \pm 0.04$  & $2.636 \pm 0.007$ & $33.22 \pm 0.70$ & $5.87\pm0.08$\\
\hline
\multicolumn{9}{c}{Giants with direct IRFM}\\
\hline
HD 45282 & $5344 \pm 54$ & $5299\pm87\,^{\mathrm{Ca10}}$ & $3.05\pm0.15$ & $-1.17 \pm 0.08$ & $-1.32 \pm 0.05$  & $0.941 \pm 0.002$ & $3.45 \pm 0.07$ & $6.55\pm0.08$\\
HD~175305 & $5099 \pm 59$ & $5059\pm80\,^{\mathrm{Hk}}$ & $2.46\pm0.15$  & $-1.35 \pm 0.07$ & $-1.35 \pm 0.03$  & $1.534 \pm 0.010$ & $7.49 \pm 0.18$ & $6.52\pm0.08$\\
HD~$218857$ & $5179 \pm 50$ & $5162\pm80\,^{\mathrm{Hk}}$ & $2.58\pm0.15$  & $-1.92 \pm 0.06$ & $-1.87 \pm 0.05$  & $1.543 \pm 0.003$ & $7.34 \pm 0.14$ & $6.00\pm0.08$\\
BPS BS 17569$-049$ & $4790 \pm 55$ & $4781 \pm 66\,^{\mathrm{Ca21}}$ & $1.04\pm0.15$  & $-2.48 \pm 0.15$ & $-2.37 \pm 0.05$  & $2.462 \pm 0.006$ & $24.71 \pm 0.59$ & $5.68\pm0.08$\\
BPS CS 22953$-003$ & $5155 \pm 47$ & $5200 \pm 66\,^{\mathrm{Ca21}}$ & $1.95\pm0.15$  & $-2.58 \pm 0.16$ & $-2.51 \pm 0.07$ & $1.795 \pm 0.003$ & $9.89 \pm 0.18$ & $5.22\pm0.07$\\
BPS~CS~22892$-052$ & $5017 \pm 161$ & $4941 \pm 66\,^{\mathrm{Ca21}}$ & $1.76\pm0.15$ & $-2.72 \pm 0.13$ & $-2.72 \pm 0.06$  &  $2.198 \pm 0.010$ & $16.61 \pm 1.04$ &  $5.05\pm0.11$\\
\hline
\multicolumn{9}{c}{Giants with asteroseismic measurements}\\
\hline
HD~3179 & $5319 \pm 81$  & $5399\pm66\,^{\mathrm{Ph}}$ & $2.25 \pm0.15$  & $-1.11 \pm 0.06$ & $-0.78 \pm 0.02$ & $1.665 \pm 0.004$ & $8.00 \pm 0.24$ & $7.02\pm0.09$\\ 
HD~13359 & $5353 \pm 66$ & $5340\pm66\,^{\mathrm{Ph}}$ &$2.58\pm0.15$  & $-1.00 \pm 0.06$ & $-0.79 \pm 0.03$  & $1.658 \pm 0.003$ & $7.84 \pm 0.19$ & $6.06\pm0.08$\\
HD~17072 & $5363 \pm 56$ & $5402\pm66\,^{\mathrm{Ph}}$ & $2.39\pm0.15$ & $-1.11 \pm 0.08$ & $-0.79 \pm 0.05$ & $1.660 \pm 0.003$ & $7.83 \pm 0.16$ & $6.96\pm0.08$\\
HD~221580 & $5330 \pm 44$ & $5280\pm66\,^{\mathrm{Ph}}$ & $2.46\pm0.15$ & $-1.17 \pm 0.10$ & $-0.97 \pm 0.03$  & $1.857 \pm 0.003$ & $9.95 \pm 0.17$ & $6.77\pm0.08$\\
TIC 168924748 & -- & $5331 \pm 66\,^{\mathrm{Ca21}}$ & $2.47 \pm 0.05\star$  & $-0.61 \pm 0.07$* & --  & -- & -- & -- \\
TIC 404605506 & -- & $5050 \pm 66\,^{\mathrm{Ca21}}$ & $2.40 \pm 0.04\star$ &  $-0.64 \pm 0.04$* & --  & --& -- & -- \\
\hline
\multicolumn{9}{c}{Other giants}\\
\hline
HD~$6229$ & $5250 \pm 62$ & $5292\pm66\,^{\mathrm{Ph}}$ & $2.35\pm0.15$  & $-0.99 \pm 0.05$ & $-0.87 \pm 0.02$  & $1.828 \pm 0.004$ & $9.91 \pm 0.24$ &$6.91\pm0.08$\\
HD~$105740$ & $4743 \pm 41$ & $4737\pm66\,^{\mathrm{Ph}}$ & $2.38\pm0.15$  & $-0.42 \pm 0.11$ & $-0.64 \pm 0.06$ & $1.527 \pm 0.005$ & $8.59 \pm 0.15$& $7.53\pm0.08$\\
HD~$115444$ & $4667 \pm 86$ & -- & $1.55\pm0.15$ & $-3.10 \pm 0.06$ & $-2.77 \pm 0.05$  & $2.359 \pm 0.010$ & $23.11 \pm 0.88$ &$5.07\pm0.09$\\
HD~$186478$ & $4804 \pm 75$ & $4707\pm66\,^{\mathrm{Ph}}$ & $1.59\pm0.15$ & $-2.58 \pm 0.08$ & $-2.37 \pm 0.03$ &  $2.511 \pm 0.007$ & $25.99 \pm 0.83$ & $5.54\pm0.09$\\
BD$+09\,2860$ & $5408 \pm 41$ & $5434\pm66\,^{\mathrm{Ph}}$ & $2.26\pm0.15$  & $-1.72 \pm 0.07$ & $-1.45 \pm 0.06$ &  $2.062 \pm 0.002$ & $12.23 \pm 0.19$ & $6.28\pm0.07$\\
BD$+09\,2870$ & $4725 \pm 80$ & $4701\pm66\,^{\mathrm{Ph}}$ & $1.38\pm0.15$  &$-2.41 \pm 0.06$ & $-2.27 \pm 0.05$ &  $2.550 \pm 0.009$ & $28.10 \pm 0.98$ & $5.61\pm0.07$\\
BD$+11\,2998$  & $5607 \pm 50$ & $5444\pm66\,^{\mathrm{Ph}}$ & $2.33\pm0.15$  & $-1.02 \pm 0.06$ & $-0.84 \pm 0.07$ &  $1.650 \pm 0.002$ & $7.08 \pm 0.13
$ & $7.23\pm0.08$\\
BD$+17\,3248$ & $5301 \pm 130$ & $5330\pm66\,^{\mathrm{Ph}}$ & $2.04\pm0.15$  & $-1.96 \pm 0.08$ & $-1.85 \pm 0.07$ & $2.165 \pm 0.006$ & $14.33 \pm 0.69$& $6.12\pm0.09$\\
BD$-18\,5550$ & $5030 \pm 76$ & $4857\pm66\,^{\mathrm{Ph}}$ & $1.36\pm0.15$ &  $-2.64 \pm 0.11$ & $-2.59 \pm 0.05$  &  $2.083 \pm 0.005$ &$14.48 \pm 0.44$ & $5.15\pm0.08$\\
CD$-30\,298$ & $5206 \pm 66$ & $5299\pm66\,^{\mathrm{Ph}}$ & $2.49\pm0.15$  & $-3.36 \pm 0.10$ & $-3.14 \pm 0.12$ &  $1.505 \pm 0.003$ & $6.95 \pm 0.18$ & $4.43\pm0.08$ \\
CD$-41\,15048$ & $5770 \pm 59$ & $5739\pm66\,^{\mathrm{Ph}}$ & $2.84\pm0.15$ & $-1.86 \pm 0.08$ & $-1.51 \pm 0.04$ &  $1.812 \pm 0.002$ & $8.05 \pm 0.16$ & $5.98\pm0.10$\\
CD$-62\,1346$ & $5435 \pm 53$ & $5424\pm66\,^{\mathrm{Ph}}$ & $2.22\pm0.15$ &  $-1.45 \pm 0.04$ & $-1.11 \pm 0.07$ & $1.829 \pm 0.003$ & $9.26 \pm 0.18$ & $6.44\pm0.08$\\
BPS CS 22186$-025$  & $5036 \pm 41$  & $5034\pm66\,^{\mathrm{Ph}}$ & [$1.69\pm0.15$] & $-2.77 \pm 0.16$ & $-2.62 \pm 0.05$ &  $2.128 \pm 0.003$ & $15.22 \pm 0.25$& $5.21\pm0.07$\\
BPS CS 22189$-009$  & $4947 \pm 141$  & $4992\pm66\,^{\mathrm{Ph}}$ & [$1.87\pm0.15$]  & $-3.18 \pm 0.17$ & $-3.22 \pm 0.12$ & $1.944 \pm 0.009$ & $12.75 \pm 0.71$& $4.50\pm0.12$\\
BPS CS 22891$-209$  & $4896 \pm 132$ & $4782\pm66\,^{\mathrm{Ph}}$ & [$1.37\pm0.15$] & $-3.01 \pm 0.16$ & $-2.93 \pm 0.07$ & $2.520 \pm 0.010$ &$25.28 \pm 1.35$ & $4.86\pm0.11$\\
BPS CS 22896$-154$  & $5274 \pm 53$ & $5293\pm66\,^{\mathrm{Ph}}$ & $2.61\pm0.15$ &  $-2.62 \pm 0.14$ & $-2.66 \pm 0.04$ & $1.636 \pm 0.002$ & $7.88 \pm 0.16$ & $5.33\pm0.08$\\
BPS CS 22897$-008$  & $4984 \pm 139$ & $4954\pm66\,^{\mathrm{Ph}}$ & $1.86\pm0.15$ & $-3.25 \pm 0.09$ & $-3.21 \pm 0.05$ &  $2.026 \pm 0.009$ & $13.82 \pm 0.76$ & $4.71\pm0.11$\\
BPS CS 22949$-048$  & $4864 \pm 119$ & $4872\pm66\,^{\mathrm{Ph}}$ & [$1.93\pm0.15$]   & $-3.11 \pm 0.14$& $-2.87 \pm 0.05$ & $1.799 \pm 0.009$ & $11.16 \pm 0.54$ & $4.69\pm0.11$\\
BPS CS 22956$-050$  & $4960 \pm 187$ & $5009\pm66\,^{\mathrm{Ph}}$ & [$1.75\pm0.15$]  & $-3.16 \pm 0.13$ & $-3.03 \pm 0.09 $ & $2.072 \pm 0.012$ & $14.71 \pm 1.07$& $4.87\pm0.14$\\
BPS CS 22966$-057$  & $5469 \pm 146$ & $5543\pm66\,^{\mathrm{Ph}}$ & $3.36\pm0.15$ & $-2.34 \pm 0.09$ & $-2.28 \pm 0.05$ &  $0.760 \pm 0.003$ & $2.67 \pm 0.14$& $5.55\pm0.11$\\
BPS CS 29491$-053$  & $4775 \pm 135$ & $4831\pm66\,^{\mathrm{Ph}}$ & $1.56\pm0.15$ & $-2.93 \pm0.13$ & $-2.75 \pm 0.09$ &  $2.342 \pm 0.013$ & $21.66 \pm 1.22$ & $5.11\pm0.11$\\
BPS CS 29502$-042$  & $5301 \pm 52$ & $5296\pm66\,^{\mathrm{Ph}}$ & $2.70\pm0.15$ & $-3.03 \pm 0.08$ & $-2.85 \pm 0.07$ &  $1.362 \pm 0.002$ & $5.68 \pm 0.11$ & $4.97\pm0.07$\\
BPS CS 29518$-051$  & $5340 \pm 91$ & $5333\pm66\,^{\mathrm{Ph}}$ & $2.29\pm0.15$ & $-2.51 \pm 0.12$ & $-2.47 \pm 0.04$ & $1.522 \pm 0.004$ & $6.74 \pm 0.23$& $5.43\pm0.09$\\
BPS CS 30325$-094$ & $5121 \pm 111$ & $5129\pm66\,^{\mathrm{Ph}}$ & $1.88\pm0.15$ & $-3.16 \pm 0.10$ & $-3.08 \pm 0.09$ & $1.941 \pm 0.006$& $11.86 \pm 0.51$ & $4.91\pm0.09$\\
BPS~CS~31082$-001$ & $5036 \pm 41$ & $5020\pm66\,^{\mathrm{Ph}}$ & $1.96\pm0.15$ & $-2.64 \pm 0.11$ & $-2.39 \pm 0.12$ &  $2.006 \pm 0.003$ & $13.23 \pm 0.22$& $5.35\pm0.07$\\
\hline
\multicolumn{9}{c}{CEMP giants}\\
\hline
HD~$26$ & $5125 \pm 40$ & -- & $2.21\pm0.15$  & $-0.77 \pm 0.05$ & $-0.67 \pm 0.02$ &  $1.778 \pm 0.003$ & $9.82 \pm 0.16$ & $7.47\pm0.08$\\
HD~$5223^\S$ & $4687 \pm 40$ & -- & $1.27\pm0.15$  & $-1.56 \pm 0.04$ & $-1.57 \pm 0.02$ & $2.232 \pm 0.005$ & $19.83 \pm 0.36$ & $5.96\pm0.08$\\
HD~$76396$ & $5064 \pm 72$ & -- &  $2.40\pm0.15$  & $-1.47 \pm 0.05$ & $-1.55 \pm 0.02$ & $1.857 \pm 0.004$ & $11.02 \pm 0.31$ & $6.41\pm0.09$\\
HD~$196944$ & $5539 \pm 41$ & -- & $2.44 \pm 0.15$  & $-2.05 \pm 0.06$ & $-2.04 \pm 0.03$ &  $2.106 \pm 0.002$ & $12.26 \pm 0.18$ & $5.80\pm0.07$\\
HD~$201626$ & $5008 \pm 40$ & -- & $2.15 \pm0.15$   & $-1.66 \pm 0.05$ & $-1.72 \pm 0.03$ & $1.779 \pm 0.003$ & $10.29 \pm 0.17$ & $6.36\pm0.08$\\
HD~$224959$ & $4852 \pm 40$ & -- & $1.96\pm0.15$  & $-2.42 \pm 0.09$ & $-2.18 \pm 0.03$ &  $1.823 \pm 0.004$ & $11.53 \pm 0.20$& $5.68\pm0.08$\\
HE~0507-1653 & $5228 \pm 139$ & -- & $2.93\pm0.15$  & $-1.17 \pm 0.07$ & $-1.30 \pm 0.12$ &  $1.491 \pm 0.006$ & $6.78 \pm 0.35$& $6.56\pm0.12$\\
\hline
\end{tabular}
\begin{tablenotes}
\item{} \textbf{Notes.} {Second column lists \teffa\ with its internal errors, i.e. the method precision. For the total uncertainties, the estimated model accuracy error of $\pm$48~K (as discussed in Sect.~\ref{sec:accuracy_teff}) must be added in quadrature. Third column lists alternative temperatures consistent with \teffa\ scale. The method by which they were derived and their corresponding sources are indicated with the following codes: IntK20: direct interferometry in \cite{karovicova2020A&A...640A..25K}. Ph: IRFM-based photometric calibrations in \cite{casagrande2021MNRAS.507.2684C} using Gaia $B_p - R_p$ colours, its errors are those of the calibration in the corresponding source. Hk: direct IRFM in \cite{hawkins2016A&A...592A..70H}. Ca10: stands for direct IRFM in \cite{Casagrande2010}. Ca21: direct IRFM in \cite{casagrande2021MNRAS.507.2684C}. Fourth column lists surface gravities derived from Mg triplet lines (values in Table~\ref{tab:logg_Mg} averaged), determined accurate in Sect.~\ref{sec:logg_accuracy}.
Fifth column lists iron abundances from neutral lines under LTE, they have been proven to be biased to low values in Sect.~\ref{sec:fe_LTE}, thus they must be used only for spectral modeling under LTE.
Sixth column lists iron abundances from ionised lines under LTE, they were determined to be equivalent to abundances derived under non-LTE in Sect~\ref{sec:fe_NLTE}, thus they should be preferred for calibration purposes, as they are closer to real.
Seventh column lists luminosities. Eighth column lists non-LTE corrected Mg abundances, to which $-0.07$~dex must be added to reproduce LTE spectra; see Sect.~\ref{sec:magnesium}. The star HD~175305, which has  both interferometric and IRFM temperature measurements is repeated for compatibility with Table~\ref{tab:titans_giants}. The symbol $\S$ indicates that lines (log ($EW/\lambda) > -5.05$) were included to derive [\ion{Fe}{i}/H] and [\ion{Fe}{ii}/H]. [\ion{Fe}{i}/H] values indicated with an asterisk (*) were not derived here but were extracted from \cite{buder2021MNRAS.506..150B}.
\logg\ values in brackets are \loggiso\ offset corrected values, see Sect.~\ref{sec:logg_accuracy}.
\logg\ values accompanied by the symbol ($\star$) were derived from asteroseismologic measurements in Sect.~\ref{sec:astero}.}
\end{tablenotes}
\end{threeparttable}
\end{table*}
\FloatBarrier

\section{Data reduction}
\label{sec:reduction}
High-resolution spectra were acquired with HERMES, FEROS, HARPS or UVES instruments, as indicated in seventh column of Table~\ref{tab:titans_giants}.
For HERMES, the Doppler correction was performed by cross-correlating the stellar spectra with a mask covering the wavelength range 4800 - 6500~\AA\ and mimicking the spectrum of Arcturus (K1.5 III). The restricted wavelength span is to avoid both telluric lines at the red end and the crowded and poorly exposed blue end of the spectra.
The HERMES spectra were reduced with an automated pipeline, merging the different orders and correcting for the blaze function of the echelle grating as well as for the Earth motion around the solar-system barycenter.

UVES, FEROS, and HARPS 1D spectra were Doppler corrected to the rest-frame wavelength scale using iSpec \citep{blanco-cuaresma2014}; see details in Paper~I. 
It was not possible to correct the most metal-poor spectra with this tool because of the very few available metal lines, which are used by the algorithm to cross-correlate with the template. 
They were thus corrected using the IRAF\footnote{IRAF is distributed by the National Optical Astronomy Observatory, which is operated by the Association of Universities for Research in Astronomy (AURA) under a cooperative agreement with the National Science Foundation: \url{http://ast.noao.edu/data/software}}
tasks \textit{fxcor} and \textit{dopcor} taking as template a spectrum of BPS~CS~31082$-001$ corrected with iSpec. 
All spectra were globally normalised by spline polynomials with iSpec.  More precise local re-normalizations around the lines of interest were applied later during the element abundance determination process.

\section{Atmospheric parameters}
\label{sec:parameters}

The atmospheric parameters \teff, [Fe/H], and \logg\ were derived by the method described in Paper~I.
It consists on iterative loops 
with the following steps: 
derivation of \teff\ through H$\alpha$ profile fitting (hereafter \teffa); [Fe/H] determination through spectral synthesis; and determination of \logg\ through isochrone fitting (hereafter \logg$_{\mathrm{iso}}$).
Each parameter is derived while keeping the other ones fixed, and the procedure is iterated until the \teffa\ variation does not exceed its fitting error. 

H$\alpha$ is virtually insensitive to typical \logg\ and [Fe/H] errors (see Sect.~\ref{sec:teff}),
therefore we used \logg\ and [Fe/H] values from the literature as priors in the first loop to get the first \teffa\ guess.
This first \teffa\ can be very different than temperature values from the literature (see Sect.\ref{sec:literature}) because custom methods are prone to biases due to their high parameter interdependence.
Therefore, \logg\ and [Fe/H] in the first loop are likely to highly vary with respect to literature values.
The temperatures constrained in subsequent loops vary little, 
no more than few tens of Kelvins;
thus loops are basically run to tune the parameters.
In Sect.~\ref{sec:logg_accuracy} we verify that \logg\ from Mg~I~b triplet lines is more reliable than the isochrone fitting outcomes;
for this reason, we run a final loop using the former method instead of the later.
Table~\ref{tab:results} lists our final parameters, the determination of which is fully described in this section, whereas their scrutiny is performed in Sect.~\ref{sec:accuracy}.

\subsection{Effective temperature}
\label{sec:teff}

We described in detail the method to derive \teffa\ in Paper~I, where we also determined the accuracy of 3D non-LTE theoretical H$\alpha$ profiles \citep{amarsi2018} in dwarf metal-poor stars. We refer the reader to \cite{giribaldi2019} for technical aspects on the normalization-fitting procedure of H$\alpha$ line profiles.
Here we employ the same method on our sample of giants.

\begin{figure*}
    \centering
    \includegraphics[width=1\linewidth]{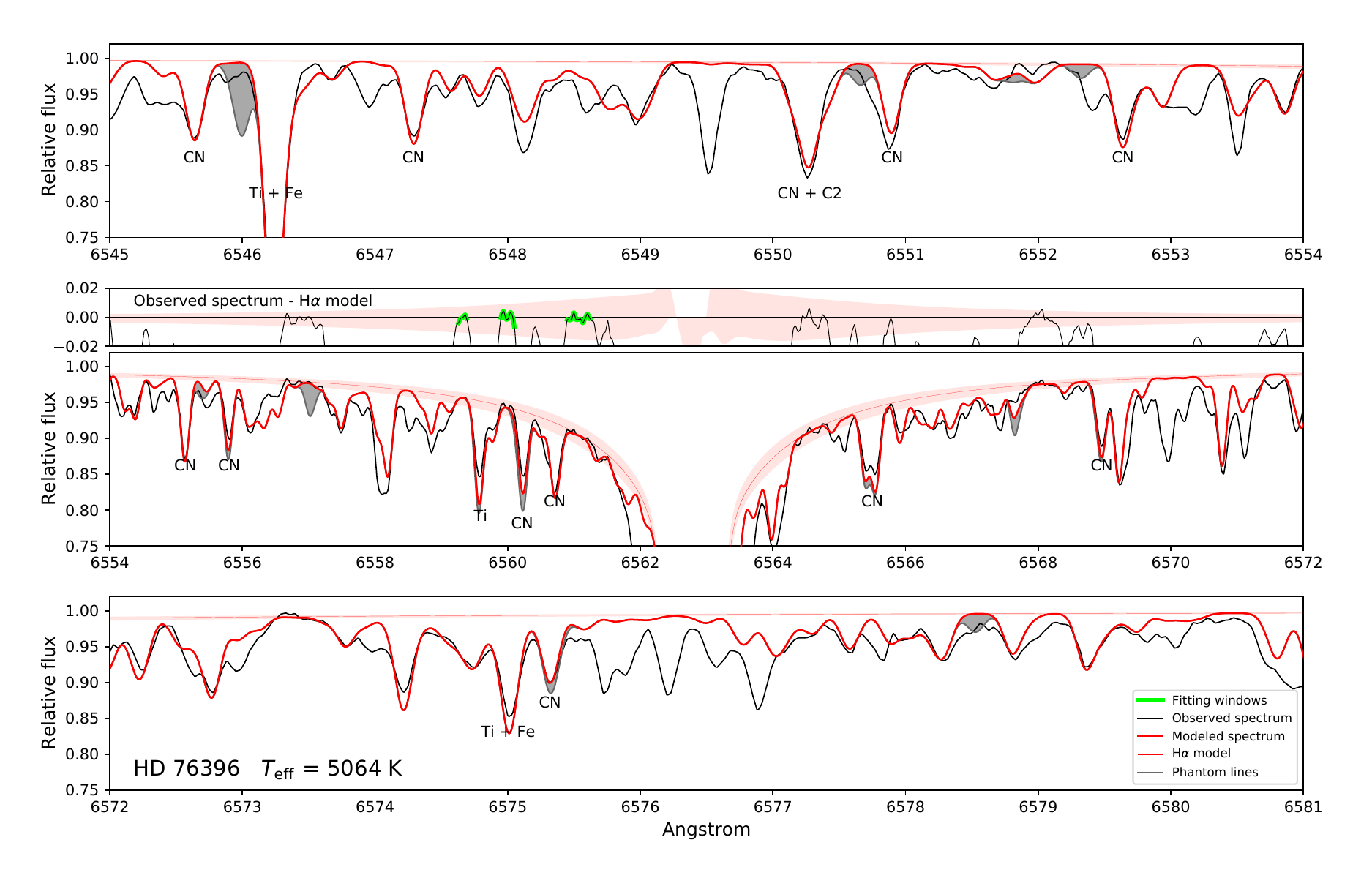}
    \caption{\tiny Observed spectra of the CEMP star HD 76396 compared with its modeled spectra. The upper and lower panels are used to ensure that a proper normalization of the observed spectrum is performed in the H$\alpha$ region illustrated in the middle panel. Pink shades represent variations up to $\pm200$~K to  provide a view of the flux sensitivity to \teff\ along the wavelength range. Gray shades indicate transitions not appearing in the observational spectrum, which were eliminated from the synthesis line lists, according to the procedure described in \cite{giribaldi2023ExA....55..117G}. The fitting windows used are highlighted in green in the residual plot in the second panel. Some atomic and molecular characteristics are labeled in the plots.}
    \label{fig:HD76396}
\end{figure*}

\begin{figure*}
    \centering
    \includegraphics[width=1\linewidth]{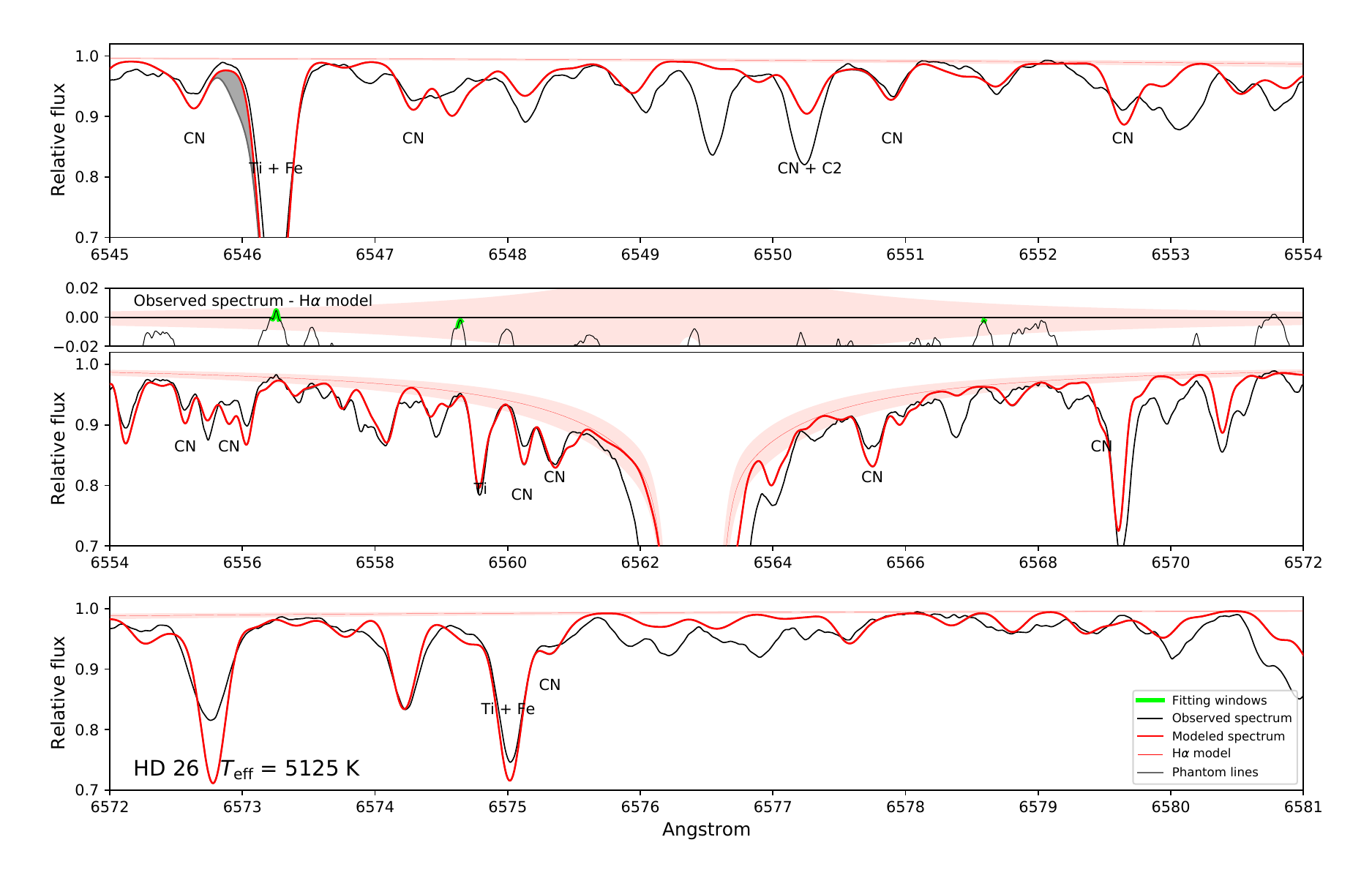}
    \caption{\tiny Observed spectra of the CEMP star HD~26 
    compared with its modeled spectra.
    The elements of the plots are the same as those in Fig.~\ref{fig:HD76396}.}
    \label{fig:HD26}
\end{figure*}

CEMP stars challenge the applicability of the H$\alpha$ fitting because they have narrow line profiles blended by CN and C$_2$ molecular features. 
To perform a proper application of this method, windows free from molecular contamination and tellurics were  manually identified in the neighborhood of the H$\alpha$ line, for each spectrum. 
We synthesize H$\alpha$ profiles at two temperatures separated by 200~K to assess the sensitivity of the windows selected, as displayed in Figs.~\ref{fig:Ha_interferometry}, \ref{fig:Ha_IRFM}, \ref{fig:Ha_asteroseismic}, \ref{fig:Ha_other_giants4}, and \ref{fig:Ha_CEMP}. 
For CEMP stars,  the windows widths were optimised using C, N, O, and Fe abundances derived from the atmospheric parameters in every loop. Figs.~\ref{fig:HD76396} and ~\ref{fig:HD26} show the performance of this procedure for the CEMP stars HD~76396 and HD~26, respectively. 
The former presents a small number of narrow fitting windows in the most sensitive wavelength ranges of its H$\alpha$ profile, note the location of wavelength regions marked in green within the pink shades in second panel of Fig.~\ref{fig:HD76396}.
The latter presents only three narrow windows, from which only one lies in a very sensitive region, see second panel of Fig.~\ref{fig:HD26}.
When the windows are too narrow, a bootstrap method is performed to robustly assess the effective temperature. Starting from the literature atmospheric parameters, the effective temperature is iteratively derived until the temperature difference of the two last ierations does not exceed the estimated uncertainty. When several spectra are available for a given star, the fitting windows are selected independently for each spectrum to avoid frequent tellurics contamination and artifacts.

As done in Paper~I, the uncertainty of \teffHa\ is given by the following expression:
\begin{equation}
    \label{eq:uncertainty}
    \centering
    \delta(T_{\mathrm{eff}}^{\mathrm{H}\alpha})^{2} = 
    \delta_{T_{\mathrm{eff}}-\mathrm{model}}^{2} +
    \delta_{T_{\mathrm{eff}}-\mathrm{fit}}^{2} +
    \delta_{T_{\mathrm{eff}}-\mathrm{inst}}^{2} +
    \delta_{T_{\mathrm{eff}}-\mathrm{log}\;g}^{2} +
    \delta_{T_{\mathrm{eff}}-\mathrm{[Fe/H]}}^{2}
\end{equation}

\noindent
where $\delta_{T_{\mathrm{eff}}-\mathrm{model}}$ is the uncertainty of the synthetic profile, 
$\delta_{T_{\mathrm{eff}}-\mathrm{fit}}$ is the uncertainty of the fitting, 
$\delta_{T_{\mathrm{eff}}-\mathrm{inst}}$ is the uncertainty induced by an instrument residual pattern,
$\delta_{T_{\mathrm{eff}}-\mathrm{log}\;g}$ is the uncertainty related to the interdependence between \teffHa\ and \logg, and 
$\delta_{T_{\mathrm{eff}}-\mathrm{[Fe/H]}}$ is the uncertainty related to the interdependence between \teffHa\ and [Fe/H].

According with the analysis in Sect.~\ref{sec:accuracy_teff}, we consider that  the temperature uncertainty due to errors on the model synthetic spectrum  can be estimated from the dispersion of the difference of \teffHa\ with \teff\ derived from IRFM and from interferometry: it amounts to  $\delta_{T_{\mathrm{eff}}-\mathrm{model}}=48$~K.
$\delta_{T_{\mathrm{eff}}-\mathrm{fit}}$ is given by the internal uncertainty of \teffHa. 
The compatibility between observational and synthetic profiles is evaluated by a temperature histogram, whose frequencies are associated to the temperatures of the synthetic profiles that best match every wavelength pixel within the fitting windows; see Figs.~\ref{fig:Ha_interferometry}, \ref{fig:Ha_IRFM}, \ref{fig:Ha_asteroseismic}, \ref{fig:Ha_other_giants4}, and \ref{fig:Ha_CEMP} where all fits of the stars in this work are compiled.
When only one spectrum is available, $\delta_{T_{\mathrm{eff}}-\mathrm{fit}}$ is given by the 1$\sigma$ dispersion of a Gaussian fitted to the histogram of temperatures (see details in Sect.~3.1 of Paper~I), whereas when more than one spectrum is available, $\delta_{T_{\mathrm{eff}}-\mathrm{fit}}$ equals to the 1$\sigma$ dispersion of the temperatures associated to every spectra.
For $\delta_{T_{\mathrm{eff}}-\mathrm{inst}}$, we adopted the value 33~K determined with UVES spectra in Paper~I. 

\begin{figure}
    \centering
    \includegraphics[width=0.85\linewidth]{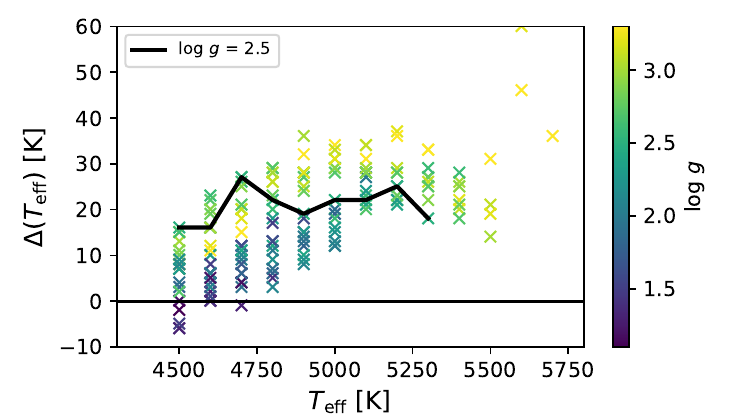}
    \includegraphics[width=0.85\linewidth]{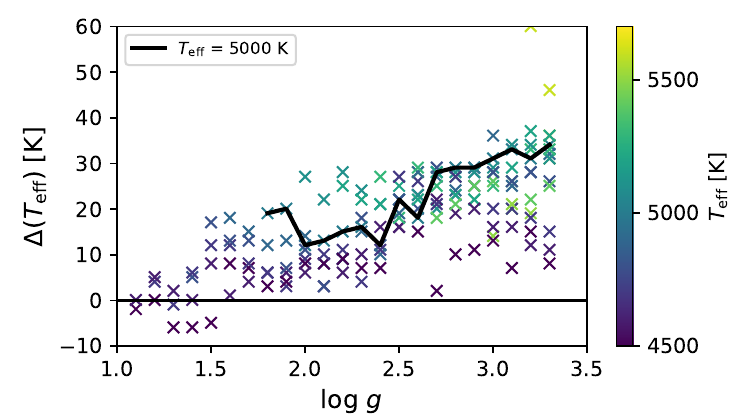}
    \caption{\tiny Temperature offset obtained by assuming a \logg\ bias of $+0.1$~dex.
    Differences are plotted as function of \teff\ (top panel), and  \logg\ (bottom panel). The cases for \teff\ $= 5000$~K and \logg\ $= 2.5$~dex are highlighted by black lines.}
    \label{fig:logg_map}
\end{figure}

We evaluated $\delta_{T_{\mathrm{eff}}-\mathrm{log}\;g}$ as follows. 
We simulated observational profiles by adding noise equivalent to $S/N = 300$ to our grid of synthetic 3D non-LTE H$\alpha$ profiles.
We derived their temperatures the same way as done with authentic observational profiles.
The exercise was done fitting profiles with [Fe/H] equal to $-1$, $-2$, and $-3$~dex, separately. Since similar results were obtained in each analysis, we present here that of [Fe/H] $= -2$~dex, to be used as proxy for the entire metallicity range covered by the stars in the present study.
The fittings were done within wavelength regions with high sensitivity to \teff.
For line profiles associated to \teff~$< 5400$~K we used the intervals [6557.8: 6561.5]~\AA\ and [6564.5: 6567.8]~\AA\ for the blue and red wings, respectively. For those associated to \teff~$\geq 5400$~K we restricted the regions to avoid the influence of intense line cores, thus the intervals used are [6557.8: 6560.0]~\AA\ and [6565.5: 6567.8]~\AA\ for the blue and red wings, respectively.
Here we did not use the actual \logg\ values of the profiles as inputs for the fittings, but modified by $+0.1$~dex.
The fittings produced biased temperatures, which are mapped as function of \teff\ and \logg\ in Fig.~\ref{fig:logg_map}.
We conclude that temperature offsets ($\Delta$\teff) due to 0.1 dex errors in \logg\ are typically within 30~K; as illustrated by  the representative case  \logg~= 2.5~dex and \teff~= 5000~K, identified with the black lines in Fig.~\ref{fig:logg_map}.
The highest offsets are present for the hottest stars.
This outcome occurs because degeneracy appears approximately for \teff\ $>5500$~K with \logg~$< 2.5$~dex, the effects of which become stronger as the \teff-\logg\ pair gets further away from the RGB evolutionary path in the Kiel diagram; this is, to the horizontal branch.
The top plot in the figure shows that for the \teff\ range in this work (hotter than 4500~K), stars with low \logg\ tend to be less sensitive to potential input \logg\  biases.
Namely, $\Delta$\teff~$\sim$15~K corresponds to \logg~$< 2.5$~dex, and $\Delta$\teff~$\sim$25~K corresponds to \logg~$\geq 2.5$~dex. We use this estimation as a practical rule for $\delta_{T_{\mathrm{eff}}-\mathrm{log}\;g}$.
In Sect.~\ref{sec:logg_accuracy} we estimate that the typical uncertainty of our surface gravities is 0.15~dex, thus the values above must be multiplied by 1.5 to obtain the corresponding errors: $\delta_{T_{\mathrm{eff}}-\mathrm{log}\;g}$ = 23~K for stars with \logg~$< 2.5$~dex and  $\delta_{T_{\mathrm{eff}}-\mathrm{log}\;g}$ = 38~K for stars with \logg~$> 2.5$~dex.

\begin{figure}
    \centering
    \includegraphics[width=0.85\linewidth]{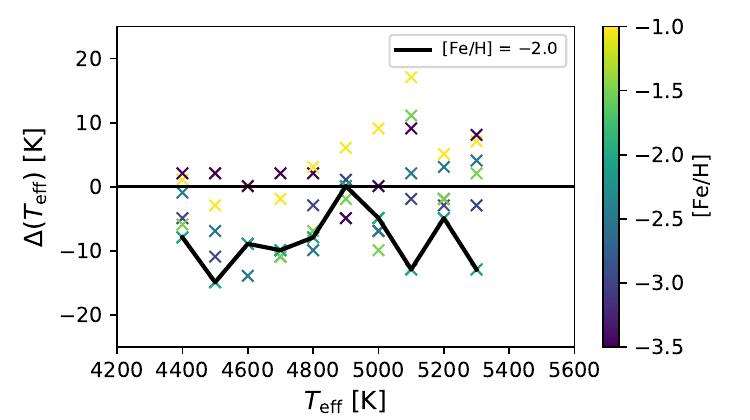}
    \includegraphics[width=0.85\linewidth]{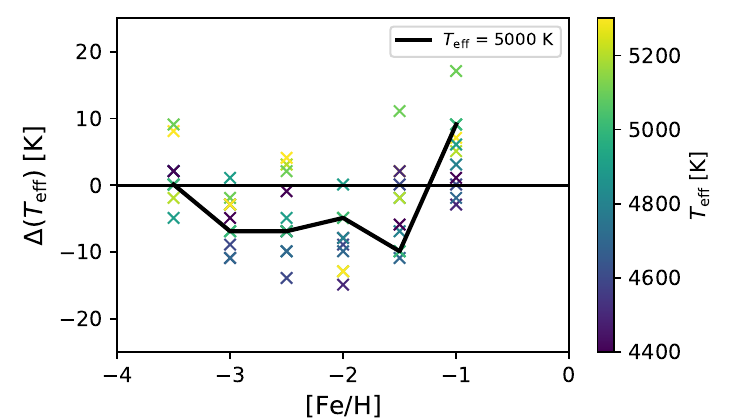}
    \caption{\tiny Temperature offset obtained by assuming a [Fe/H] bias of $+0.07$~dex.
    Differences are plotted as function of \teff\ (top panel), and [Fe/H] (bottom panel).  The cases for \teff\ $= 5000$~K and [Fe/H]] $= -2.0$~dex are highlighted by black lines.}
    \label{fig:feh_map}
\end{figure}

To evaluate $\delta_{T_{\mathrm{eff}}-\mathrm{[Fe/H]}}$ we ran the same procedure as above, but replacing \logg\ by [Fe/H].  
Figure~\ref{fig:feh_map} shows $\Delta$\teff\ induced by adding $+0.07$~dex to the actual metallicity of the simulated observational profiles.
This quantity, assumed as the typical metallicity uncertainty, is obtained from the top plot in Fig.~\ref{fig:Fe_dif}, where the dispersion ($\pm$0.14~dex) is attributed evenly to both [\ion{Fe}{i}/H] and [\ion{Fe}{ii}/H] measurements.
$\Delta$\teff\ in Fig.~\ref{fig:feh_map} is slightly shifted to negative values, typically not lower than $-10$~K. See for example, the cases of [Fe/H]~= $-2.0$~dex and \teff~= 5000~K represented by the black lines. The \teff\ dispersion in the plots is, however, dominated by the spectral noise, as $\delta_{T_{\mathrm{eff}}-\mathrm{fit}}$ associated to the fits in this test is typically $\pm$40~K. Therefore, $\delta_{T_{\mathrm{eff}}-\mathrm{[Fe/H]}}$ is hereafter neglected.  

\subsection{Metallicity  and microturbulence}
\label{sec:feh}
To derive metallicity, the spectral fitting was done using the iSpec package \citep{blanco-cuaresma2014} 
running the radiative transfer code Turbospectrum \citep{turbospectrum} with spherical MARCS model atmospheres \citep{gustafson2008} considering the atomic parameters excitation potential and oscillator strength in \cite{heiter2021A&A...645A.106H}.
As proceeded in Paper~I, we first derived the broadening parameters (microturbulence $v_{mic}$ and macroturbulence $v_{mac}$).
For that, the projected rotational velocity $v\mathrm{sin}~i$ was fixed to 1.6~km~s$^{-1}$ while $v_{mic}$, $v_{mac}$, and [Fe/H] were permitted to vary until the best global fit was obtained for all Fe lines. 
This give us the first $v_{mic}$ estimate, which is later tuned as explained below.
Several line lists from the literature were compiled to derive these parameters: \ion{Fe}{i} and \ion{Fe}{ii} for 
giants from \cite{jofre2014}, the "ASPL" and "MASH" lists of \ion{Fe}{i} from \cite{dutra-ferreira2016A&A...585A..75D}, and the \ion{Fe}{ii} line list from \cite{melendez_barbuy2009A&A...497..611M}.
Since our method does not assume ionization equilibrium, the abundance biases associated to some of these line lists and discussed in the papers mentioned above do not affect our results.
Actually, The capacity of our method to recover accurate \teff\ and \logg\ was evaluated by comparing its outcomes against those inferred by interferometric measurements in Paper~I.

As demonstrated in Sect.~\ref{sec:accuracy}, our set of \teff\ and \logg\ are statistically accurate, therefore the accuracy of our set of iron abundances is mainly subject to potential systematic errors resulting from our \ion{Fe}{ii} lines modeling\footnote{Neglecting systematic errors from spectral acquisition and reduction.}, 
carried out with 1D, spherically symmetric model atmospheres, assuming LTE.
Iron abundances were determined from \ion{Fe}{ii} lines, although abundances from \ion{Fe}{i} lines were also derived to quantify the offsets induced by the LTE modeling;
hereafter we refer to [\ion{Fe}{ii}/H] when mentioning metallicity.
Only weak lines were considered in our procedure to minimise 1D modeling the defects.
Namely, we restricted lines with reduced equivalent width (REW = log$(EW/\lambda)$\footnote{$EW$ is the equivalent width in \AA\ and $\lambda$ is the wavelength in \AA.}) lower than $-5$ to make sure we work in the linear part of the curve of growth.
For CEMP stars, we extended this upper limit to  $-4.80$ because very few weak and unblended lines were available. 

\begin{figure}
    \centering
    \includegraphics[width=0.95\linewidth]{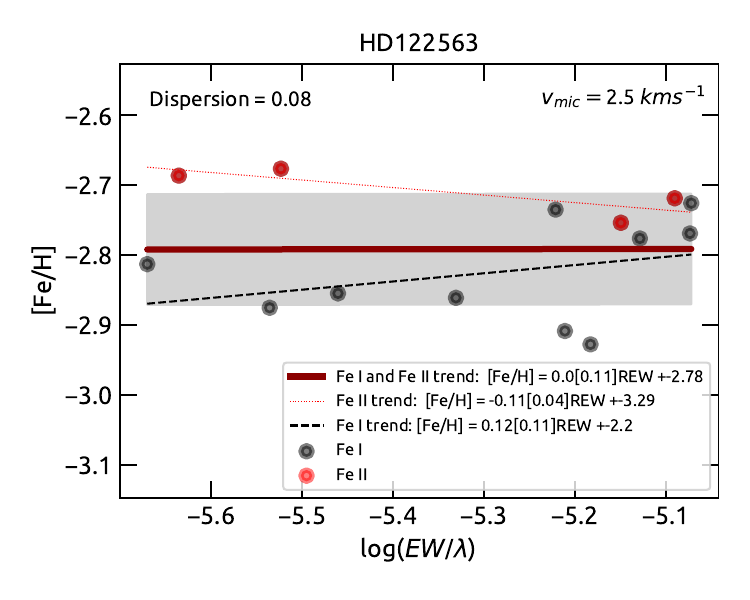}\\
    \includegraphics[width=0.95\linewidth]{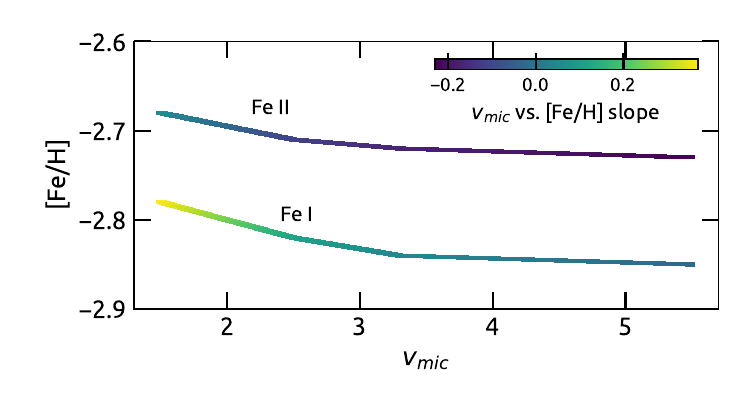}
    \caption{\tiny {\it Top panel:} For the star HD~122563, line-to-line Fe abundances computed with $v_{mic} = 2.5\, kms^{-1}$ as function of the reduced equivalent width.
    Gray and red dots represent \ion{Fe}{i} and \ion{Fe}{ii} lines, respectively.
    The solid dark red line is the trend of all \ion{Fe}{i} and \ion{Fe}{ii} measurements.
    The black dashed and red dotted lines are the trends of \ion{Fe}{i} and \ion{Fe}{ii} measurements, respectively.
    The shade represents the dispersion.
    {\it Bottom panel:} 
    For HD~122563, metallicity as function of $v_{mic}$ color-coded by the slope of the trends in the $v_{mic}$-[Fe/H] plane.
    }
    \label{fig:slopes}
\end{figure}

We verified the absence of correlation between [Fe/H] and REW for the determined $v_{mic}$ values using \ion{Fe}{i} and \ion{Fe}{ii} lines altogether.
In Sect.~\ref{sec:fe_LTE} we find that \ion{Fe}{i} abundances are generally lower than \ion{Fe}{ii} abundances under LTE.
This implies that $v_{mic}$ may vary depending of which group of lines are used for its determination: either neutral, ionised, or neutral and ionised lines.
On the other hand, it reinforces the justification of using only weak lines to derive Fe abundances, as they are little sensitive to $v_{mic}$, as already mentioned above.
We exemplify this concept with the star HD~122563 in Fig.~\ref{fig:slopes}. 
Top panel shows no correlation between [Fe/H] and REW when both \ion{Fe}{i} and \ion{Fe}{ii} are considered (solid red line). 
Only a small correlation appears when [Fe/H] is computed from \ion{Fe}{i} lines (black dashed line). 
The bottom panel in the figure show that this slope may be eliminated by increasing $v_{mic}$ to $\sim 3.30\,kms^{-1}$ or more.
With $v_{mic} = 3.30\,kms^{-1}$, the average [\ion{Fe}{i}/H] abundance would decrease from $-2.81$ to $-2.84$~dex, whereas  
[\ion{Fe}{ii}/H] would remain equal with $-2.71$~dex.
The total uncertainty of [Fe/H] is estimated in Sect.~\ref{sec:fe_LTE}.

\subsection{Surface gravity from Mg triplet lines}
\label{sec:Mg_triplet}

\begin{figure*}
    \centering
    \includegraphics[width=0.8\linewidth]{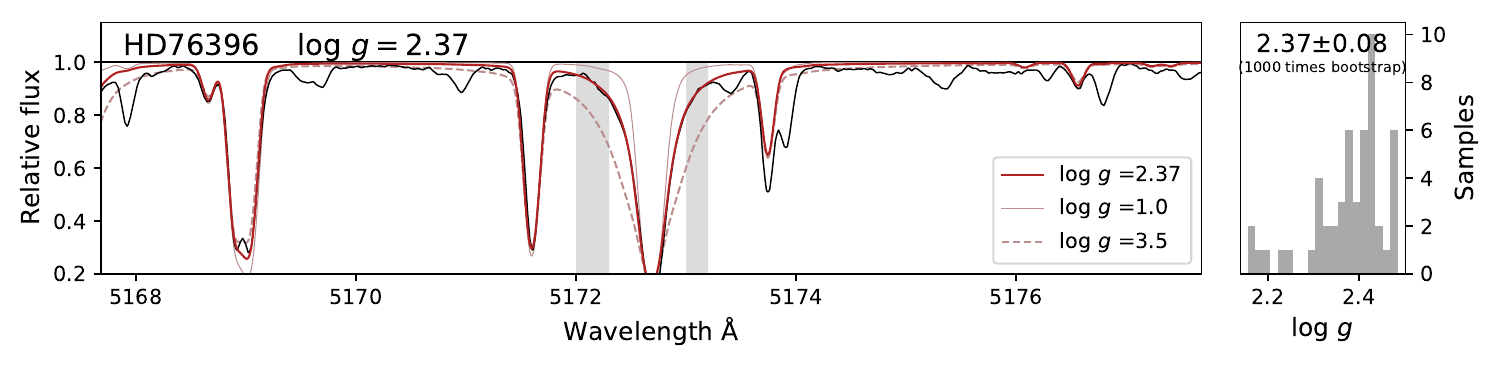}
    \includegraphics[width=0.8\linewidth]{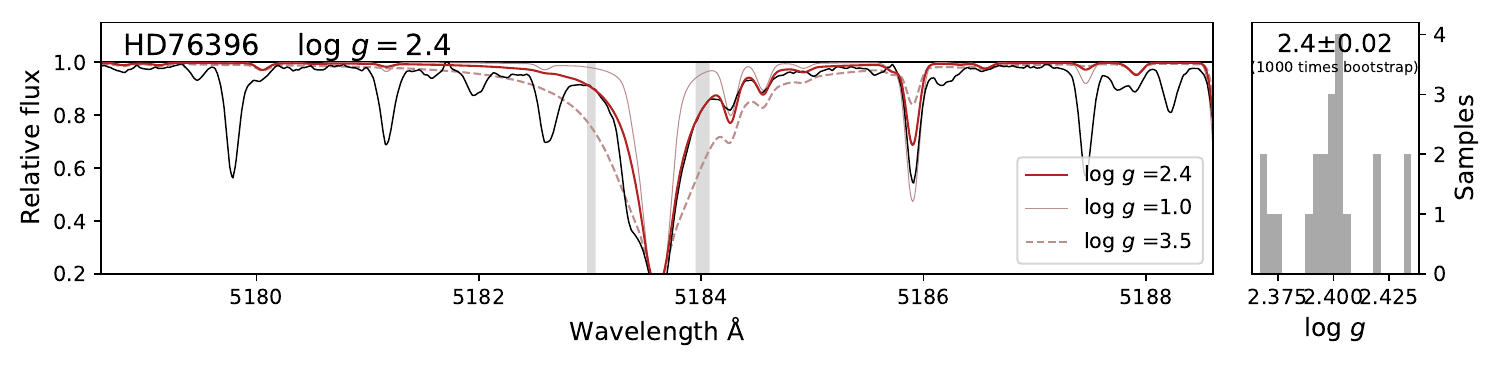}
    \caption{\tiny Magnesium profile fits of the CEMP star HD~76396.
    The observational profile is represented by the black line. The thick red line represents the fitted profile and shaded areas indicate the fitting windows.
    Right panels show histograms of the \logg\ values associated to all pixels within the shaded windows.
    The most probable \logg\ and its error are obtained by bootstrapping, see main text.
    }
    \label{fig:HD76396_Mg}
\end{figure*}

Conventional isochrones have a standard (solar) chemical composition, scaled to the considered stellar metallicity. They are not adapted to determine \logg\ of stars with peculiar surface compositions, because the isochrone position is sensitive to the photospheric CNO abundances, as illustrated in Figs.~24 and 26 of \cite{karinkuzhi2021A&A...645A..61K}. Indeed, a modified composition can affect the opacities, and thus the bolometric magnitude and the effective temperature.

Anticipating this problem, we derived surface gravities from the 5172 and 5183~\AA\ magnesium triplet lines, which, unlike the line 5167~\AA, are reasonably free from blends for most metal-poor stars.
To determine the surface gravity from Mg lines, we  synthesised spectra adopting the following parameters for each star: (i) \teffa and (ii) [\ion{Fe}{ii}/H] as listed in Table~\ref{tab:results}, and (iii) \logg\ in the range [0.5-3.5] with a step of 0.5 dex. These grids were interpolated in \logg\ with a step of 0.01 dex. We considered two fitting windows located far from the line cores, in order to avoid chromospheric effects. Wavelength regions with line blends were avoided.

For stars with [Fe/H] $\lesssim -2$, Mg lines are relatively narrow (coverage less than 1~\AA) and display asymmetries that 1D LTE models cannot reproduce. For them, we fixed the fitting windows to [5172.0-5172.5]~\AA\ and [5172.8-5173.4]~\AA\ for the line at 5172~\AA, and to [5182.8-5183.45]~\AA\ and [5183.8-5184.1]~\AA\ for the line at 5183~\AA.
For CEMP stars, we set the limits of fitting windows to the regions without blends, always avoiding the line core (at a wavelength distance of at least $\pm0.3$~\AA).
An example of these windows for the CEMP star HD~76396 is illustrated on Fig.~\ref{fig:HD76396_Mg}.

The fitting procedure is similar to that of H$\alpha$. Each wavelength pixel inside the fitting windows is associated with the \logg\ of the most compatible interpolated synthetic profile. 
This generates a dispersion of \logg\ values, the histogram of which represents a probability distribution; see for example right panels in Fig.~\ref{fig:HD76396_Mg}.
Since in many cases only few wavelength bins are possible to fit, we do not directly fit Gaussians to the histograms, as done with H$\alpha$ lines (see right panels in Figs.~\ref{fig:Ha_interferometry}, \ref{fig:Ha_IRFM}, \ref{fig:Ha_asteroseismic}, \ref{fig:Ha_other_giants4}, and \ref{fig:Ha_CEMP}). We estimate the most probable \logg\ and its error by bootstrap. This is, the \logg\ dispersion associated to the wavelength bins is randomly re-sampled (bootstrapped dataset), and its median and standard deviation are computed.
This procedure is repeated 1000 times, and the mean of all computed medians is considered the most probable \logg, whereas the mean of all computed standard deviations is considered its error.
The total uncertainty of \logg\ is estimated in Sect.~\ref{sec:logg_accuracy}.

\subsection{Asteroseismic surface gravity}
\label{sec:astero}

\begin{table*}
\caption{Giants with asteroseismic measurements.}
\label{tab:astero}
\centering
\tiny 
\begin{threeparttable}
\renewcommand{\arraystretch}{1.8}
\begin{tabular}{lccccccc}
\hline\hline
Star & TIC & $\nu_{\mathrm{max}}$ [$\mu$Hz]  & \logg$_\mathrm{seis}$ & \loggmg &\loggiso  \\
\hline
HD~3179 & 98626543 & $36.1 \pm2.6$ & $2.49\pm0.04$ & $2.25 \pm 0.15$ & $2.90 \pm 0.07$ \\
HD~13359 & 469975334 & $36.6 \pm 2.9$  & $2.50 \pm 0.04$ & $2.58 \pm 0.15$ & $2.95 \pm 0.06$ \\
HD~17072 & 259862594 & $37.3 \pm 2.2$  & $2.50 \pm 0.03$ & $2.39 \pm 0.15$ & $2.96 \pm 0.06$ \\
HD~221580 & 206287785 & $34.9 \pm 2.3$  & $2.47\pm0.03$ & $2.46 \pm 0.15$  &$2.80 \pm 0.04$ \\
TIC 168924748 & 168924748 & $34.8 \pm 3.4$  & $2.47\pm0.05$ & -- & $2.95 \pm 0.04$ \\
TIC 404605506 & 404605506 & $30.4 \pm 2.3$  & $2.40\pm0.04$ & -- & $2.74 \pm 0.06$ \\
\hline
\end{tabular}
\renewcommand{\arraystretch}{1.}
\begin{tablenotes}
\item{} \textbf{Notes.} {Values in third column were extracted from \cite{Hon2021ApJ...919..131H}. In fifth column, \loggmg\ is the weighted average of the values obtained from the lines 5172 and 5183~\AA, its errors are estimated in Sect.~\ref{sec:logg_accuracy}.
} 
\end{tablenotes}
\end{threeparttable}
\end{table*}

We cross-matched the ESO archive with the TESS catalogue \citep{Stassun2018AJ....156..102S} searching for stars with asteroseismic  measurements of 
maximum frequency ($\nu_{\mathrm{max}}$) in \cite{Hon2021ApJ...919..131H}.
We identified four stars with archived spectra of good $S/N$: HD~3179, HD13359, HD~17072, and HD~221580. Their \teffa, \logg$_{\mathrm{iso}}$, \loggmg, and [Fe/H] were derived as described in previous sections. 
We also included the stars TIC~168924748 and TIC~404605506 in this analysis, although we did not find archival spectra of them. 
They have IRFM \teff\ \citep{casagrande2021MNRAS.507.2684C} listed in the GALAH DR3 catalogue \citep{buder2021MNRAS.506..150B}, which is determined here to be consistent with \teffa\ (Sect.~\ref{sec:accuracy_teff}), thus, along with the four stars above, they are appropriate to determine the accuracy of the \loggiso\ and \loggmg\ scales.
These stars appear as dwarfs in our initial cross-match according to their parameters in \cite{Stassun2018AJ....156..102S}, where their \logg\ values are 4.49 and 4.48~dex, respectively. However, the GALAH catalogue indicates they are giants, the \logg\ values of which given Table~\ref{tab:titans_giants} were taken as preliminary.
We verified in the \textsc{StarHorse} catalogue \citep{Starhorse2022A&A...658A..91A} that these stars are most likely red giants: it provides \logg\ values 2.46 and 2.80~dex for each TIC~168924748 and TIC~404605506, respectively.

We determine asteroseismic surface gravity (\loggseis) by the expression:

\begin{equation}
    \centering
    \mathrm{log}\,g_{\mathrm{(seis)*}} = \mathrm{log}\,g_\odot + \mathrm{log}\, \left( \frac{\nu_{\mathrm{max\,*}}}{\nu_{\mathrm{max\,\odot}}} \right) + \frac{1}{2} \mathrm{log} \left(\frac{T_\mathrm{eff\,*}}{T_\mathrm{eff\,\odot}}\right)
\end{equation}

\noindent
where we adopted the values $\nu_{\mathrm{max\,\odot}} = 3090\,\mu$Hz, 
$\Delta\nu_{\odot} = 135.1\, \mu$Hz,  log$g_{\odot} = 4.44$~dex, and \teff$_{\odot} = 5777$~K \citep{huber2011ApJ...743..143H}.
\loggseis\ is compiled in Table~\ref{tab:astero} along with  $\nu_{\mathrm{max}}$ derived by \cite{Hon2021ApJ...919..131H}.
We used $\Delta\nu_{\odot}$ along with the \teffa\ errors in Table~\ref{tab:results} and $\nu_{\mathrm{max}}$ errors in Table~\ref{tab:astero} to obtain \loggseis\ errors.
We included in Table~\ref{tab:astero} \loggmg\ and \loggiso\ for comparison.

\subsection{Luminosity and radius}
We computed luminosities ($L$) by the relation $L_{*} = L_0^{-0.4(X+BC_X)}$, where $L_{*}$ is the luminosity of the star, $L_0$ is the zero point luminosity $3.0128 \times 10^{28}$~W \citep{mamajek2015arXiv151006262M}, and $X$ and $BC_X$ are the absolute magnitude in a determined photometric band and its corresponding bolometric correction.
Absolute magnitudes were computed from apparent magnitudes (extinction corrected as described in Paper~I) and distance estimates of \cite{bailer-jones2018AJ....156...58B}, where the Gaia parallax zero-point correction \citep[+0.021 mas,][]{lindegren2021A&A...649A...4L} was considered.
Bolometric corrections were computed for each Gaia magnitude $G$, $B_P$, and $R_P$, extinction corrected, with the routine \textit{bcutil.py}\footnote{\url{https://github.com/casaluca/bolometric-corrections}} \citep{casagrande2014,casagrande2018MNRAS.479L.102C}.
Luminosities in Table~\ref{tab:results} are average values obtained with every Gaia magnitude band, values relative to the Sun in logaritmic scale (log~$L/L_{\odot}$) are listed.
Its uncertainties were computed by adding in quadrature the errors induced by the \teff, \logg, and [Fe/H] errors given in the table. 
Stellar radius ($R$) was computed from log~$(L/L_{\odot})$ and \teff\ in the table by means of the Stefan-Boltzmann relation; values relative to the Sun ($R/R_{\odot}$) are listed. Its errors were expanded from log~$(L/L_{\odot})$ and \teff\ errors into the formula.

\section{Accuracy tests}
\label{sec:accuracy}

\subsection{Accuracy of H$\alpha$ effective temperature}
\label{sec:accuracy_teff}

\begin{figure}
    \centering
    \includegraphics[width=0.8\linewidth]{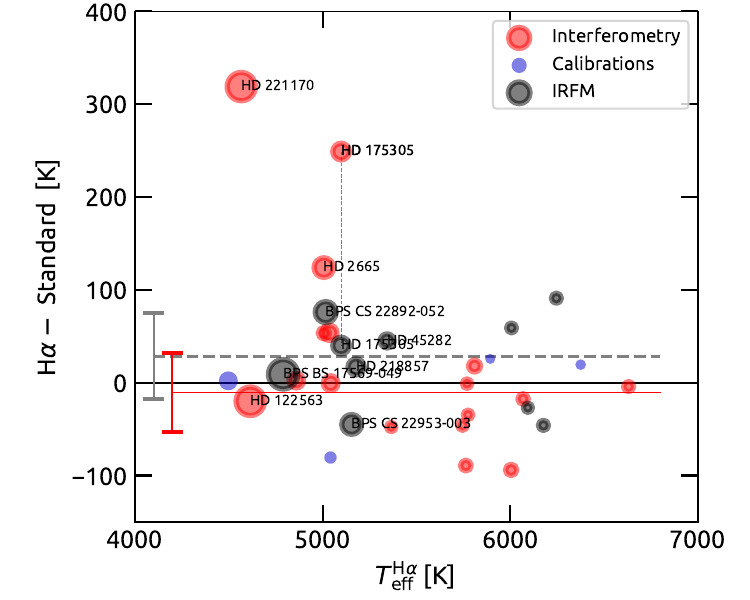}
    \includegraphics[width=0.8\linewidth]{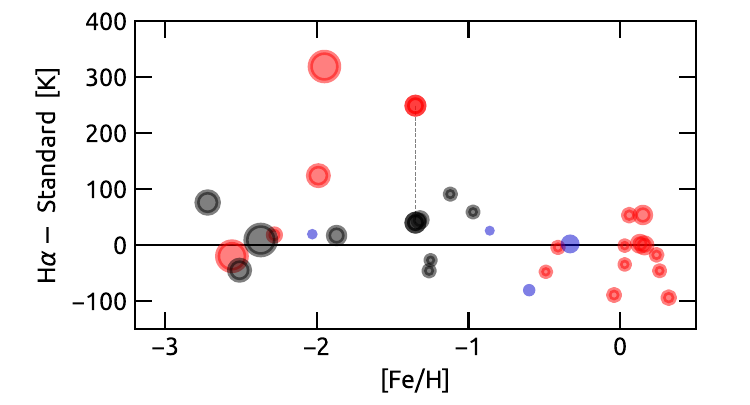}
    \includegraphics[width=0.8\linewidth]{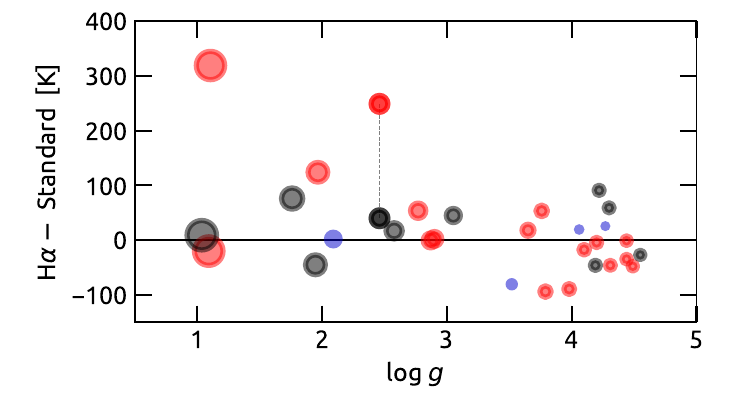}
    \includegraphics[width=0.8\linewidth]{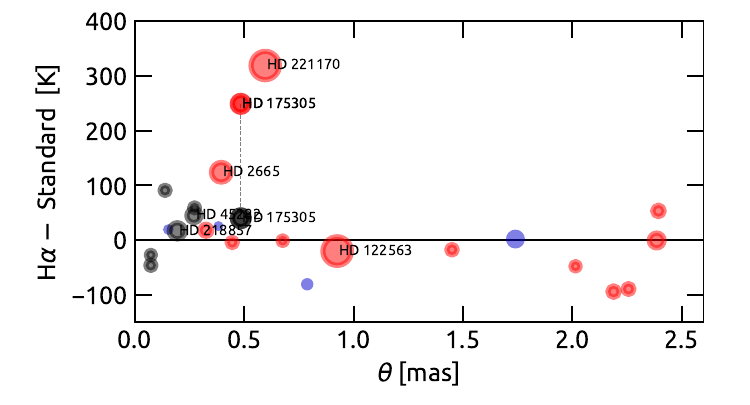}
    \caption{\tiny 
    Temperature difference as function of the atmospheric parameters in Table~\ref{tab:results} and the angular diameter in Table~\ref{tab:titans_giants}. The plots include the Gaia benchmarks (see Fig.~7 of Paper~I). 
    The stars for which \teffHa\ were derived in this work can be identified by their name labels in the top and bottom panels.
    The stars are color-classified according to the method used to infer their standard \teff\ in the literature: direct application of interferometry (red), indirect interferometry based in color calibrations (blue), and IRFM (gray).
    The symbol sizes are inversely proportional to the \logg\ values. The two temperature measurements of HD~175305 (see Table~\ref{tab:titans_giants}) are connected by a vertical dashed line.
    The red (dashed grey) line at $-11$~K ($+28$~K) indicates the median of the differences between H$\alpha$ temperatures and those derived from interferometry (from IRFM), computed omitting HD~221170, HD~175305, and HD~2665. Their dispersion of 42~K (46~K) is indicated by the error bar with the same color.}
    \label{fig:interferometry}
\end{figure}

Assessing model-based \teff\ determinations can be done with the help of stars with angular diameters ($\theta$) inferred via interferometry.
Their \teff\ are considered to carry marginal model influence, therefore to lie on a scale close to accurate.
The set of Gaia benchmarks compiled in \citet{Heiter} and \citet{jofre2014} includes 34 nearby stars with interferometric \teff. Among them only four have [Fe/H] $< -1$~dex, from which one is a red giant, HD~122563. 
This paucity was later mitigated by the incorporation of a sub-sample of ten stars with $-2\lesssim$ [Fe/H] $\lesssim -1$~dex in \citet{hawkins2016A&A...592A..70H}, where two are red giants: HD~175305 and HD~218857. Their temperatures were however determined by means of the so-called InfraRed Flux Method \citep[IRFM,][]{Blackwell1977,Blackwell1979,Blackwell1980}, 
rather insensitive to model inaccuracies as well, as it makes use of photometry in the Rayleigh-Jeans spectral region.

We demonstrated in Paper~I that \teff\ determinations from H$\alpha$ fitting and from interferometrc measurements are compatible for a wide range of atmospheric parameters of F-, G-, and K-type stars. Furthermore, we demonstrated 
an excellent agreement with IRFM in the metallicity range $-3 \lesssim$ [Fe/H] $\lesssim -1$~dex  for dwarf and turnoff stars.
The validation of \teff\ determinations using H$\alpha$  was however not tested for metal-poor giants, given the paucity of benchmark stars of this category.
\cite{karovicova2020A&A...640A..25K} recently provided interferometric \teff\ determinations for ten stars with [Fe/H] $<-0.7$~dex.
Two of them are Gaia Benchmarks, 
and the remaining are new standards, six giants and two dwarfs.

Concerning IRFM, \cite{casagrande2021MNRAS.507.2684C} expanded the applicability range of the implementation in \cite{Casagrande2010} to the RGB. Some of these stars have been studied in the context of the "First Stars" large programme\footnote{Large Program 165.N-0276, P.I.: R. Cayrel.}
and are available in the ESO archive.

Figure~\ref{fig:interferometry} shows the comparison between \teffa\ and what we call "standard scale", which represents either interferometric temperatures in red symbols, temperatures based on   $\theta$ determined by
calibrations in blue symbols \citep{cohen1999AJ....117.1864C, kervella2004}, or IRFM temperatures in gray symbols.
The stars tested in the present work are indicated by their labels in the top and bottom panels; the additional points are the benchmark stars of Paper~I.
For the sake of completeness of the comparisons in the following discussion we include the subgiant metal-poor star HD~140283 with \teffa\ $= 5810 \pm 32$~K derived in Paper~I. Its interferometric \teff\ is $5792 \pm 55$~K \citep{karovicova2020A&A...640A..25K} and IRFM \teff\ is $5777 \pm 55$~K \citep{Casagrande2010}.

The agreement between H$\alpha$ and interferometric temperatures is excellent for HD~140283 and HD~122563, with negligible differences of $\sim$20~K. 
For HD~2665 the difference is larger (124~K), but the interferometric and H$\alpha$ temperatures are still in agreement within 1$\sigma$.
For HD~175305 and HD~221170 the differences are 249 and 319~K, respectively, between 1 and 2$\sigma$ errors. 
However, their reported interferometric errors are relatively large (2-3\%, see Table~\ref{tab:titans_giants}).
HD~175305 also has an IRFM \teff\ \citep{hawkins2016A&A...592A..70H}, which is in much better agreement 
(40~K difference)
with \teffa\ (grey symbol linked to the red one by the dashed line in Fig.~\ref{fig:interferometry}).
Therefore, the interferometric temperature of HD~175305 seems to be underestimated.
We now discuss whether it could also be the case for
HD~221170, which unfortunately does not have an accurate IRFM determination to evaluate this possibility.
Some clue is however provided in 
\cite{casagrande2014}, where systematics in interferometric temperatures are observed towards low $\theta$ values.
In the analysis related to their Fig.~4, these authors explain that such systematics may arise for 
$\theta \lesssim 0.9$~mas
due to insufficient power of beam combiners for sampling the visibility curve of the star disc at its border, which requires high spatial frequencies.
The bottom panel of our Fig.~\ref{fig:interferometry} 
shows similar systematics to those in Fig.~4 of \cite{casagrande2014}.
Examples of how the sampling at high frequency can dramatically improve angular diameter measurements were provided by \cite{white2013MNRAS.433.1262W} in their comparison of the outcomes from the Michigan Infrared Combiner \citep[MIRC,][]{monnier2004SPIE.5491.1370M}, the Classic combiner, and the Precision Astronomical Visible Observations \citep[PAVO,][]{ireland2008SPIE.7013E..24I} combiner in the Center for High Angular Resolution Astronomy (CHARA) array \citep[][]{ten_brummelaar2005ApJ...628..453T}.
\cite{casagrande2014} also comment on the outcomes for the solar twin 18 Sco \citep{porto1997} from PAVO by \cite{bazot2011A&A...526L...4B} and from Classic by \cite{Boyajian2012ApJ...746..101B}, 
where the latter presents significant systematics.
Although the angular diameter measurements of HD~2265, HD~175305, and  HD~221170 were acquired with PAVO \citep{karovicova2020A&A...640A..25K}, which offers sampling at the highest frequency, 
it is possible that their visibility curves are still biased to high counts at high frequencies. If so, they
have affected $\theta$ determination and could be at the origin of the discrepancies between \teffa\ and interferometric temperatures of the three stars above. 

We find that \teffa\ and IRFM~\teff\ agree for all stars in our red giant sample within 1$\sigma$ individual errors. 
It includes HD~140283 and HD~175305 with interferometric \teff\ compared above; we obtain for them \teffa\ - IRFM \teff\ values of +33 and +40~K, respectively.
As Fig.~\ref{fig:interferometry} shows, no difference exceeds 76~K, which corresponds to a star with a spectrum of moderately low quality in our sample, BPS CS 22892-052 with S/N $\sim$139. The median of the differences between \teffa\ and the IRFM \teff\ 
is $+28$~K (grey horizontal line
in the upper panel of Fig.~\ref{fig:interferometry}), close to the IRFM \teff\ zero-point uncertainty of 20~K estimated in \cite{Casagrande2010}.
For stars with interferometric data, we obtain a median difference of $-11$~K when the outliers HD~2265, HD~175305, and HD~22117 are dismissed.

We conclude that H$\alpha$ and IRFM temperatures are compatible for giants and dwarfs and show no significant systematics with any atmospheric parameter nor with angular diameter, therefore the outcomes of both methods are equivalent and can be averaged to improve imprecise \teffa\ determinations due to low $S/N$ spectra, such as the one of BPS~CS~22892$-052$; see Table~\ref{tab:results}.
Interferometric \teff\ should be used with caution for stars with angular diameters lower than 1~mas, since in some cases it can be underestimated by a few hundred Kelvins. For such stars, effective temperatures are best derived using H$\alpha$ fitting or IRFM.

We associate the largest dispersion in Fig.~\ref{fig:interferometry}, $\pm46$~K, as the official uncertainty of the H$\alpha$ 3D non-LTE model. However, we emphasise that the IRFM and interferometric \teff\ errors of the stars in the comparison are similar or even larger than this quantity, thus they dominate the dispersion of the comparison.
For this reason, the internal uncertainties in Table~\ref{tab:results} are good estimate to be expanded for deriving dependent parameters and abundances. 

\subsection{Validation of IRFM-based Gaia photometric calibrations}
\label{sec:phot_calib}

\begin{figure*}
    \centering
    \includegraphics[width=0.33\linewidth]{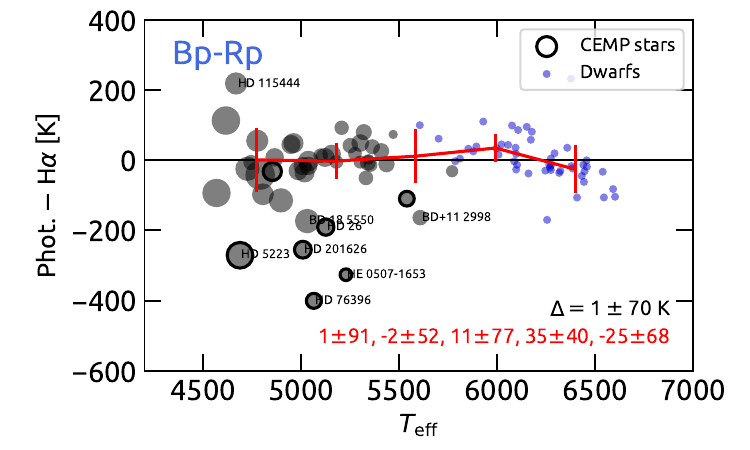}
    \includegraphics[width=0.33\linewidth]{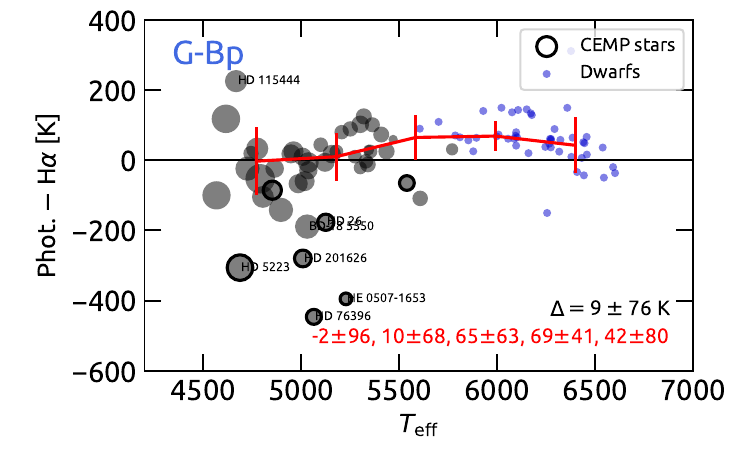}
    \includegraphics[width=0.33\linewidth]{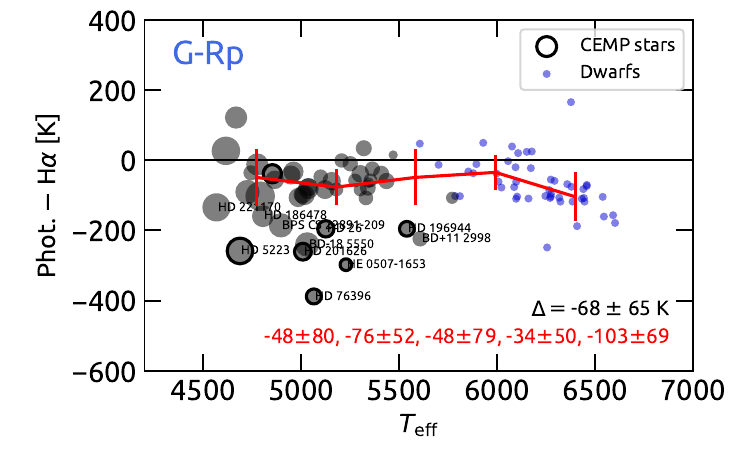}
    \includegraphics[width=0.33\linewidth]{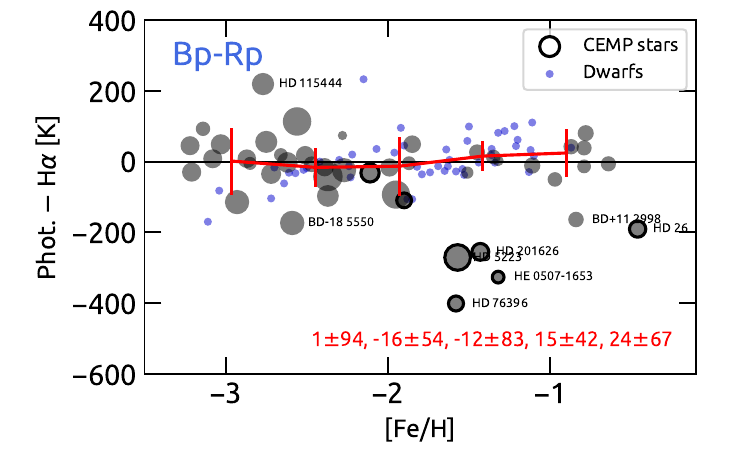}
    \includegraphics[width=0.33\linewidth]{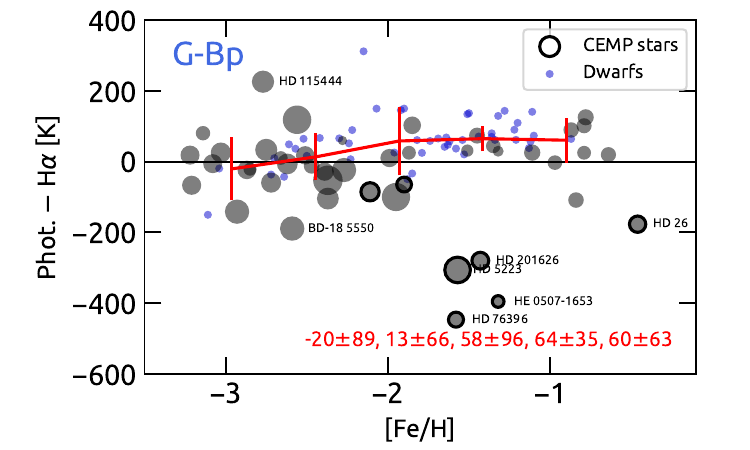}
    \includegraphics[width=0.33\linewidth]{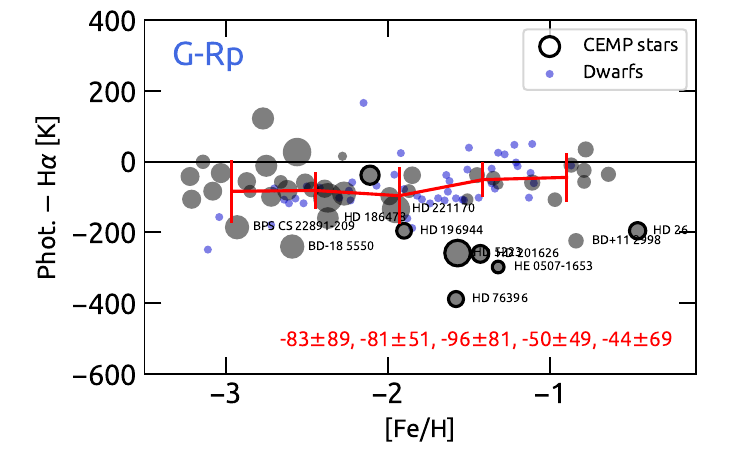}
    \includegraphics[width=0.33\linewidth]{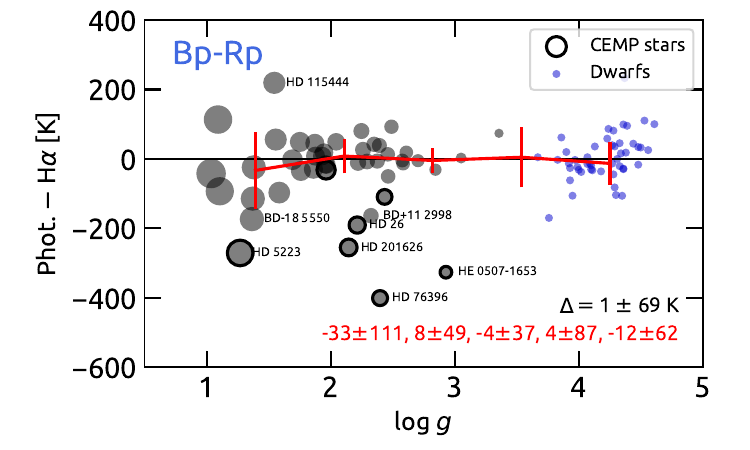}
    \includegraphics[width=0.33\linewidth]{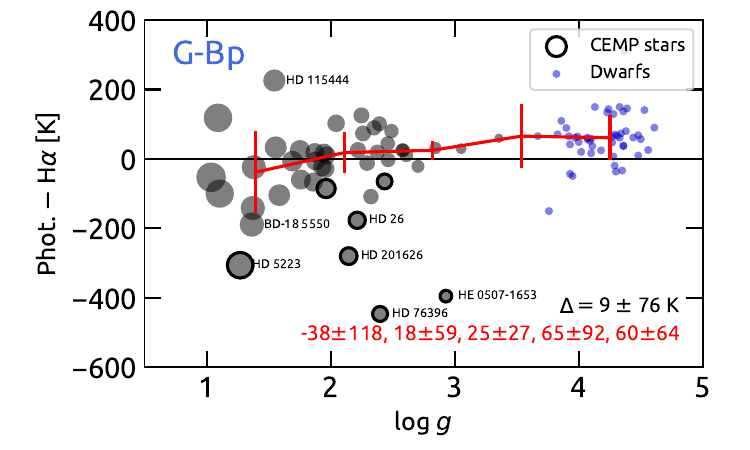}
    \includegraphics[width=0.33\linewidth]{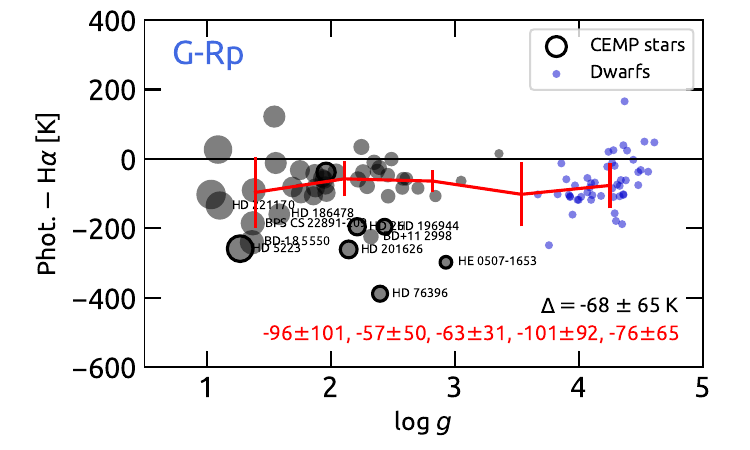}
    \caption{\tiny. Comparison between effective temperatures from the color-\teff\ relations of \cite{casagrande2021MNRAS.507.2684C} and those derived by H$\alpha$ in this work. Each column corresponds to one Gaia colour, and each row shows the distribution of the differences with respect to \teff, [Fe/H], and \logg, respectively. \titan~I dwarfs are plotted in blue, and CEMP stars are highlighted with dark contours. The symbol sizes are inversely proportional to \logg. Red lines connect medians computed in equally spaced bins, and vertical bars correspond to 1$\sigma$ dispersions; their values are indicated in red characters. Medians and corresponding 1$\sigma$ dispersions for only giants (gray symbols) dismissing CEMP stars are given in red characters.
    Giants with temperature differences larger than 2$\sigma$ dispersions are labeled by their catalogue numbers.}
    \label{fig:phot_test}
\end{figure*}

\cite{casagrande2021MNRAS.507.2684C} provide colour-\teff\ calibrations based on the IRFM for about $590\,000$ stars in the GALAH DR3 catalogue with Gaia and 2MASS photometry. 
The zero-points were finely tuned with solar twins,
thus accurate \teff\ are expected for stars with parameters close to solar.
The calibrations with the Gaia colours $BP-RP$, $G-BP$, and $G-RP$ are of special interest  for automatic \teff\ determinations of the large amount of giants in the Gaia catalogue. 
They are expected to provide precise \teff\ for metal-poor stars with preliminary [Fe/H] and \logg\ \citep[as those in][]{andrae2022arXiv220606138A}, as they show small response to typical offsets in these parameters. 
Typically, $\pm$40-50~K is the combined effect of $\Delta$\logg~$=0.2$, $\Delta$[Fe/H]~$=0.1$, and $\Delta E(B-V)$~$=0.1$.
 On the other hand, the performance of this calibration 
at the metal-poor range has not been rigorously quantified because of the lack of an extended sample of metal-poor reference stars; see Fig.~6 of 
\cite{casagrande2021MNRAS.507.2684C}.

Here we test the performance of this calibration in the parameter range covered by our sample of metal-poor red giants.
We derived colour-calibrated temperatures\footnote{
In the following, "colour-calibrated temperature" refers to the temperature based on the colour calibration of IRFM-derived temperatures \citep{casagrande2021MNRAS.507.2684C}.}  from Gaia colours running the script \emph{colte}\footnote{\url{https://github.com/casaluca/colte}}, using [\ion{Fe}{ii}/H] and \logg\ in Table~\ref{tab:results}, and $E(B-V)$ in Table~\ref{tab:titans_giants}; the last was extracted from either \emph{Stilism} \citep{Capitanio2017} when available or from \cite{SFD} otherwise. 
For obtaining dereddened colours, the script \emph{colte} converts $E(B-V)$ to the Gaia system using extinction coefficients of either \cite{schlafly2011ApJ...737..103S} or \cite{cardelli1989ApJ...345..245C} and \cite{odonell1994ApJ...422..158O}.
We chose to use the former in the analysis below; we verified that using the latter we obtain temperatures that differ by only few Kelvins. 
Figure~\ref{fig:phot_test} compares the colour-calibrated temperatures with \teffa\ as a function of the atmospheric parameters. For completeness, dwarf stars from Paper~I are included in the analysis and  represented in blue. 
We observe that the calibrations using $BP-RP$ provide the best agreement with \teffa. 
As shown by the plots on the left column of Fig.~\ref{fig:phot_test}, there is perfect agreement across the entire \teff-[Fe/H]-\logg\ parameter space.
The calibrations with the $G-BP$ color agree with \teffa\ for giants, whereas they produce temperatures $\sim$60~K hotter for dwarfs.
The calibrations with the $G-RP$ color produce temperatures $\sim$70~K cooler than \teffa\ for giants, whereas for dwarfs they can be $\sim$100~K cooler.
We remark that although small offsets for dwarfs between \teffa\ and the colour-calibrated temperature appear for $G-BP$ and $G-RP$ colours, we demonstrated in Paper~I that 
IRFM temperatures (the base of these color calibrations) are fully compatible with \teffa. 

As far as CEMP stars are concerned, Fig.~\ref{fig:phot_test} illustrates the large discrepancies between their colour calibrated temperatures and \teffHa, with differences larger than 2$\sigma$. Actually, the C$_2$ and CN absorption bands produce strong flux absorption which are not taken into account by the photometric calibrations considering solar-scaled chemical abundances.
Standard photometric calibrations can thus produce strong biases when applied to CEMP stars.

\subsection{Comparison of literature temperatures with accurate \teff}
\label{sec:literature}
\begin{figure}
    \centering
    \includegraphics[width=0.85\linewidth]{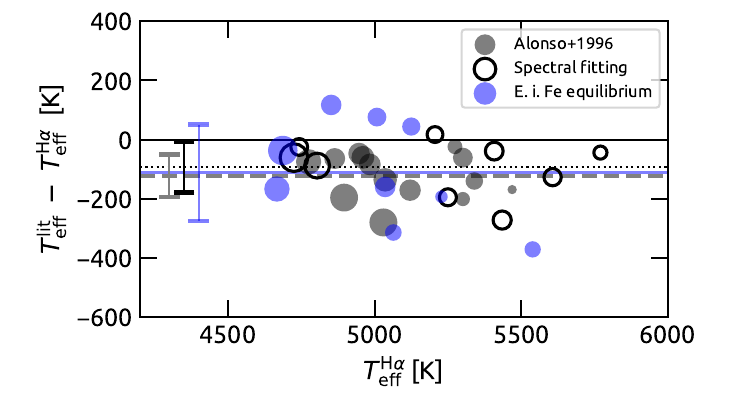}
    \includegraphics[width=0.85\linewidth]{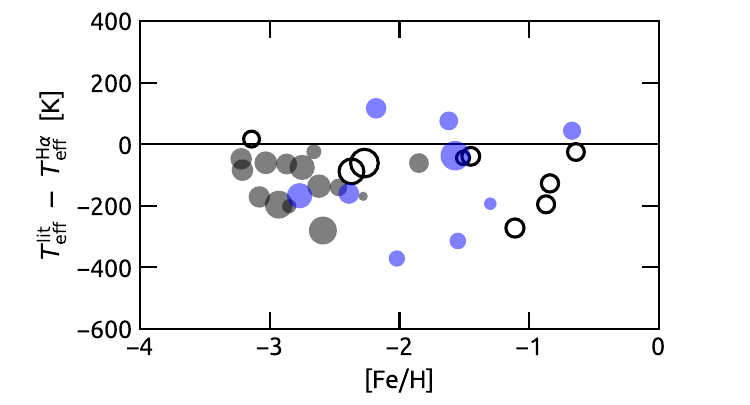}
    \includegraphics[width=0.85\linewidth]{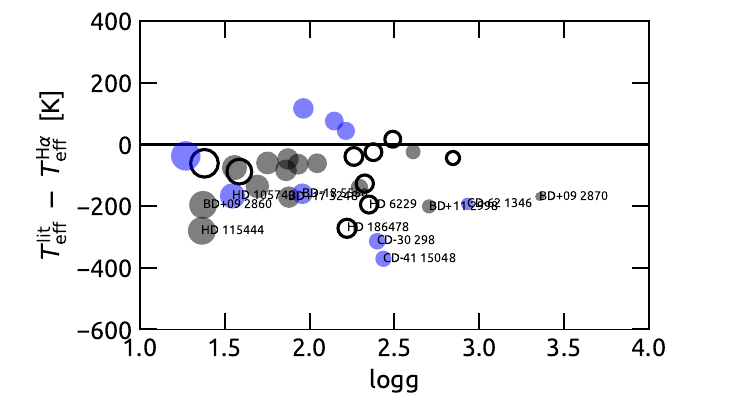}
    \caption{\tiny Comparison between \teffa\ and preliminary temperatures from the literature ($T_{\mathrm{eff}}^{\mathrm{lit}}$) compiled in Table~\ref{tab:titans_giants}, as function of the atmospheric parameters.
    The size of the symbols is inversely proportional to \logg\ in Table~\ref{tab:results}. The symbols distinguish the method by which $T_{\mathrm{eff}}^{\mathrm{lit}}$ was derived: Gray circles for color-\teff\ relations of \cite{alonso1996A&AS..117..227A}, black contour circles for spectral fitting, and blue circles for excitation and ionization equilibrium of Fe lines.
    The gray dashed, black dotted, and blue lines represent the mean offsets of each scale: $-93$, $-122$, and $-105$~K, respectively.
    Error bars are 1$\sigma$ dispersions and correspond to symbol colors: 72, 86, and 169~K, respectively.
    Bottom panel includes labels with the names of the stars with discrepancies higher than 150~k.}
    \label{fig:teff_lit}
\end{figure} 

Disregarding the first three sections of Table~\ref{tab:titans_giants}, which contain standard stars for accuracy tests, 
several stars have temperatures derived with the color-\teff\ relations of \cite{alonso1996A&AS..117..227A} in their respective source papers:  \cite{cayrel2004A&A...416.1117C} and \cite{carney2008AJ....135..196C}.
Preliminary temperatures of a few stars were derived by spectral fitting \citep{arentsen2019A&A...627A.138A,koleva2012A&A...538A.143K,beers2017ApJ...835...81B}, and few by others assuming
excitation ionization equilibrium of Fe lines
\citep{johnson2002ApJS..139..219J,hansen2018ApJ...858...92H,karinkuzhi2021A&A...645A..61K}.

Figure~\ref{fig:teff_lit} shows the offsets of the literature temperatures with respect to \teffa, determined accurate in Sect.~\ref{sec:accuracy_teff}.
No general correlations with the atmospheric parameters are observed. 
Average differences, indicated by the horizontal lines, show that the three scales underestimate \teff\ by $\sim$100~K.
\citet[][Fig.~3]{Casagrande2010} diagnosed the same bias for the scale of \cite{alonso1996A&AS..117..227A} with dwarf stars, 
here we verify it remains the same for red giants.
\citet[][Fig.~11]{Casagrande2010} also found temperature underestimations similar to ours for the spectral fitting method when revising the catalog of \cite{valenti2005}.
Preliminary temperatures derived by the excitation and ionization equilibrium of Fe lines display the largest dispersion.
Among them, temperatures of four stars (HD~26, HD~5223, HD~201626, and HD~224959) differ with \teffa\ by less than 120~K.
These stars, however, have literature \logg\ and [Fe/H] that differ from ours by up to 0.7 and 0.4~dex, respectively, in the worse cases (HD~5223 and HD~224959).
Thus, it is plausible that, due to the strong parameter interdependence that this method involves, biases permute among \teff, \logg, and [Fe/H] in a complicated manner that we cannot rule with our small and selection-biased sample.
In Sects.~\ref{sec:fe_LTE} and \ref{sec:fe_NLTE} we demonstrate that the ionization equilibrium of Fe is not satisfied under LTE for giants. Thus, we indicate that this assumption is likely the main  source of the biases displayed here.

\subsection{Accuracy of surface gravity}
\label{sec:logg_accuracy}

\begin{figure}
    \centering
    \includegraphics[width=0.85\linewidth]{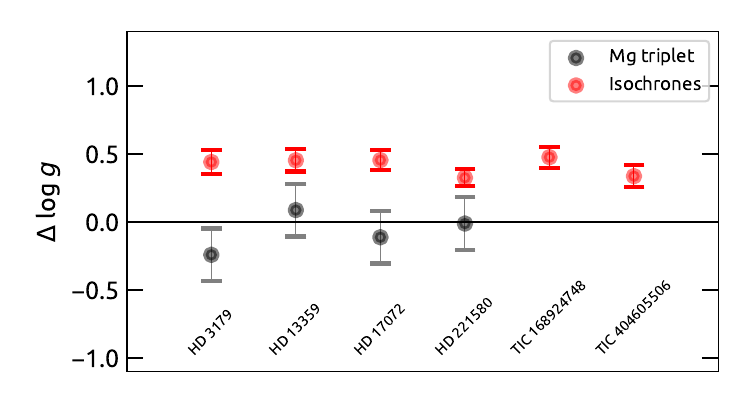}
    \caption{\tiny Surface gravity difference for each star in Table~\ref{tab:astero}. 
    \loggiso\ and \loggmg\ relative to \loggseis\ are displayed in red and gray, respectively.}
    \label{fig:logg_compare}
\end{figure}

Custom methods to constrain \logg, such as forcing the ionization balance of Fe lines and isochrone fitting, may lead to unreliable outcomes for giant stars.
In Sect.~\ref{sec:fe_LTE} and \ref{sec:fe_NLTE} we demonstrate that the problem of the former is that \ion{Fe}{i} and \ion{Fe}{ii} in red giants are generally unbalanced under LTE.
When isochrone fitting is applied, it is generally assumed that \logg\ (or luminosity) varies with \teff, age, metallicity, and mass. 
Other parameters such as the mixing length, $\alpha$ element enhancement, atomic diffusion, and the helium abundance, for instance, are also relevant in the RGB \citep[e.g.][]{song2018ApJ...869..109S,cassisi2017EPJWC.16004002C,cassisi2020A&ARv..28....5C}, and may vary case-to-case in complicated and diverse manners not accounted in ready isochrone grids as those used in this work. 
As a consequence, the accuracy of \logg\ from isochrone fitting for field stars, is uncertain and likely imprecise.

We evaluate the accuracy of \loggiso\ and \loggmg\ using \loggseis\ in Table~\ref{tab:astero} as the reference scale.  
Figure~\ref{fig:logg_compare} shows that \loggmg\ and \loggseis\ are compatible within 1$\sigma$ errors for three stars out of four with measurements available, whereas \loggiso\ is significantly deviating to higher values for every star. Accordingly, we accept \loggmg\ as the accurate surface gravity determination of our sample.
Table~\ref{tab:results} lists averaged values from both Mg triplet lines employed (5172 and 5183~\AA) among our recommended parameters, its errors are estimated below.
We note that the stars in this test are constrained within a narrow \logg\ range between 2.40 and 2.50~dex. Thus, the accuracy of our \logg\ in Table~\ref{tab:results} is guaranteed within this range and near surroundings, say $\pm0.5$~dex approximately. 
The accuracy of our recommended \logg\ values lower than $\sim$2~dex requires comparison with outcomes from fundamental methods on eclipsing binaries \citep[e.g.][]{2019MNRAS.484..451H,Ratajczak2021MNRAS.500.4972R,miller2022MNRAS.517.5129M,maxted2023MNRAS.522.2683M} or well calibrated asteroseismology.

\begin{figure}
    \centering
    \includegraphics[width=0.85\linewidth]{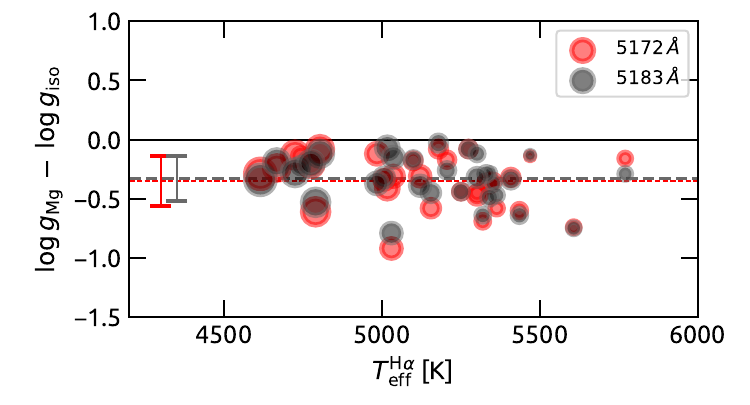}
    \includegraphics[width=0.85\linewidth]{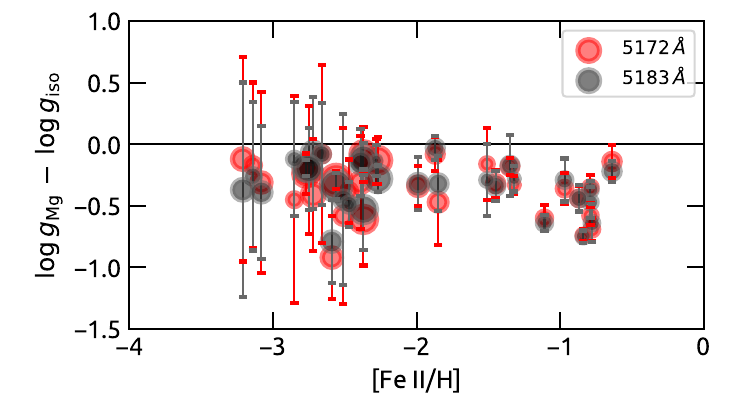}
    \includegraphics[width=0.85\linewidth]{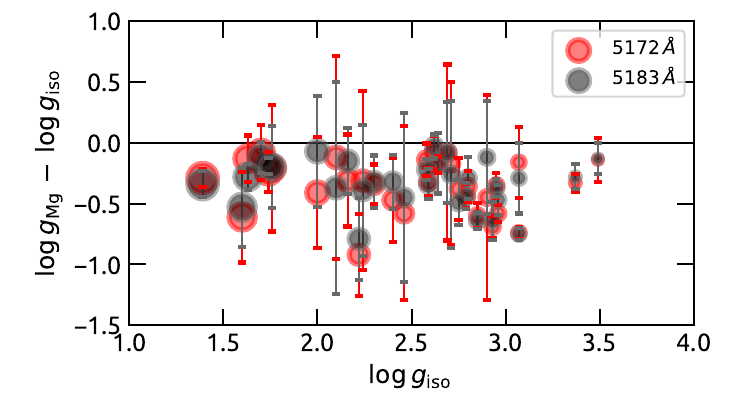}
    \caption{\tiny Surface gravity differences as function of the atmospheric parameters. The size of the symbols is inversely proportional to \loggiso. The points are coded in red or grey depending on the \ion{Mg}{I} line (5172\AA~ or 5183~\AA) used to compute \loggmg\ $-$ \loggiso.
    In the top panel, the offsets and $\pm1\sigma$ dispersions are represented by the dashed lines and error bars, the values of which are $-0.35 \pm0.21$ for the line at 5172~\AA\ (in red) and  $-0.33 \pm0.19$ for the line at 5183~\AA\ (in gray). Individual \logg\ values are listed in Table~\ref{tab:logg_Mg}.
    Error bars in the middle and bottom plots are the quadratic sum of the errors of \loggiso\ and those of \logg$_{\mathrm{(Mg)}}$.}
\label{fig:Mg_calibration}
\end{figure}

The determination of \loggiso\ is straightforward and can  still be useful when the \loggmg\ determination is unfeasible, as far as its offsets are well characterised. This is the case of the stars BPS~CS~22186$-025$, BPS~CS~22189$-009$, BPS~CS~22891$-209$, BPS~CS~22949$-048$, and BPS~CS~22956$-050$, the Mg lines of which are too narrow to be used as \logg\ indicator.
In Fig.~\ref{fig:Mg_calibration}, we compare the \loggmg\ values obtained from each Mg line with \loggiso; Table~\ref{tab:logg_Mg} lists all values in the comparison. We remind here that the latter may provide reasonable outcomes only for non-enriched stars, therefore CEMP stars are excluded from Fig.~\ref{fig:Mg_calibration}. The plots show no obvious correlation with any atmospheric parameter, but display systematic offsets. Even though the internal precision of the individual \loggmg\ measurements worsens as metallicity decreases, as shown in the middle plot, the dispersions of the distributions remain roughly constant across the considered parameter range.
As shown on the top panel of the figure, both Mg lines provide similar \loggmg, their offsets with respect to \loggiso\ being nearly identical. 
Its average, $-0.34$~dex, is here determined as the correction factor for \loggiso.
To determine \loggmg\ error, we assume that the dispersion of the difference between \loggiso\ and averaged \loggmg\ values, 0.17~dex, is composed by the addition in quadrature of the corresponding errors. 
Since the typical \loggiso\ error is 0.08~dex, the typical \loggmg\ error results as 0.15~dex.

\subsection{Fe ionization balance under LTE}
\label{sec:fe_LTE}
The  so-called "excitation and ionization balance of Fe lines" has been largely used as a standard method to determine atmospheric parameters; see a few recent examples in \cite{hema2018ApJ...864..121H,Hill2019A&A...626A..15H}, and \cite{karinkuzhi2021A&A...645A..61K}. It assumes the 
identity of the abundances derived from neutral and ionised Fe lines. 
Such an assumption leads to reliable \logg\ only under LTE validity.
\citet{tsantaki2019MNRAS.485.2772T} selected \ion{Fe}{ii} lines so that the iron ionization balance is satisfied in their sample of F, G and K solar-metallicity dwarfs; whether this configuration can be extrapolated to other spectral types and metallicities remains to be verified.

Ionised Fe lines have been observed to be virtually insensitive to departures from LTE \citep[e.g.][]{fabrizio2012PASP..124..519F,Lind2012MNRAS.427...50L,mashonkina2011A&A...528A..87M,Sitnova2015ApJ...808..148S} in F-, G- and K-type stars.
Moreover, \cite{amarsi2016} have shown that \ion{Fe}{ii} abundances remain nearly the same when derived from 1D LTE or 3D non-LTE models. In the same work, it has been observed that 3D non-LTE \ion{Fe}{i} abundances closely approach to those of \ion{Fe}{ii}. Therefore, \ion{Fe}{ii} lines can be safely used when fast 1D LTE calculations are performed.

\begin{figure}
    \centering
    \includegraphics[width=0.85\linewidth]{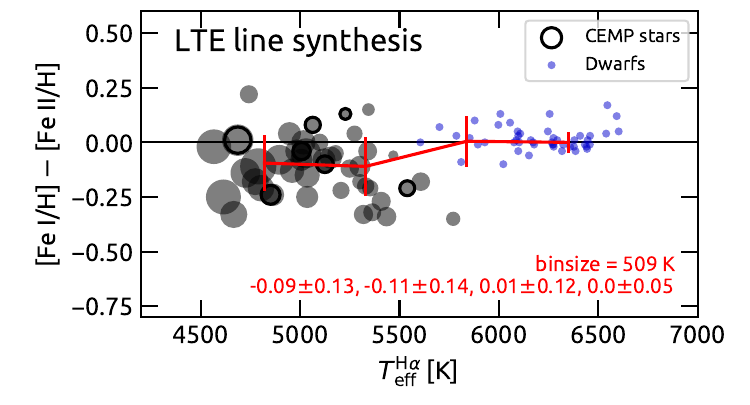}
    \includegraphics[width=0.85\linewidth]{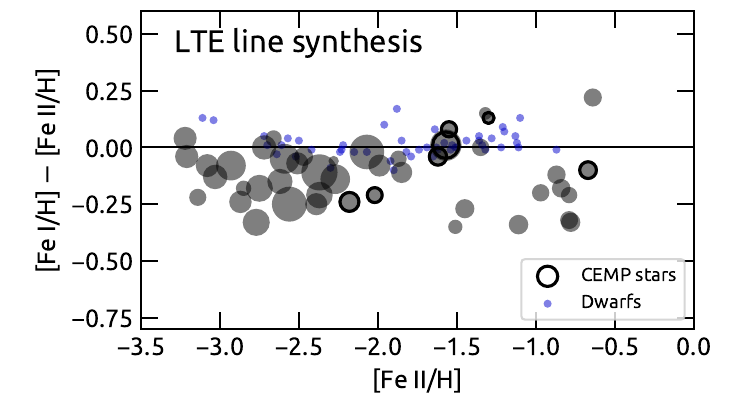}
    \includegraphics[width=0.85\linewidth]{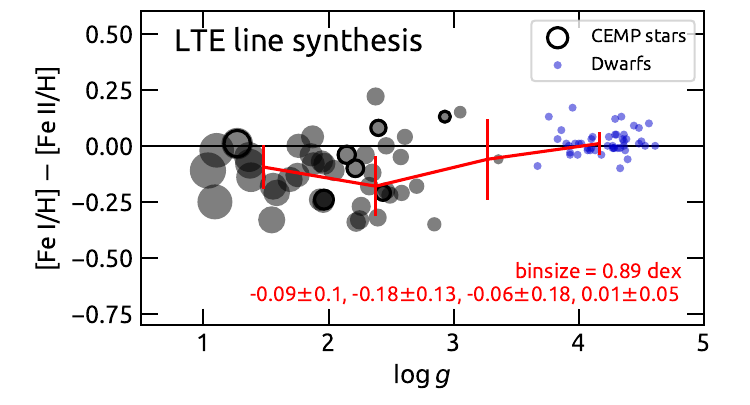}
    \caption{\tiny Deviation from ionization balance as function of the atmospheric parameters.
    Dwarf stars from Paper~I are plotted in blue, and CEMP stars are highlighted with dark contours. The symbol sizes are inversely proportional to \logg.
    Red lines in top and bottom panels connect medians computed in equally spaced bins of \teff\ and \logg, respectively; medians and bin-sizes are indicated in the plots. The vertical bars correspond to 1$\sigma$ standard deviations; corresponding values are indicated in the plots as well.}
    \label{fig:Fe_dif}
\end{figure}

Given the relatively large sample of metal-poor stars with accurate \teff\ and \logg\ compiled in Paper~I and here, we examine the compatibility of \ion{Fe}{i} and \ion{Fe}{ii} abundances under LTE along the \teff-\logg-[Fe/H] parameter space. 
The following analysis is relevant because, in many cases, only few \ion{Fe}{ii} lines may be available in observational spectra  due to severe blending.
Figure~\ref{fig:Fe_dif} shows the comparison between both abundance sets.
The ionization unbalance appears 
only for red giants, this is, stars with \logg~$\lesssim 3.5$~dex.
The distributions of the abundance differences with \teff\ and \logg\ are quite similar (see top and bottom plots of Fig.~\ref{fig:Fe_dif}). 
Therefore, it can be inferred that \ion{Fe}{i} lines allow reliable abundance determinations only for main sequence stars up to the very beginning of the subgiant branch.
Then on, for giants, \ion{Fe}{i} lines certainly lead to abundance underestimations between $-0.10$ and $-0.20$~dex (as shown by the red lines in the \teff\ and \logg\ ranges covered by the gray symbols in Fig.~\ref{fig:Fe_dif}).
We remark that these offsets are only valid for Fe abundances derived from accurate \teff\ and \logg.
The ionization unbalance displays no correlation with respect to [Fe/H] in the analysed parameter space (see middle panel in Fig.~\ref{fig:Fe_dif}).
Assuming LTE ionization balance for red giants will thus possibly lead to biased \logg\ determinations, for this reason this method should be avoided.

The results above confirm those of \cite{karovicova2018}. \cite{Sitnova2015ApJ...808..148S}  found that ionization equilibrium departures appear in dwarfs for [Fe/H]~$\lesssim -1$~dex (see Fig.~7 in the paper), which contradicts our results.
In their analysis, \ion{Fe}{i} abundances match \ion{Fe}{ii} when non-LTE models are used.
We cannot assert what the source of their relatively low LTE \ion{Fe}{i} values is. However, we note that their temperature scale is a possible source. 
They derived temperatures with the color-calibrations of \cite{ramirez2005ApJ...626..465R}, 
which lead to temperatures $\sim$100~K cooler than the IRFM scale of \cite{Casagrande2010} (see Fig.~5 in the paper), which is compatible with ours; see Sect.~\ref{sec:accuracy_teff}. 
We determine the total precision of our [Fe/H] determinations by accepting that the dispersion in Fig.~\ref{fig:Fe_dif} is the even contribution of the errors of [\ion{Fe}{i}/H] and [\ion{Fe}{ii}/H] added in quadrature.
Since the dispersion is $\sim$0.13~dex, our [Fe/H] precision results as $\pm$0.09~dex. 
This is a little larger than the typical internal precision $\pm$0.05~dex, 
as it accounts for \teff\ and \logg\ errors.

\subsection{Fe ionization balance under non-LTE}
\label{sec:fe_NLTE}

Spectral synthesis with iron in non-LTE was done using Turbospectrum~2020\footnote{\url{https://github.com/bertrandplez/Turbospectrum_NLTE}} \citep{gerber2023} for three MARCS model atmospheres close to the parameters of 
the stars HD~122563, BD+11~2998, and BPS~CS~29502$-042$, which have  atmospheric parameters representative of the sample.
We have first extracted the iron departure coefficients based on the iron model atom developed in \cite{bergemann2012} and \cite{semenova2020} corresponding to the three atmospheric models. 
Then, using the curated line list from the Gaia-ESO Survey \citep{heiter2021A&A...645A.106H}, we computed synthetic spectra in the range 4900-6500~\AA\ with a step of 0.005~\AA, both in LTE and non-LTE, varying the iron abundance by steps of 0.1 dex.

\begin{figure}
    \centering
    \includegraphics[width=0.85\linewidth]{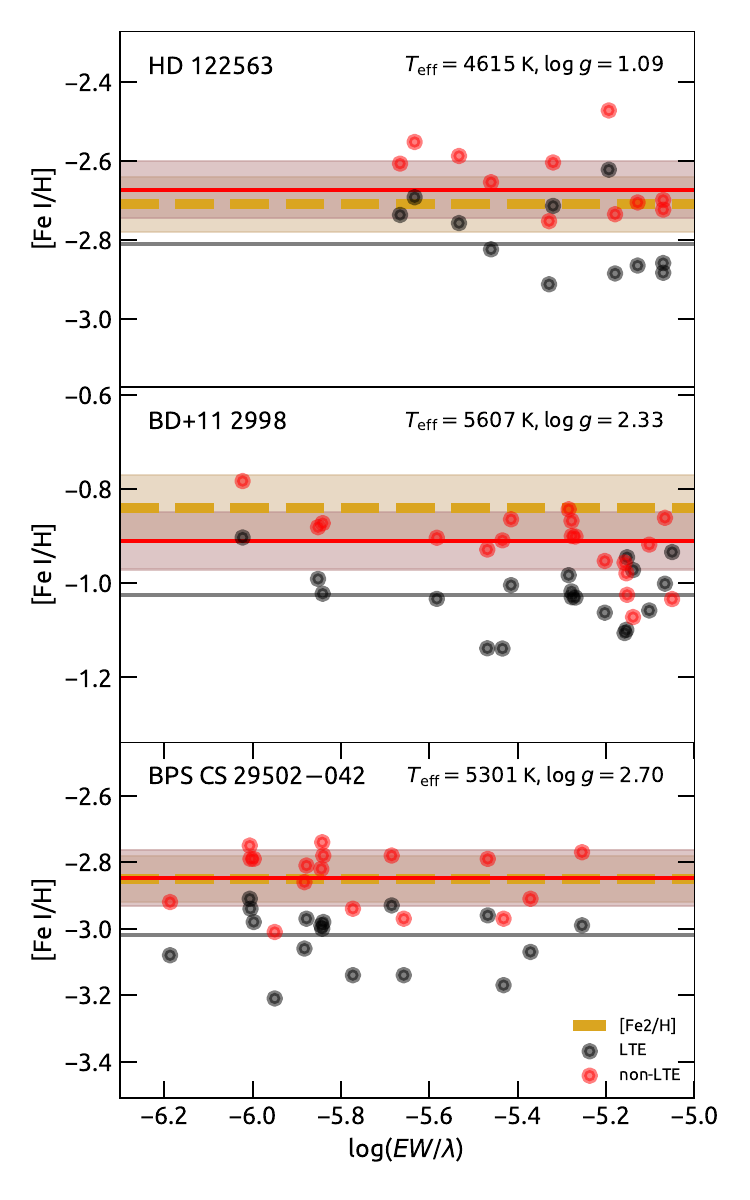}
    \caption{\tiny Inferred iron abundance against the ratio between equivalent width and wavelength for selected \ion{Fe}{I} lines.
    Horizontal lines represent the averages of the symbols with same colors. Dashed lines represent [\ion{Fe}{ii}/H] abundances.
    Shades represent 1$\sigma$ dispersions around average values for non-LTE \ion{Fe}{i} (red color tone), and \ion{Fe}{ii} (yellow color tone).}
    \label{fig:NLTE_LTE}
\end{figure}

With these spectra we compare non-LTE with LTE abundances performing a differential line-by-line analysis.
This is, we used the same input parameters, 
the same line list for non-LTE and for LTE models, and we ran the same fitting algorithm. 
Our analysis only considers lines with log$(EW/\lambda) \leq -5$. As expected, for \ion{Fe}{ii}, we got the same outcomes with LTE and non-LTE models.  
Individual abundances of \ion{Fe}{i} lines obtained in LTE and non-LTE are compared in Fig.~\ref{fig:NLTE_LTE}. 
We note that taking into account non-LTE effects shifts the \ion{Fe}{i} line abundances up by $\sim$0.12~dex and brings them in agreement with the \ion{Fe}{ii} line abundances within 1$\sigma$.
Accordingly, non-LTE \ion{Fe}{i} and LTE \ion{Fe}{ii} can be averaged or, one can be used in absence of the other for stars in the parameter range here analysed.

\section{Mg, C, N, and O abundances}
\label{sec:abundance}
\begin{figure}
    \centering
    \includegraphics[width=0.85\linewidth]{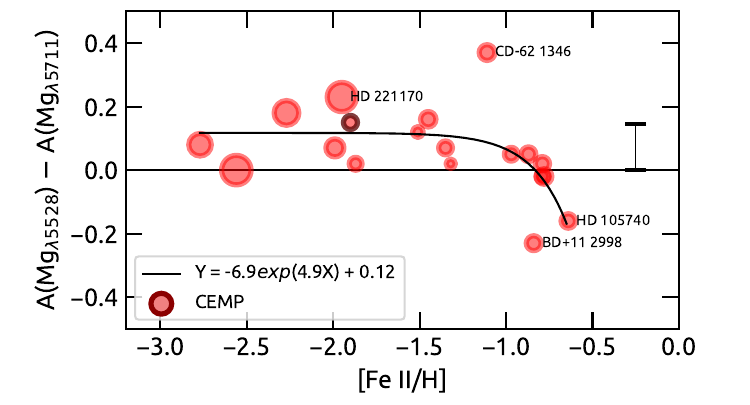}
    \includegraphics[width=0.85\linewidth]{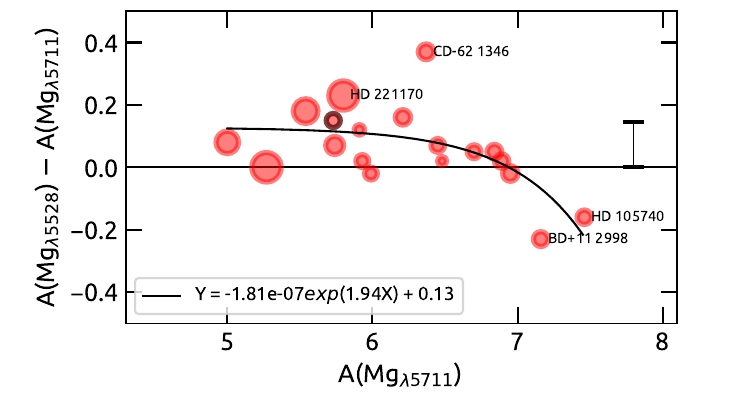}
    \includegraphics[width=0.85\linewidth]
    {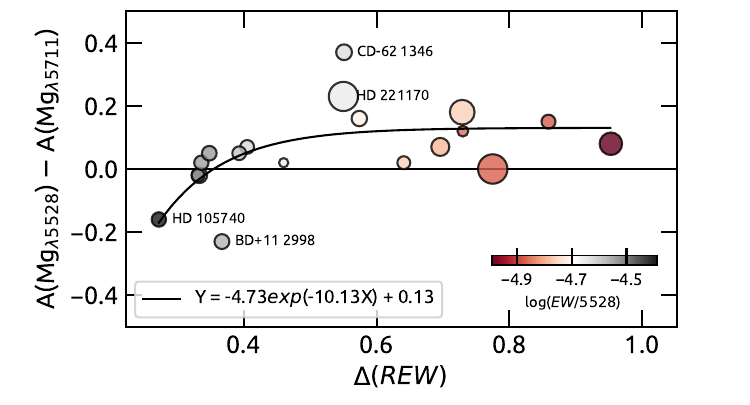}
    \caption{\tiny Consistency of Mg abundance from the lines 5528 and 5711~\AA. Top panel displays abundance differences as function of metallicity. Mid panel displays differences as function of the Mg abundance of the 5711~\AA\ line. Bottom panel, displays differences as function of the difference between log($EW_{5528\AA}/5528$) and log($EW_{5711\AA}/5711$). 
    Symbols are colour-coded according to the reduced equivalent width of the line 5528~\AA\ in logarithmic scale.
    Symbols sizes are inversely proportional to \logg.
    Exponentials are fitted to the distributions in all panels.
    Stars with the largest discrepancies are tagged by their identifiers.
    1$\sigma$ errors ($\pm0.07$) around the average ($+0.07$) are represented by the error bars.}
    \label{fig:Mg_dif}
\end{figure}

Element abundances were derived assuming the following atmospheric parameters: \teffa, [\ion{Fe}{ii}/H], and \logg\ in Table~\ref{tab:results}.

\subsection{Magnesium}
\label{sec:magnesium}
Magnesium was derived primarily from the line at 5711~\AA. 
This line has been shown to be little affected by 3D non-LTE effects, both theoretically \citep{mashonkina2013A&A...550A..28M,bergemann2017ApJ...847...15B} and observationally \citep{giribaldi2023A&A...673A..18G}.
In case this line was not available, either because it was too weak or because it was not covered by the spectrum, we used the line at 5528~\AA\ instead.
In Fig.~\ref{fig:Mg_dif} we show abundance discrepancies between these lines as function of [Fe/H], A(Mg)\footnote{A(Mg) = log($N_{\mathrm{Mg}}/N_{\mathrm{H}}$) + 12.}, 
and the difference of the reduced equivalent widths\footnote{Equivalent widths were computed by Gaussian fitting.} of the Mg lines $\Delta(REW) = $ log$(EW/5528)\,-$ log$(EW/5711)$.
The bottom panel colour-codes the symbols according to a logarithm scale of the reduced equivalent width of the line at 5528~\AA.
It shows that, in general, discrepancies remain moderate, around 0.1~dex. This happens for most stars with [Fe/H]~$\lesssim -1$~dex, as top panel in the figure shows. At low $\Delta(REW)$ the abundances from the line at 5528~\AA\ tend to be lower than those from the line at 5711~\AA\ (darkest gray symbols).
This is the case of stars with the highest metallicities and Mg abundances: HD~105740 and BD+11~2998, as top and mid panel show.
Similar outcomes were also observed in dwarfs \citep{giribaldi2023A&A...673A..18G}. 
We remark that in the present case, abundance discrepancies remain small even for high $\Delta(REW)$ values ($\gtrsim 0.6$)  because their Mg lines are unsaturated.
As red tone symbols in bottom panel of the figure show, the $REW$ of the strongest line 5528~\AA\ remains lower than $-4.8$, therefore within in the linear section of the curve of growth.

We use the exponential fits in the mid panel to correct the offsets that the Mg determinations from the 5528~\AA\ line carry, thus Mg of all \titan, dwarfs and giants, lie in the same scale.
Finally, we adopt the non-LTE corrections of \cite{mashonkina2013A&A...550A..28M} for the line 5711~\AA\ to all Mg abundances. This is +0.07~dex, which is an average of the Drawinian
rates scaled values ("D0.1") for [Fe/H]~$= -1$ and [Fe/H]~$= -2$~dex, given in Table~5 in the paper.
Corrections were also provided by \cite{merle2011MNRAS.418..863M}, but not for all the atom parameters.
We remark that the abundance discrepancies shown in Fig.~\ref{fig:Mg_dif} are only valid when line synthesis is applied. Abundance determination via $EW$ measurement likely produces different outcomes.

The uncertainties of the abundances are computed as described in \cite{giribaldi2023A&A...673A..18G}.
Namely, grids of A(Mg) offsets induced by typical offsets of \teff, \logg, [Fe/H], and \logg\ were produced. 
For that, we first synthesised spectra around the lines 5528 and 5711~\AA.
The spectra were produced with four values of \teff\ between 4500-6000~K, three values of \logg\ between 1-3~dex, four values of [Fe/H] between $-3$ to $-0.5$~dex, seven values of A(Mg) between 4.8-7.2~dex, and two values of $v_{mic}$ between 1-2~kms$^{-1}$.
Subsequently, we computed abundances varying \teff, \logg, [Fe/H], and $v_{mic}$, one at a time, by their typical errors.
These are $\pm50$~K in \teff, $\pm0.15$~dex in \logg, $\pm0.13$~dex in [Fe/H], and $\pm0.3$~kms$^{-1}$ in $v_{mic}$.
Therefore, four grids of offsets with 672 spectra for each Mg line are available.
We interpolated these grids with the atmospheric parameters of each star in Table~\ref{tab:results} to compute their A(Mg) errors. 
Since the grids provide offsets induced by $\pm50$~K, the A(Mg) errors corresponding to the \teff\ errors in Table~\ref{tab:results} are obtained by computing the required quantity. For example, for a \teff\ error of $\pm100$~K, the error provided by the grid must be multiplied by two.
The typical A(Mg) error induced by \teffa\ errors in Table~\ref{tab:results} is $\pm0.05$~dex, whereas those induced by \logg, [Fe/H], and $v_{mic}$ are of the order of $10^{-3}$, thus they were neglected; Fig.~\ref{fig:Mg_offsets} shows the distribution of the offsets as function of the atmospheric parameters.
This error, $\pm0.05$~dex, is hence considered the typical A(Mg) error associated to the errors of all atmospheric parameters combined. This quantity is slightly smaller than the standard deviation of the abundance difference in Fig.~\ref{fig:Mg_dif}, this is $\pm0.07$~dex. Assuming this quantity is the result of the addition in quadrature of A(Mg) errors induced by the atmospheric parameters and the spectral noise and normalization, we deduce the latter is also $\pm0.05$~dex.
Therefore, we add the latter in quadrature to the A(Mg) error induced by \teffa\ to compute the total A(Mg) error given in Table~\ref{tab:results}.
 
\subsection{Carbon, nitrogen, and oxygen}

\begin{figure*}
    \centering
    \includegraphics[width=0.85\linewidth]{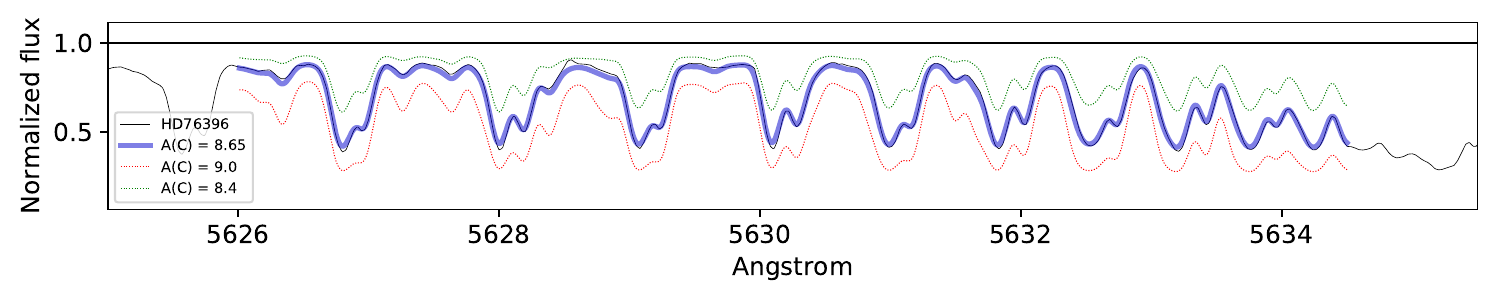}
    \includegraphics[width=0.85\linewidth]{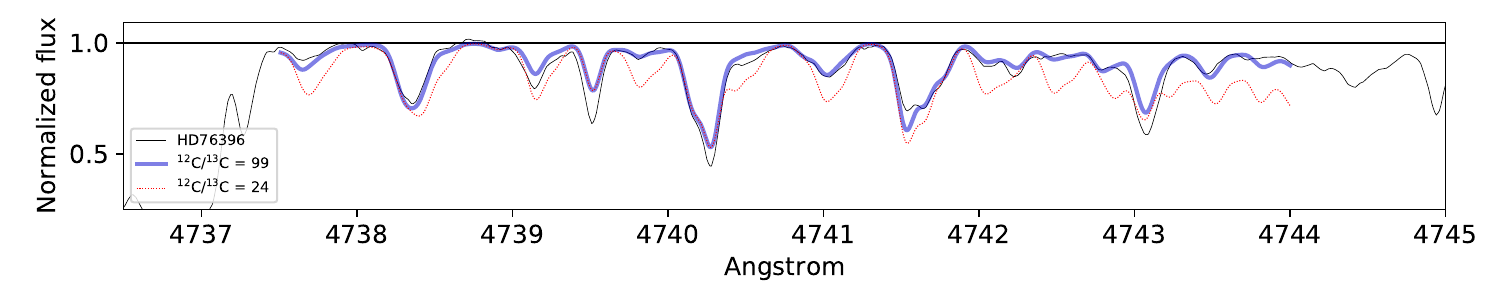}
    \includegraphics[width=0.85\linewidth]{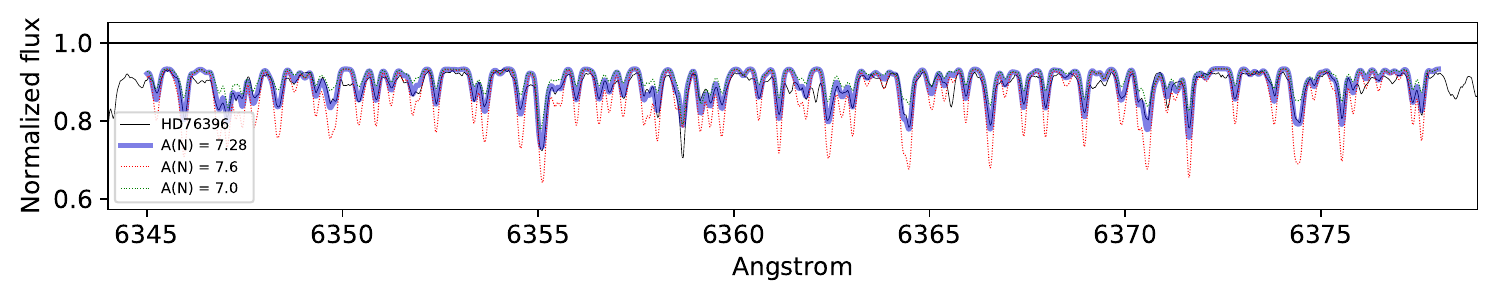}
    \caption{\tiny Spectral fits for the determination of C in the C$_2$ band (Top panel), of $^{12}$C to $^{13}$C ratio (Mid panel), and of N in the CN band (Low panel). Black lines represent the observational spectrum of HD~76396, blue lines represent the best modeled spectrum with abundances given in the legends, and green and red dotted lines represent modeled spectra with other abundances.}
    \label{fig:CNO}
\end{figure*}

C, N, and O abundances were derived following the strategy in \cite{karinkuzhi2021A&A...645A..61K}. LTE spectral synthesis were performed using  Turbospectrum with spherical MARCS model atmospheres \citep{gustafson2008} considering the linelist   
of \cite{heiter2021A&A...645A.106H}.
Oxygen was determined by fitting the triplet at 7771-7775~\AA.
C and N abundances were not determined by individual line fitting as for Fe, Mg and O, but by fitting molecular band regions. For this purpose we adapted a minimum $\chi^2$ routine, where grids of synthetic spectra varying the abundance of a given element were interpolated.
We derived C$_2$ from the 5626-5634.5~\AA\ region (avoiding the band-head at 5635~\AA), as it appears less saturated than the 5155-5164~\AA\ region, although similar abundances were obtained.
The $^{12}$C$/^{13}$C ratio was derived by fitting the 4737.5-4744~\AA\ region, and the $^{13}$C line at 8016.429~\AA\ which appears reasonably free from blends.
Nitrogen abundance was derived by fitting either the 6305-6330~\AA\ or the 6345-6378~\AA\ region once C was determined.
Figure~\ref{fig:CNO} shows an example of the spectral fitting for the determination of C, N, O, and $^{12}$C$/^{13}$C ratio for HD~76396.
Table~\ref{tab:CNO} lists CNO abundances derived for CEMP stars and their errors induced by the \teff, \logg, and [Fe/H] errors in Table~\ref{tab:results}, separately.

\begin{figure}
    \centering
    \includegraphics[width=0.9\linewidth]{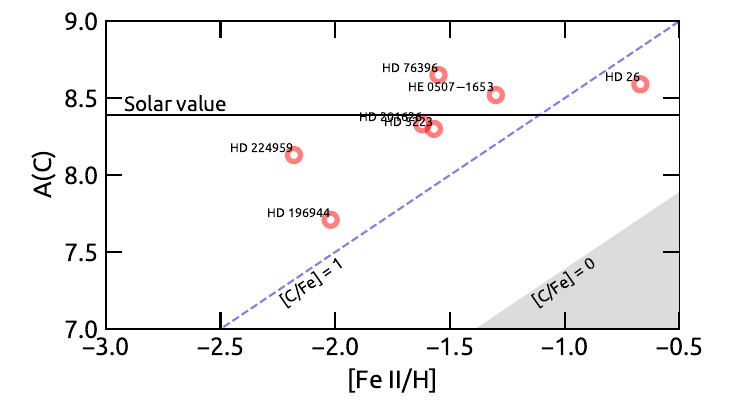}
    \caption{\tiny  CEMP star carbon abundances, A(C), as function of [Fe/H]. The solar A(C) corresponds to the Sun photospheric determination in \cite{grevesse2007SSRv..130..105G}.}
    \label{fig:C_abun}
\end{figure}

As shown on Fig.~\ref{fig:C_abun}, all our CEMP stars are indeed enriched in carbon, most of them with [C/Fe]$ > 1$.
They are consistent with the higher-metallicity, upper carbon plateau (at A(C) $\sim 8.25$) as identified in \cite{Spite-2013}, as expected since this region seems to contain mainly CEMP-s (and -sr) objects \citep{Hansen-2015, Bonifacio-2018}.
On the contrary, for lower metallicities ([Fe/H] $< -3.4$), the carbon abundance drops 
below A(C) = 7.6 \citep{Bonifacio-2018}
and this region seems to contain mainly CEMP-no objects.

\begin{table*}
\caption{Element abundances of the CEMP stars}
\label{tab:CNO}
\centering
\tiny 
\begin{threeparttable}
\begin{tabular}{lccccc}
\hline\hline
Name & A(C) & A(N) & A(O) & 
$^{12}$C/$^{13}$C \\
\hline
HD~26 & $8.61\pm0.02\pm0.01\pm0.02$ & $8.45\pm0.05\pm0.01\pm0.04$ & $8.62\pm0.05\pm0.07\pm0.00$ & 9\\
HD~5223 & $8.34\pm0.03\pm0.01\pm0.01$ & $7.49\pm0.08\pm0.01\pm0.02$ & $7.90\pm0.06\pm0.06\pm0.01$ & 9\\
HD~76396 & $8.66\pm0.05\pm0.00\pm0.01$ & $7.28\pm0.12\pm0.01\pm0.01$ & $8.13\pm0.07\pm0.04\pm0.01$ & $>99$\\
HD~196944 & $7.72\pm0.04\pm0.02\pm0.01$ & $7.87\pm0.05\pm0.01\pm0.02$ & $7.42\pm0.03\pm0.04\pm0.01$ & 5 &\\
HD~201626 & $8.34\pm0.03\pm0.01\pm0.01$ & $7.35\pm0.06\pm0.01\pm0.02$ & $7.99\pm0.05\pm0.03\pm0.01$& 32 \\
HD~224959 & $8.20\pm0.03\pm0.01\pm0.01$ & $7.32\pm0.08\pm0.01\pm0.01$ & $7.85\pm0.04\pm0.06\pm0.00$ & 11.5 &\\
HE~0507-1653 & $8.57\pm0.10\pm0.01\pm0.02$ & $7.99\pm0.11\pm0.01\pm0.03$ & $8.13\pm0.17\pm0.06\pm0.01$ & 32 & \\
\hline
\end{tabular}
\begin{tablenotes}
\item{} \textbf{Notes.} {Errors of the element abundances are due to \teff\ errors, \logg\ errors, and [Fe/H] errors listed in Table~\ref{tab:results}, from the left to the right, respectively.
} 
\end{tablenotes}
\end{threeparttable}
\end{table*}

\section{Conclusions}

\label{sec:conclusions}
We provide accurate and precise atmospheric parameters (\teff\, \logg, and [Fe/H]) and Mg, C, N, and O abundances for 47 metal-poor red giant stars 
(Table~\ref{tab:results}). 
We refer to this sample of benchmark metal-poor stars, along with the metal-poor dwarfs of \cite{giribaldi2021A&A...650A.194G}, as the \titan\ reference stars.
In this sample, 34 stars have the most precise parameters. Namely, they have total uncertainties of 
40-80~K in \teff, 0.15~dex in \logg, 0.09~dex in [Fe/H], and 0.07~dex in A(Mg).
We tested the accuracy of the derived atmospheric parameters with most of the (few) standard metal-poor stars available up to date, from which we could either obtain new HERMES spectra or recover spectra from the ESO archives. We summarize below the arguments in support of the \teff, \logg, and [Fe/H] determination accuracy.

We derived effective temperatures (\teffa) by fitting observational H$\alpha$ line profiles with 3D non-LTE synthetic models \citep{amarsi2018}.
The evaluation of the accuracy of \teffa\ is graphically summarized in Fig.~\ref{fig:interferometry}, where the reference temperatures are either from interferometry (red symbols) or IRFM (dark symbols).
The three scales are consistent and can be considered equivalently accurate. This is consistent with results obtained for dwarfs in \cite{giribaldi2021A&A...650A.194G}. 
We adopt the dispersion of the comparison in the figure, $\pm$46~K, as the  uncertainty of the 3D non-LTE H$\alpha$ model for giants.
The outlier stars HD~2665, HD~221170, and HD~175305, have relatively low interferometric \teff\ values, which are likely effects of small angular diameter offsets; see Sect.~\ref{sec:accuracy_teff} for details.

Based on the outcome above, we evaluate the accuracy of the color-dependent \teff\ relations of \cite{casagrande2021MNRAS.507.2684C} currently available for Gaia photometry. The evaluation is summarized in Fig.~\ref{fig:phot_test}, where we observe excellent agreement for $BP - RP$ and $G - BP$ colours, disregarding CEMP stars.
In Table~\ref{tab:results} we list a second set of temperatures derived by other methods found to correspond with \teffa. They can be averaged with \teffa\ to obtain more precise \teff\ of the \titan. 
We selected, when available, temperatures derived by the most direct to the less direct method, following the next hierarchy: 1) interferometry\footnote{Interferometric \teff\ of the outliers HD~2665, HD 221170, and HD~175305 were not included due to possible systematics.}, 2) IRFM, and 3) photometric calibrations using $BP - RP$. 

We provide surface gravities derived by fitting Mg~I~b triplet lines at 5172 and 5183~\AA. We verified their accuracy with  stars with asteroseismic gravities, see Fig~\ref{fig:logg_compare}.
Surface gravities from Yonsey-Yale isochrones \citep{kim2002,yi2003} resulted constantly higher by $\sim$0.35~dex along the parameter space analysed, see Fig.~\ref{fig:Mg_calibration}.
Still, they can be used to derive accurate \logg\ in the absence of Mg triplet lines, by subtracting the given offset.

We provide iron abundances from neutral and ionised species. We show in Sect.~\ref{sec:fe_LTE} that, provided accurate \teff\ and \logg\ fix inputs, LTE model atmospheres predict ionization unbalance with [\ion{Fe}{i}/H] abundances constantly lower than [\ion{Fe}{ii}/H] by $\sim$0.15~dex, within  the parameter space analysed for giants, see Fig.~\ref{fig:Fe_dif}.
We confirm in Sect.~\ref{sec:fe_NLTE} that non-LTE models restore the ionization balance within typical uncertainties ($\sim$0.05~dex), whereas [\ion{Fe}{ii}/H] under LTE and non-LTE are virtually equal, see Fig.~\ref{fig:NLTE_LTE}.

We provide Mg abundances for all stars in the sample, and CNO abundances for the CEMP stars.
Mg abundances of all stars were made consistent to the scale to the weakest line at 5711~\AA, and were corrected from non-LTE effects.
We show that Mg abundances from the lines 5528 and 5711~\AA\ under LTE, may differ substantially depending the stellar metallicity and the Mg abundance itself, see Fig.~\ref{fig:Mg_dif}. The highest discrepancies are found for the highest metallicity ([Fe/H] $\gtrsim -1$~dex) and abundances (A(Mg) $\gtrsim 7$~dex), as for dwarfs \citep[][Fig.~3]{giribaldi2023A&A...673A..18G}.
[C/Fe] rates of all CEMP stars are higher than 1, except for HD~26, the most metal rich star in the sample and spectroscopic binary as well.
Heavy elements of CEMP stars will be presented in Giribaldi et al. in prep.

A few \titan\ seem to have started their path to the horizontal branch (HB), these are BD+17~3248, BD+09~2860, CD$-41$~15048, CD$-62$~1346, BD+11~2998, and HD~196944, the latter being a CEMP star; they are tagged in the HR diagram of Fig~\ref{fig:kiel}. 
BD+17~3248 seems a genuine HB star, it have been already labeled as such, although with significantly higher \teff\ and lower \logg\ values \citep{2010AJ....140.1694F}. The other five stars shown signs of binarity: CD$-41$~15048, BD+09~286, BD+11~2998 show proper motion accelerations from Hipparcos and Gaia's astrometric data \citep{2019A&A...623A..72K, 2021ApJS..254...42B}; CD$-62$~1346 shows significant variation of its radial velocities \citep{2019A&A...626A.128E}; and HD~196944 is certainly a spectroscopic binary \citep{karinkuzhi2021A&A...645A..61K}.
Our observational \teff\ and luminosities show that CEMP stars are rather located to the right part of the normal red giant branch stars, which is supported by the STAREVOL evolutionary tracks specifically computed for CEMP stars \citep{siess2008A&A...489..395S}.

We recommend the use of [\ion{Fe}{i}/H] in Table~\ref{tab:results} only for reproducing observational \ion{Fe}{i} lines via synthesis under LTE, whereas [\ion{Fe}{ii}/H] are accurate quantities required for evolutionary analysis of stellar populations. Similarly, we recommend to use A(Mg) in Table~\ref{tab:results} corrected by $-0.07$ to reproduce observational lines under LTE, as the quantities listed are corrected from non-LTE effects.

Finally, we recommend to avoid the use of the excitation and ionization equilibrium of Fe and spectral fitting under LTE to simultaneously derive \teff, \logg, and [Fe/H] in metal-poor giants. 
They will likely provide biased determinations, as demonstrated in this work.
Temperature outcomes of these two spectroscopic methods are consistent with those from the color-\teff\ relations of \cite{alonso1996A&AS..117..227A}, however all are biased to cooler values; see also \cite{Casagrande2010} where the same offset as here was obtained. 
Users may confidently upgrade their methods for fast \teff\ determination using the Gaia colour-\teff\ relations with  $BP - RP$ and $G - BP$ \citep{casagrande2021MNRAS.507.2684C} for giants, and preferably with $BP - RP$ for dwarfs. Alternatively, the differential spectroscopic approach \citep[e.g.][]{melendez2012A&A...543A..29M,reggiani2016A&A...586A..67R} can be used with with one or more \titan\ as standards.

\begin{acknowledgements}
R.E.G. acknowledges the support by Fonds de la Recherche Scientifique (F.R.S.-FNRS) and the Fonds Wetenschappelijk Onderzoek-Vlaanderen (FWO) under the EOS programme (numbers O.0004.22 and O022818F). R.E.G. acknowledges Ana Consuelo Giribaldi Santamaria for her incessant encouragement. SVE thanks the Fondation ULB for its support. T.M. is granted by the BELSPO Belgian federal research program FED-tWIN under the research profile Prf-2020-033\_BISTRO.
This work is based on observations collected
at the European Southern Observatory under ESO programmes: 074.B-0639(A), 68.D-0546(A), 165.N-0276(A), 0103.A-9013(A), 0104.A-9005(A), 097.A-9024(A), 68.B-0618(A), 65.N-0534(A), 167.D-0173(A), 076.B-0055(A), 072.C-0488(E), 090.B-0605(A), 076.B-0055(A), 072.B-0585(A), 083.B-0281(A), 0103.D-0310(A), 68.D-0546(A), 0104.D-0059(A), 079.D-0567(A), 078.B-0238(A), 080.D-0333(A).
Use was made of the Simbad database, operated at the CDS, Strasbourg, France, and of NASA’s Astrophysics Data System Bibliographic Services. 
This research used Astropy,\footnote{http://www.astropy.org} a community-developed core Python package for Astronomy \citep{astropy:2018}.
This work presents results from the European Space Agency (ESA) space mission Gaia. Gaia data are processed by the Gaia Data Processing and Analysis Consortium (DPAC). Funding for the DPAC is provided by national institutions, in particular the institutions participating in the Gaia MultiLateral Agreement (MLA). The Gaia mission website is \url{https://www.cosmos.esa.int/gaia}. The Gaia archive website is \url{https://archives.esac.esa.int/gaia}.
The Mercator telescope is operated thanks to grant number G.0C31.13 of the FWO under
the "Big Science" initiative of the Flemish government. Based on observations obtained with the HERMES spectrograph, supported by the Fund for
Scientific Research of Flanders (FWO), the Research Council of K.U.Leuven, the Fonds National de la Recherche Scientifique (F.R.S.- FNRS), Belgium, the
Royal Observatory of Belgium, the Observatoire de Genève, Switzerland and the Thüringer Landessternwarte Tautenburg, Germany.
\end{acknowledgements}

\newpage
\begin{appendix}

\section{ }
\begin{figure}[]
    \centering
    \includegraphics[width=0.48\linewidth]{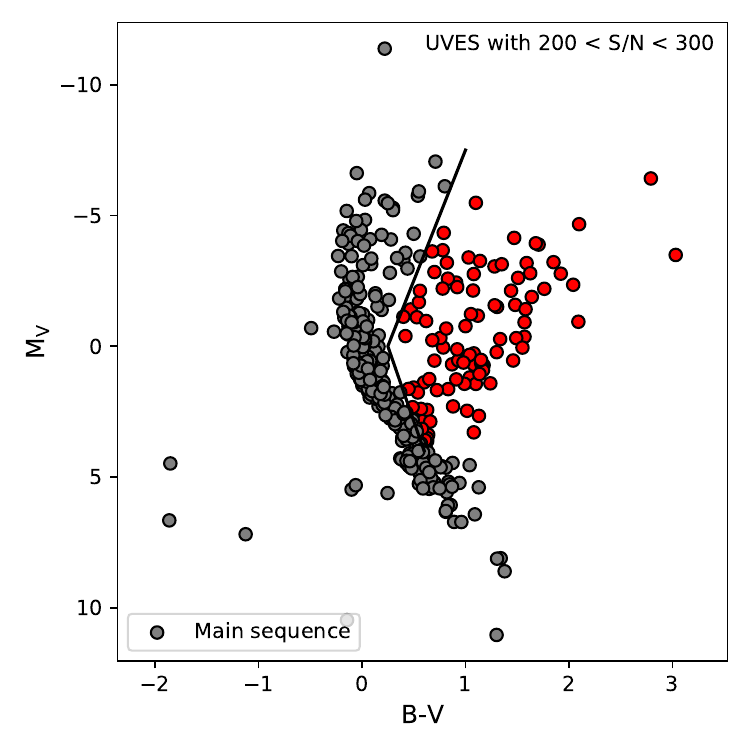}
    \includegraphics[width=0.48\linewidth]{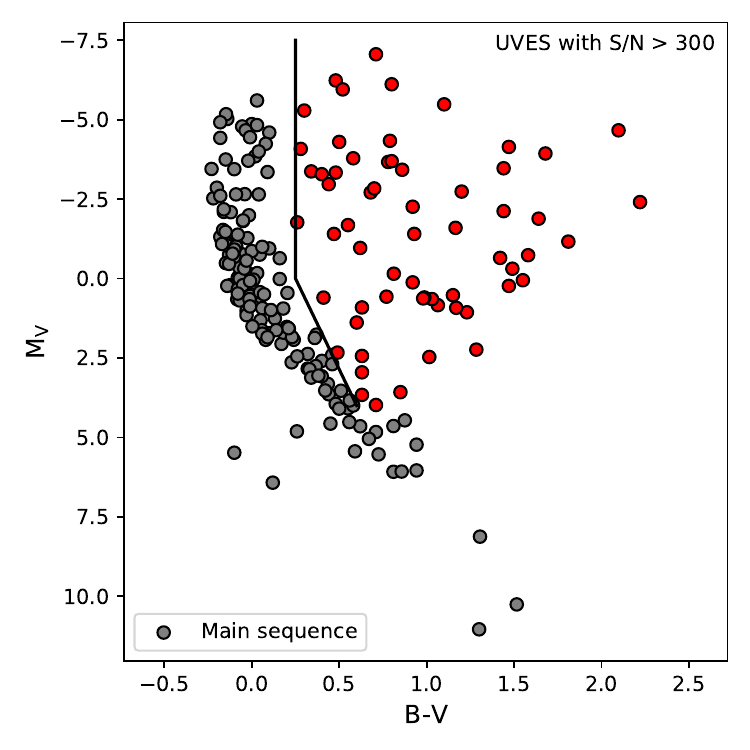}
    \caption{Examples of the initial selection of red giants with UVES spectra in the ESO archive.
    Left panel displays stars with spectra of $200 \leq S/N \leq 300$, whereas right panel displays stars with $S/N > 300$.
    The limits set are to separate main sequence stars (gray symbols) from red giant candidates (red symbols) are arbitrary.}
    \label{fig:giants_selection}
\end{figure}

\begin{figure}[]
    \centering
    \includegraphics[width=1\linewidth]{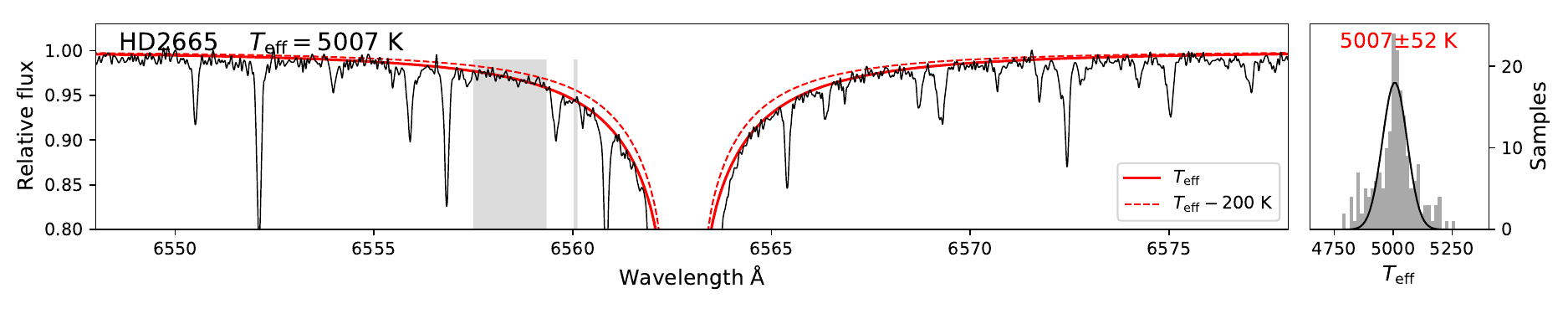}
    \includegraphics[width=1\linewidth]{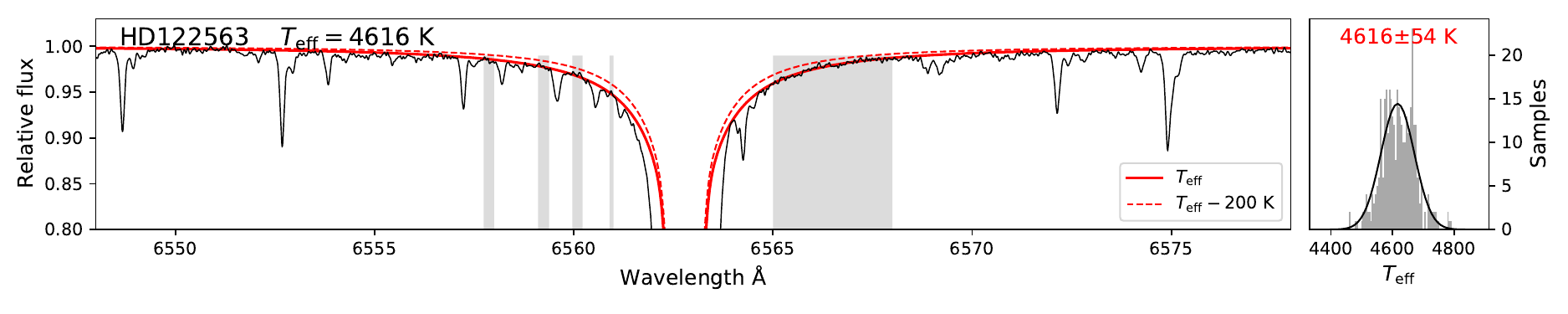}
    \includegraphics[width=1\linewidth]{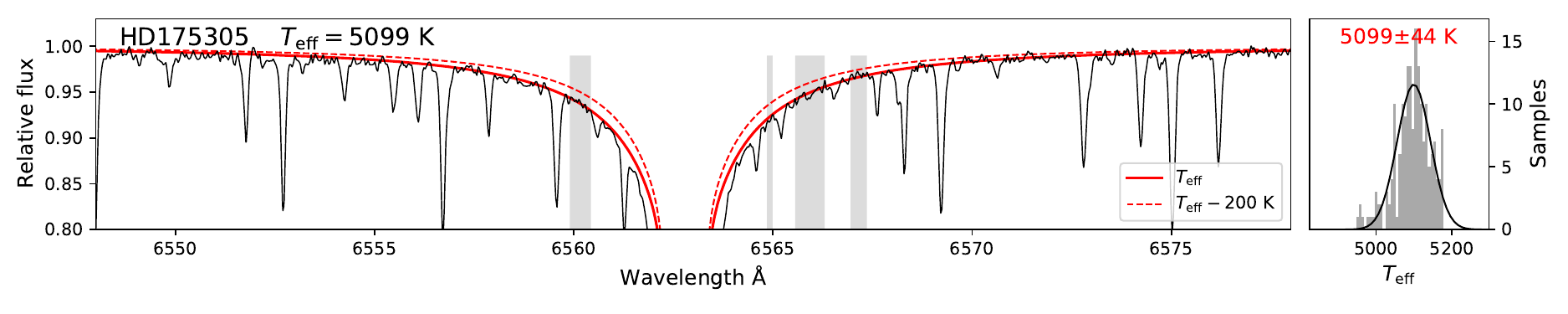}
    \includegraphics[width=1\linewidth]{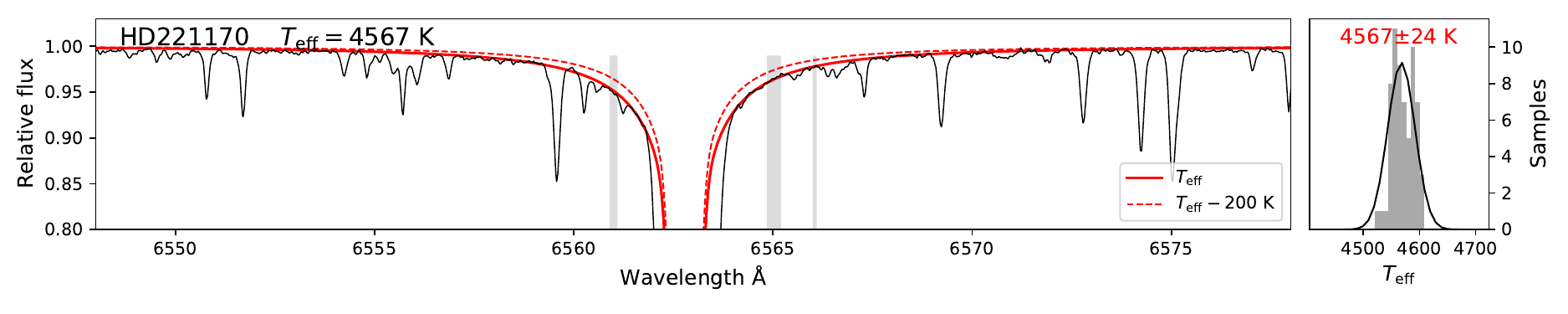}
    \caption{H$\alpha$ profile fits of stars with interferometric measurements. 
    {\it Left panels:} Observational spectra are in black, whereas fitted synthetic H$\alpha$ lines are represented by red continuous lines. 
    A synthetic profile with $-200$~K is represented by the red dashed line to provide a view of the sensitivity of the flux along the wavelength axis.
    Shades indicate the windows of fits without metal or telluric lines.
    {\it Right panels:} Histograms of the temperatures associated to each wavelength bin inside of the windows of fits. 
    Gaussians fitted to the histograms are represented in black, whose peak centers and 1$\sigma$ dispersions are indicated in red. They correspond to the determined \teffa\ and their fitting errors.
    } 
    \label{fig:Ha_interferometry}
\end{figure}

\begin{figure}[]
    \centering
    \includegraphics[width=1\linewidth]{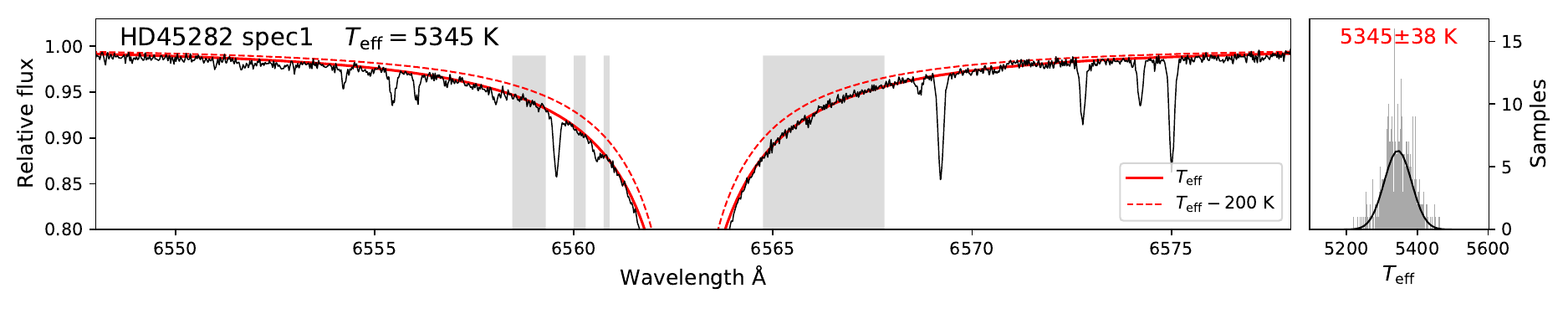}
    \includegraphics[width=1\linewidth]{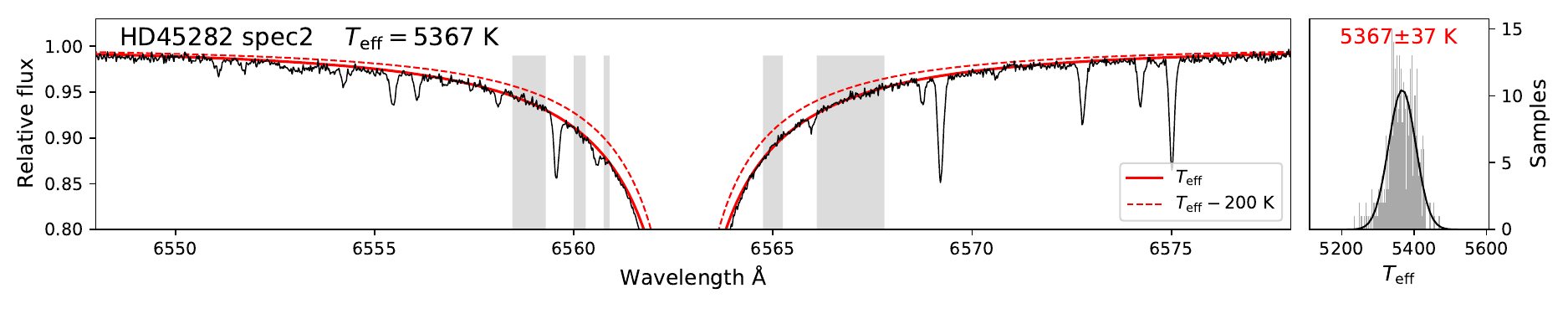}
    \includegraphics[width=1\linewidth]{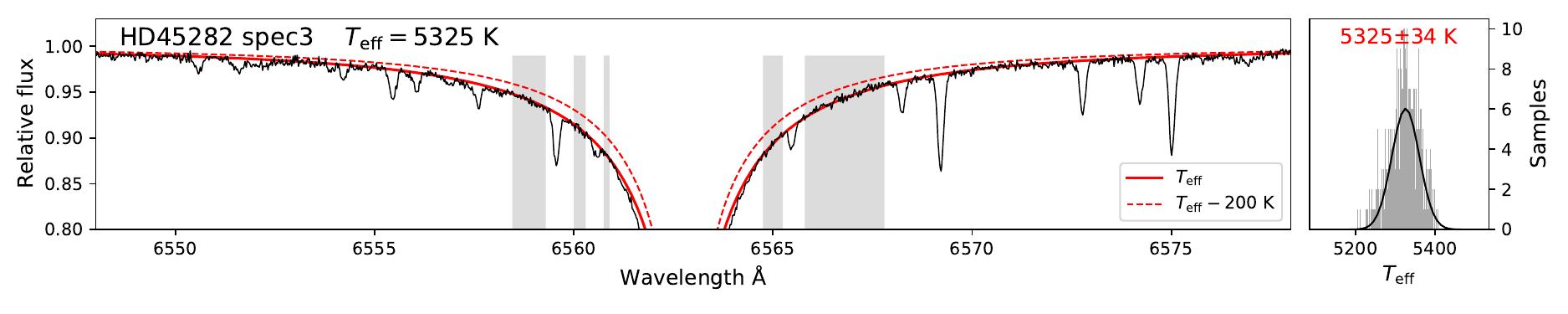}
    \includegraphics[width=1\linewidth]{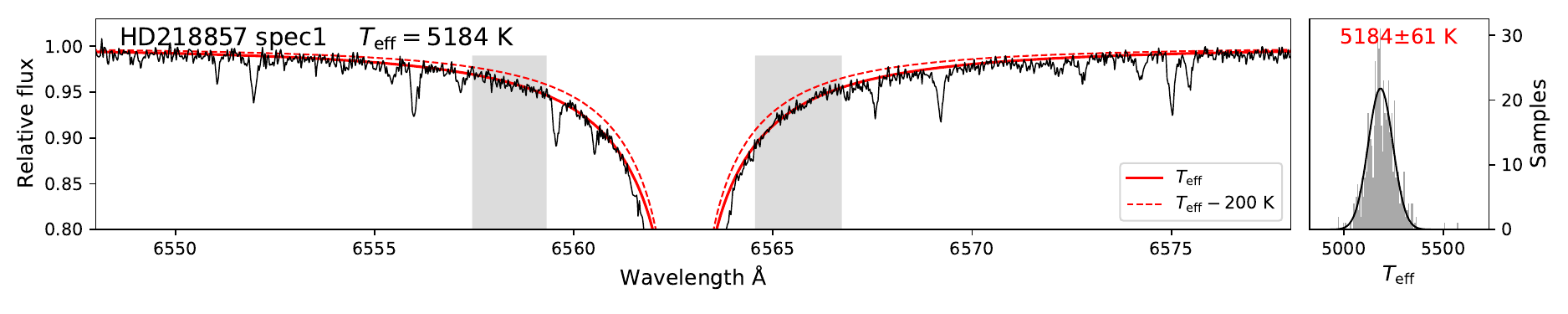}
    \includegraphics[width=1\linewidth]{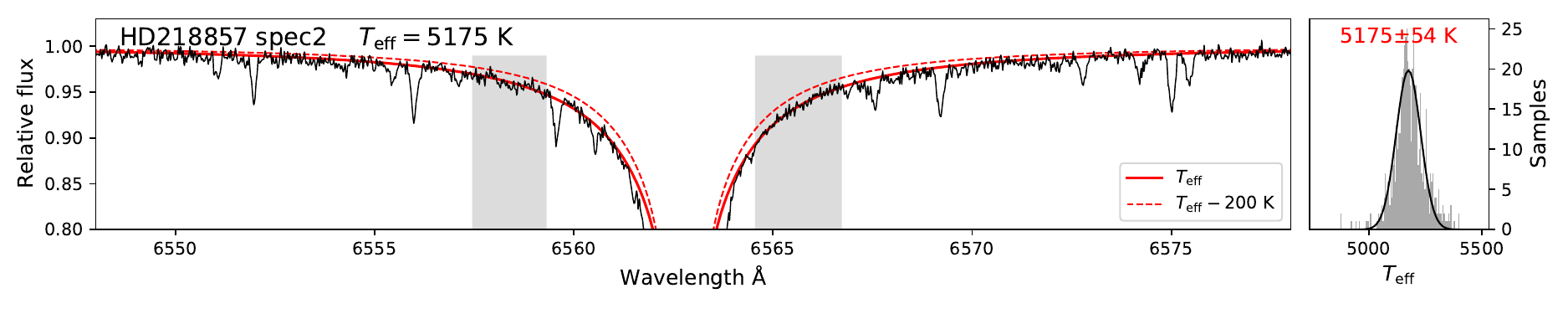}
    \includegraphics[width=1\linewidth]{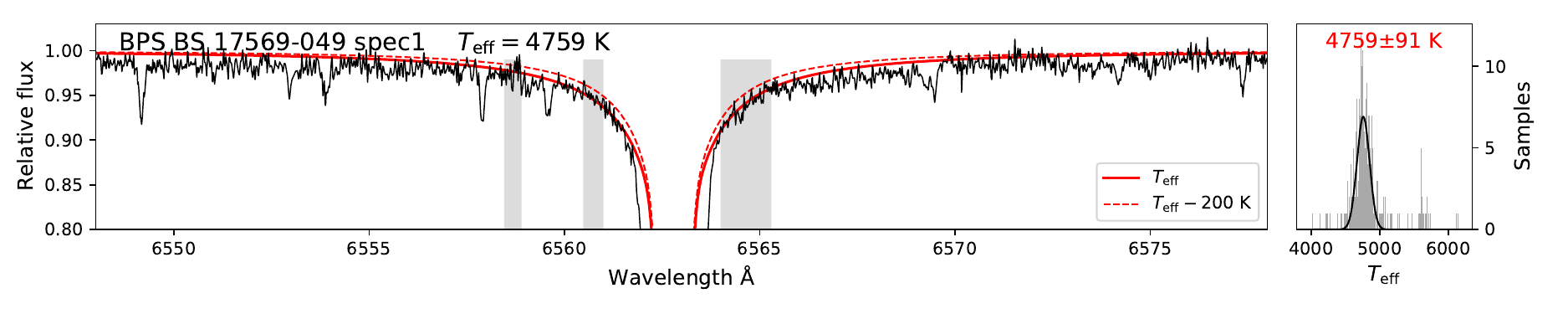}
    \includegraphics[width=1\linewidth]{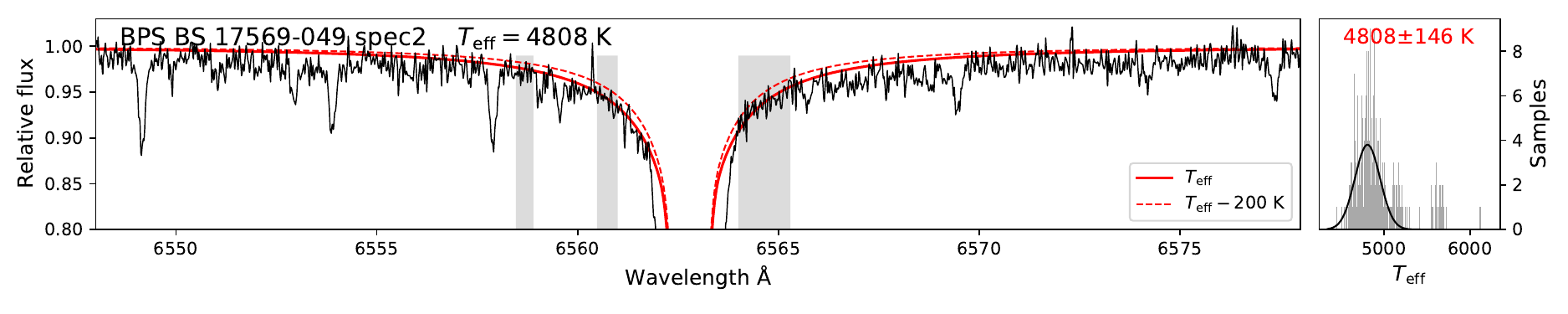}
    \includegraphics[width=1\linewidth]{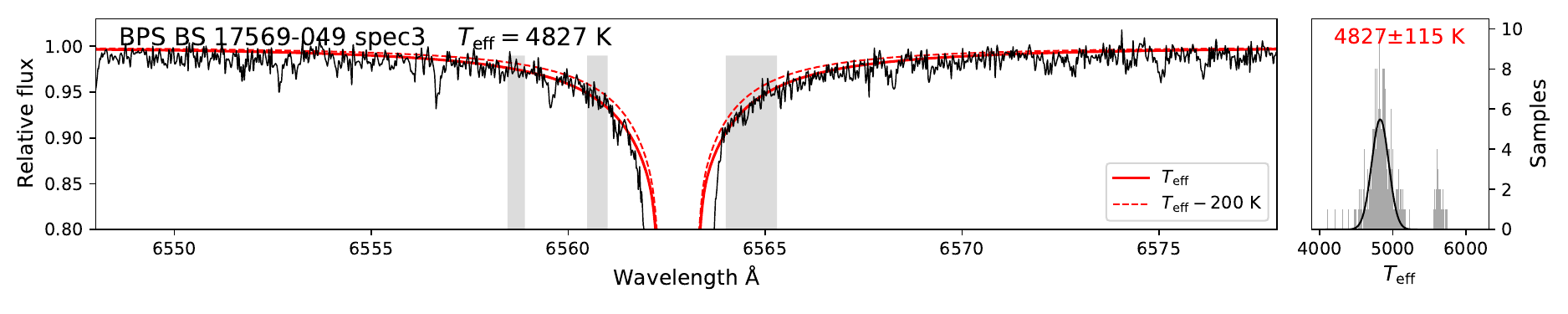}
    \includegraphics[width=1\linewidth]{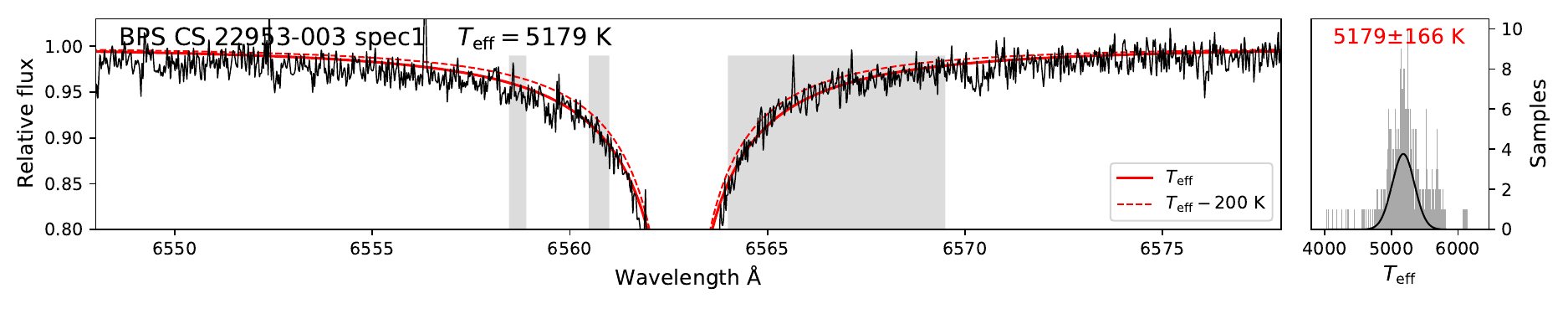}
    \includegraphics[width=1\linewidth]{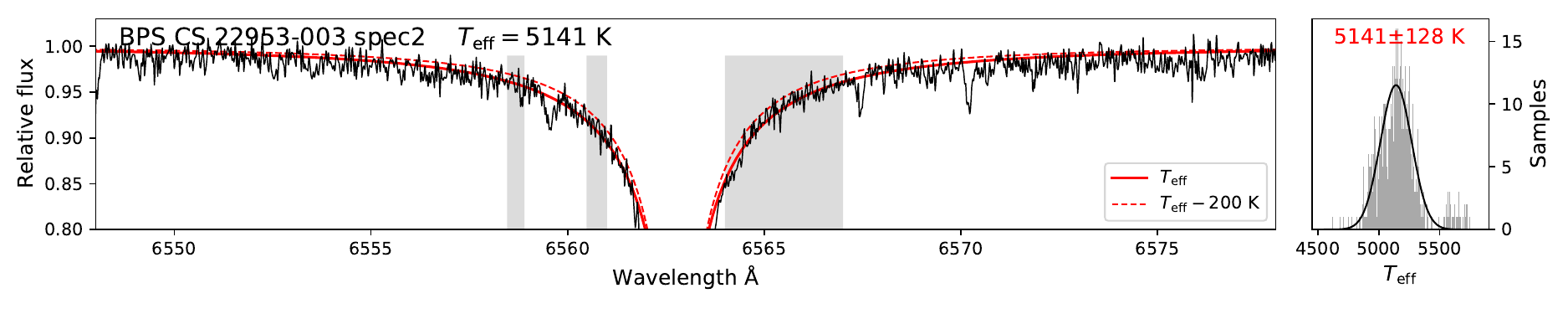}
    \includegraphics[width=1\linewidth]{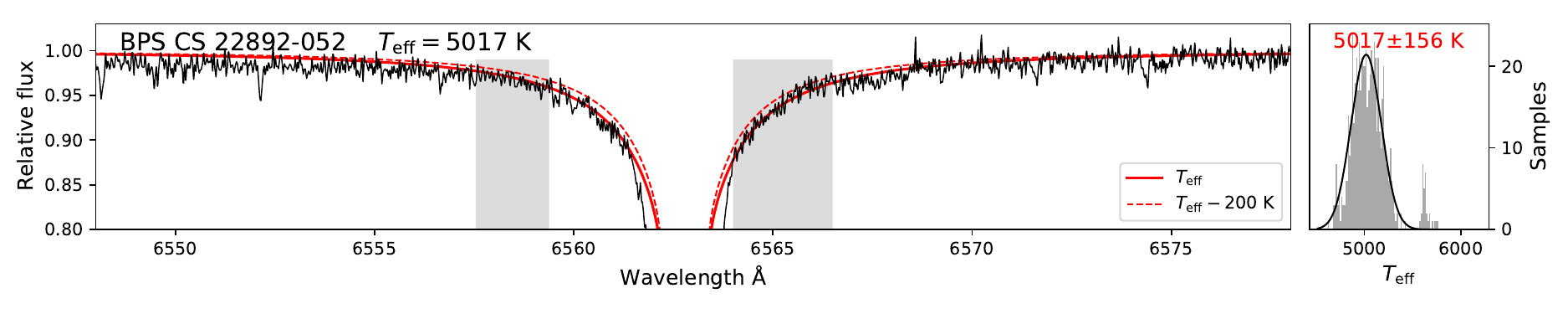}
    \caption{Similar to Fig.~\ref{fig:Ha_interferometry} for stars with direct IRFM measurements.}
    \label{fig:Ha_IRFM}
\end{figure}

\begin{figure}[]
    \centering
    \includegraphics[width=1\linewidth]{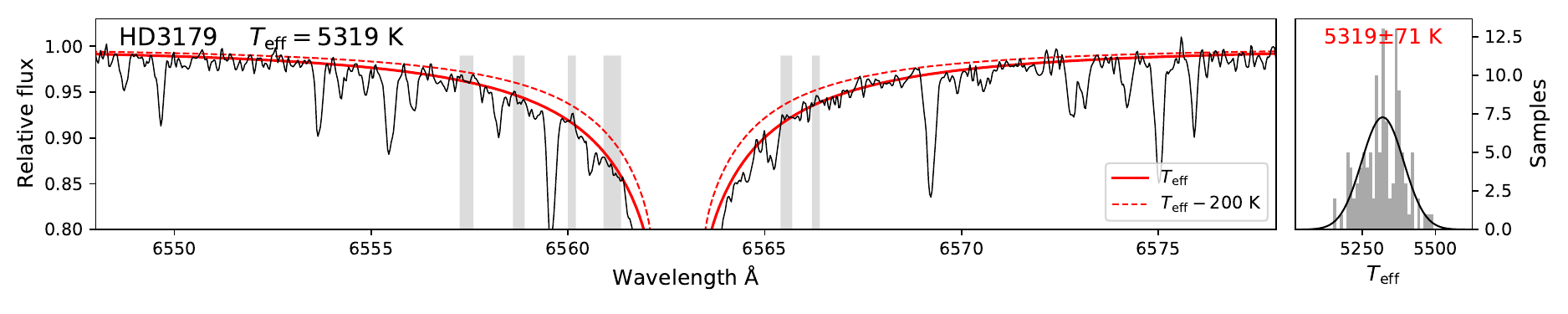}
    \includegraphics[width=1\linewidth]{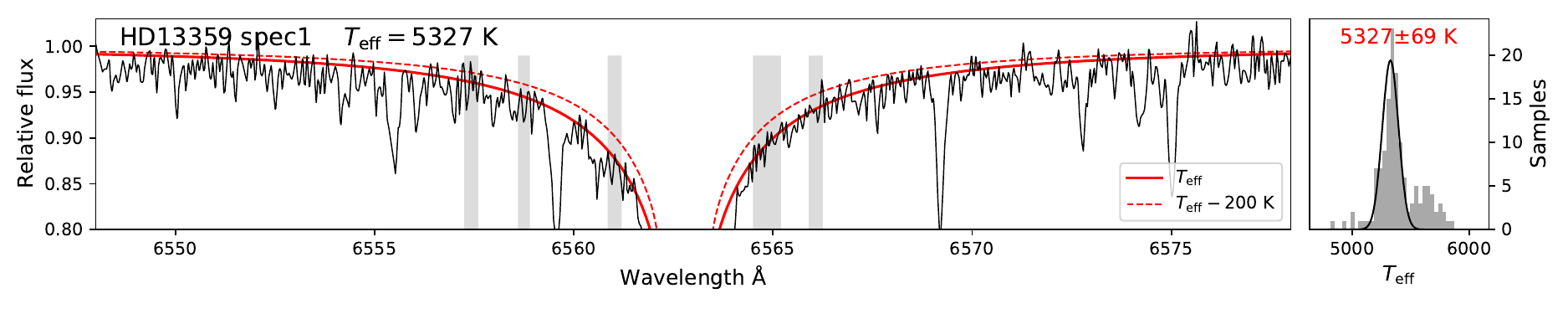}
    \includegraphics[width=1\linewidth]{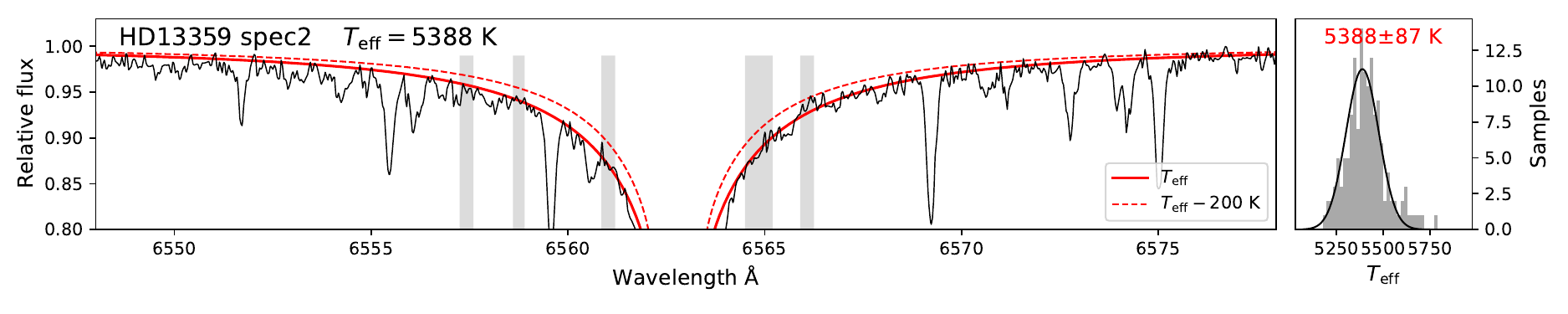}
    \includegraphics[width=1\linewidth]{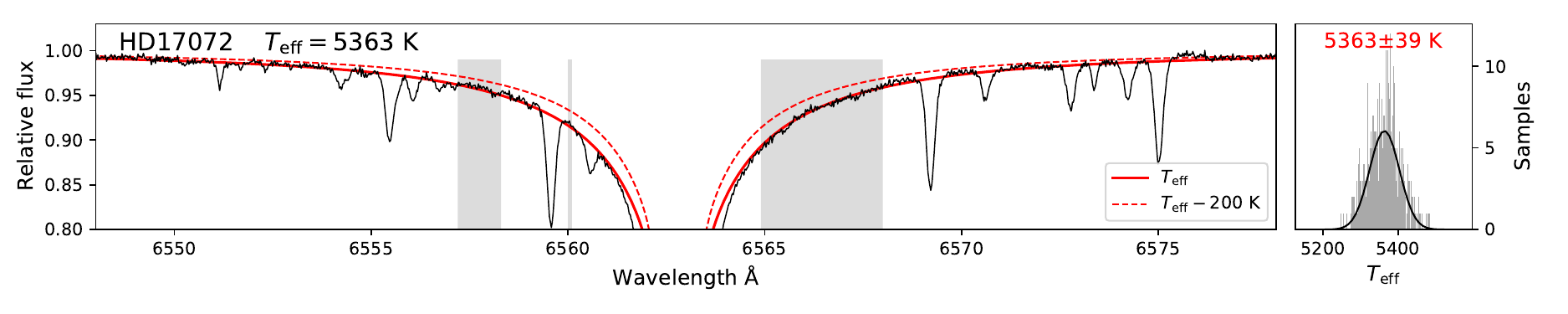}
    \includegraphics[width=1\linewidth]{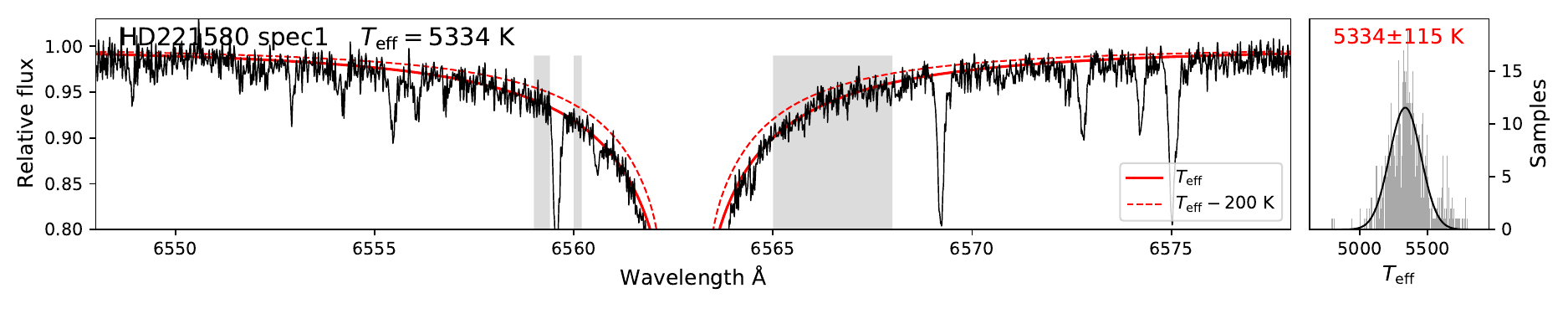}
    \includegraphics[width=1\linewidth]{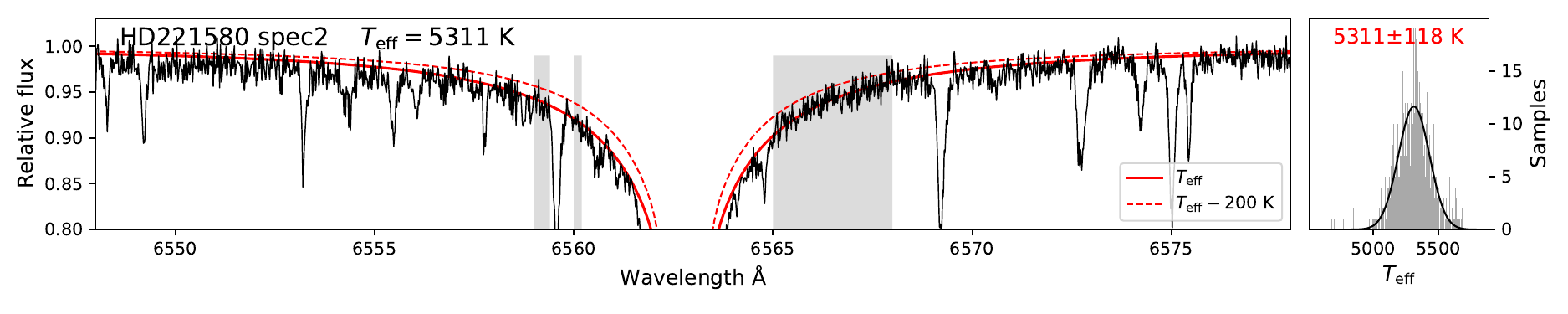}
    \includegraphics[width=1\linewidth]{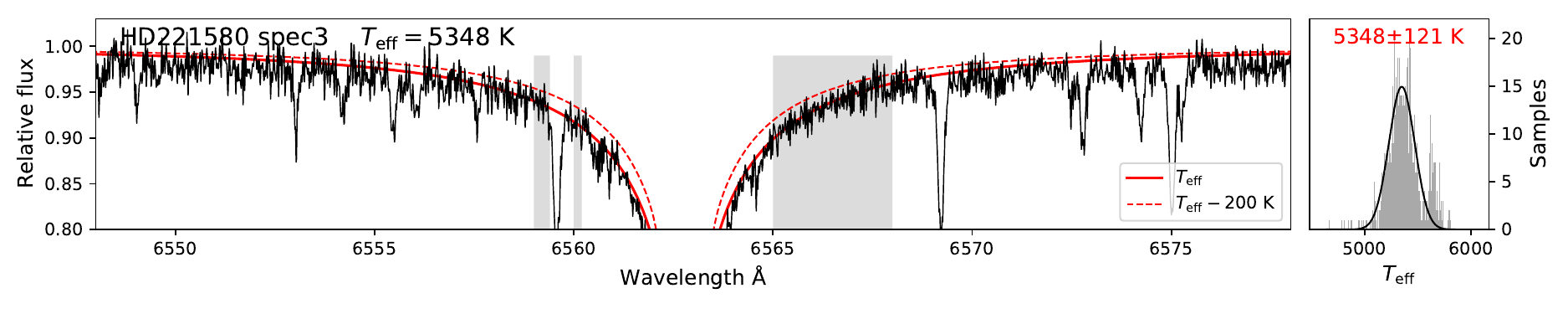}
    \caption{Similar to Fig.~\ref{fig:Ha_interferometry} for stars with asteroseismic measurements.}
    \label{fig:Ha_asteroseismic}
\end{figure}

\begin{figure}[]
    \centering
    \includegraphics[width=1\linewidth]{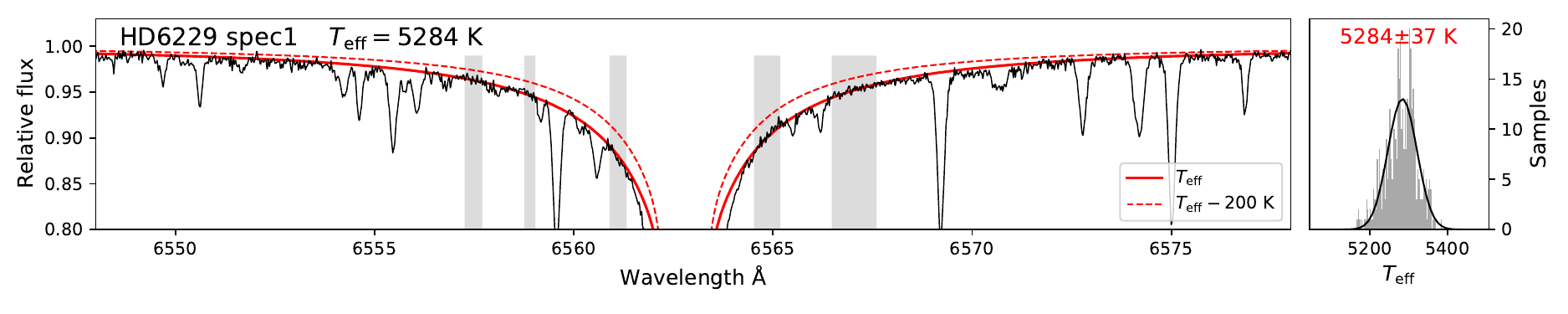}
    \includegraphics[width=1\linewidth]{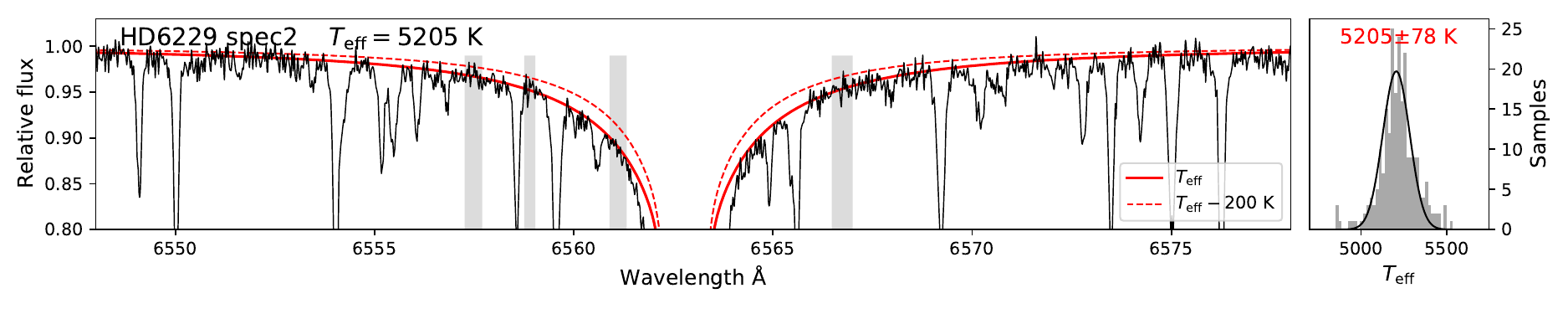}
    \includegraphics[width=1\linewidth]{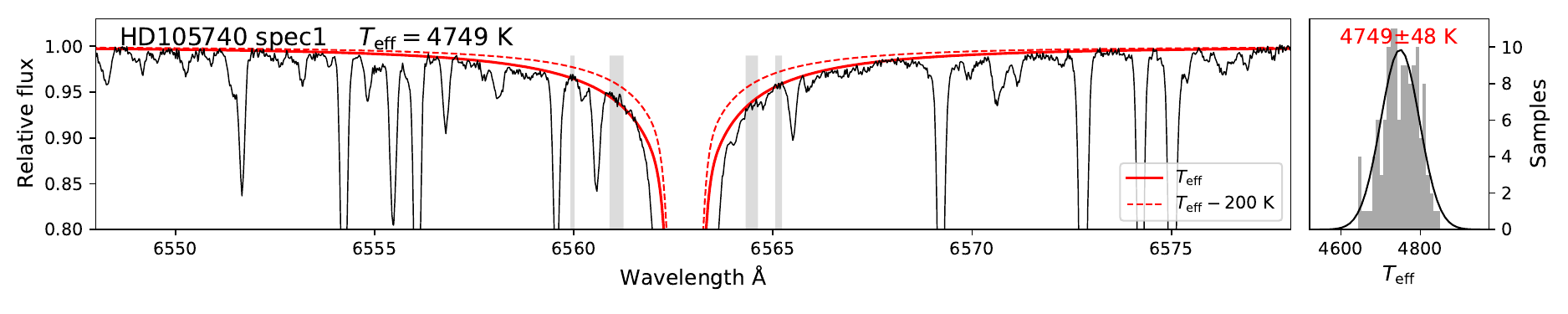}
    \includegraphics[width=1\linewidth]{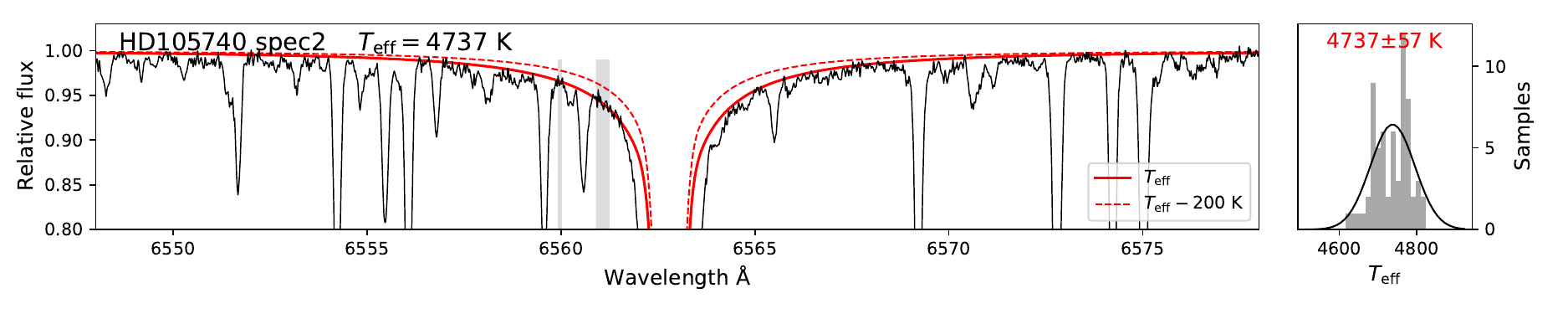}
    \includegraphics[width=1\linewidth]{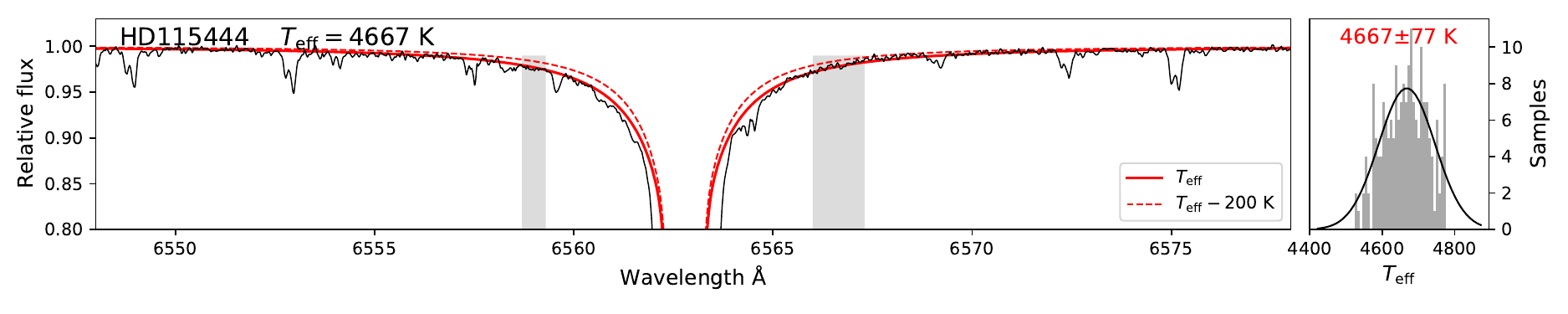}
    \includegraphics[width=1\linewidth]{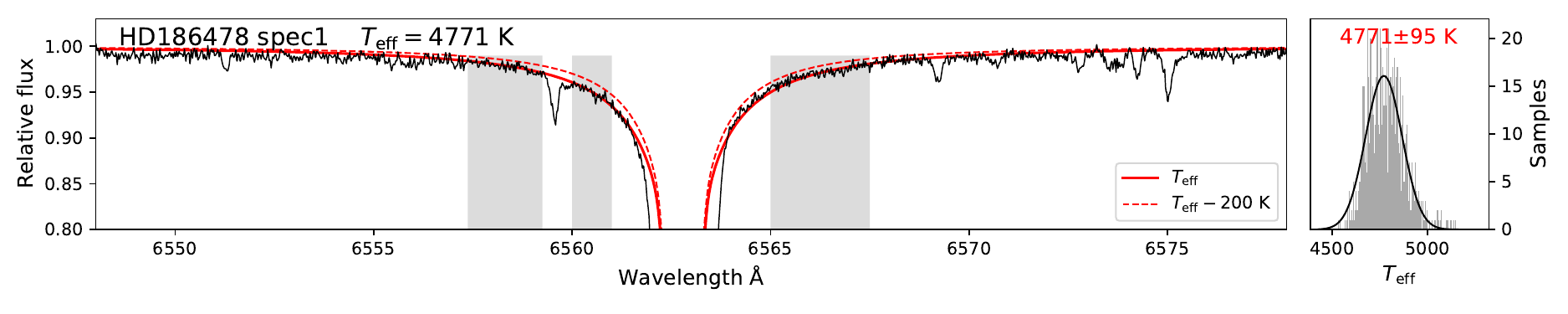}
    \includegraphics[width=1\linewidth]{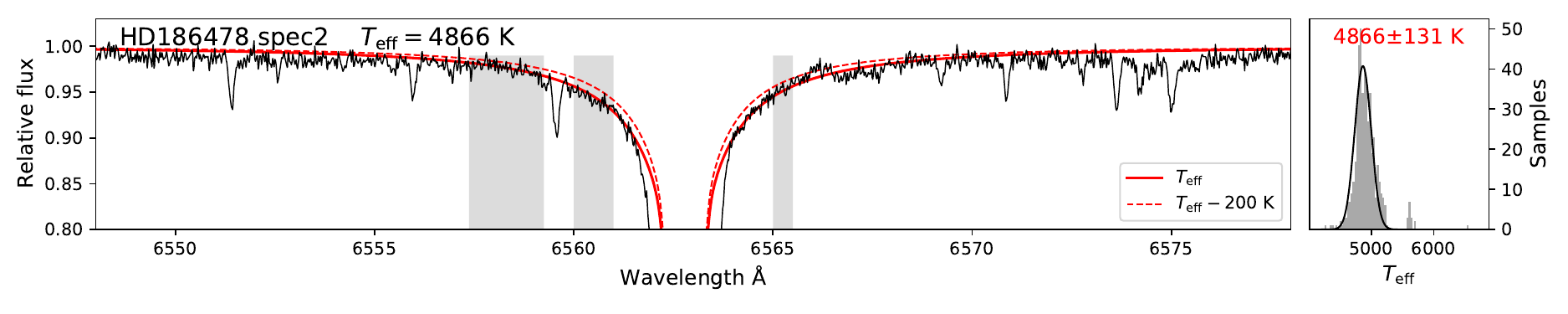}
    \includegraphics[width=1\linewidth]{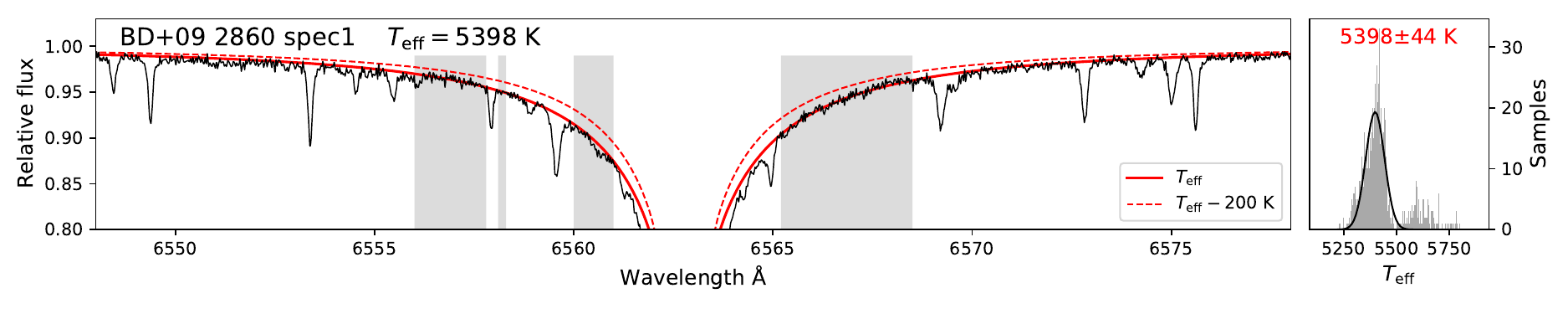}
    \includegraphics[width=1\linewidth]{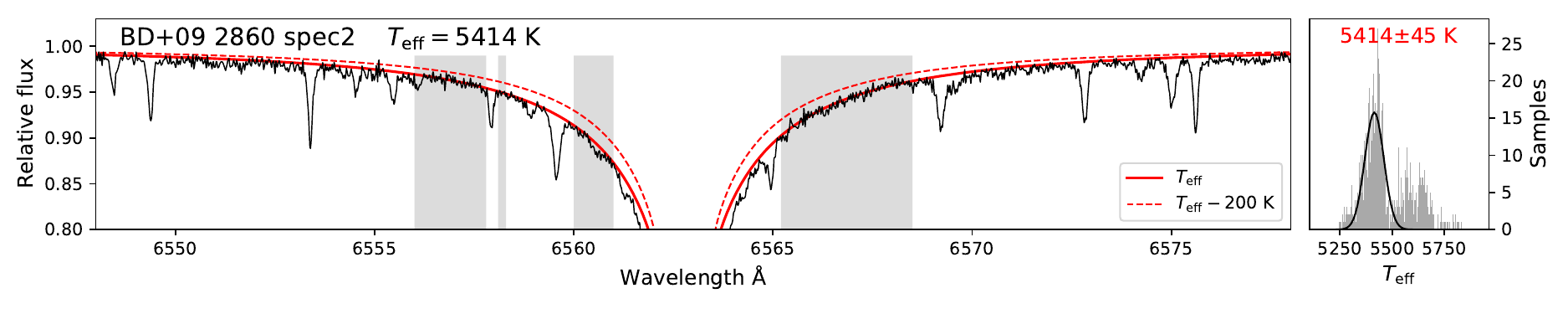}
    \includegraphics[width=1\linewidth]{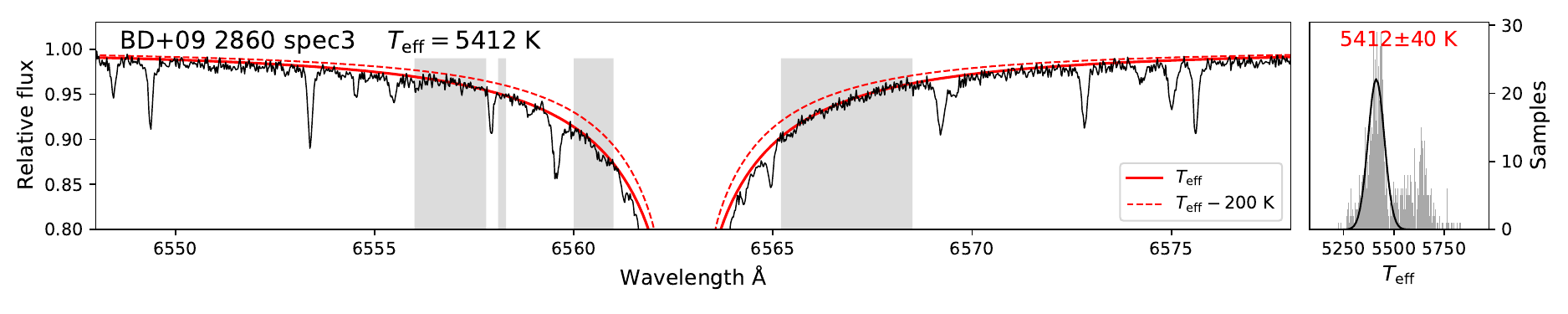}
    \includegraphics[width=1\linewidth]{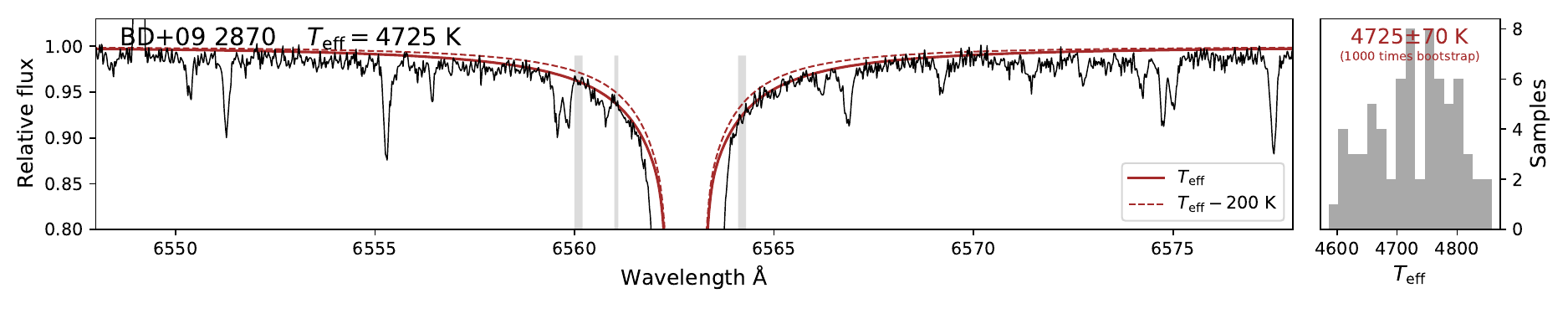}
    \includegraphics[width=1\linewidth]{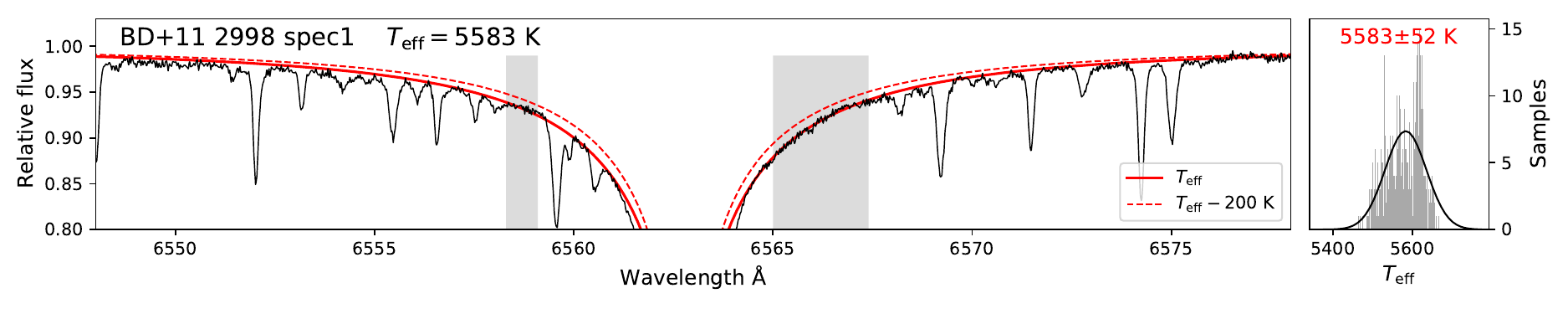}
    \includegraphics[width=1\linewidth]{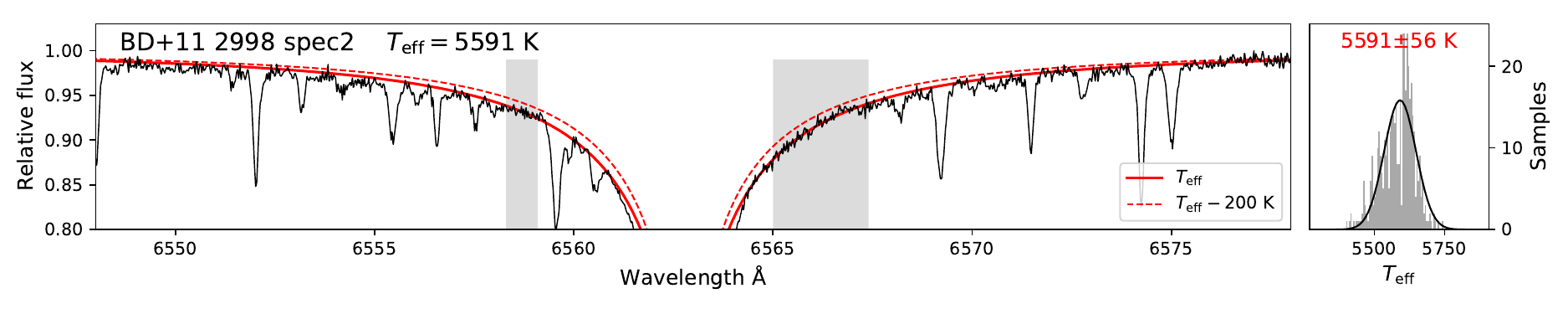}
    \label{fig:Ha_other_giants}
\end{figure}

\begin{figure}[]
    \centering
\includegraphics[width=1\linewidth]{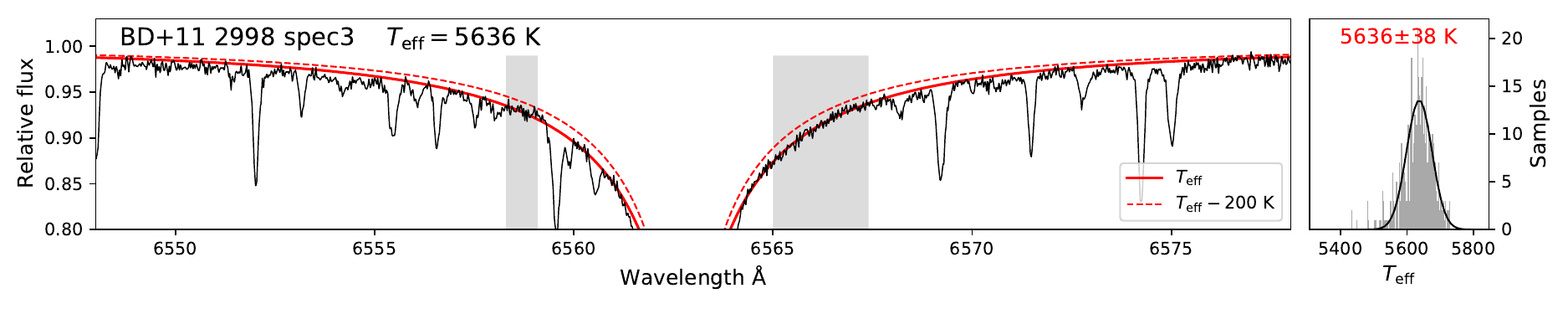}
\includegraphics[width=1\linewidth]{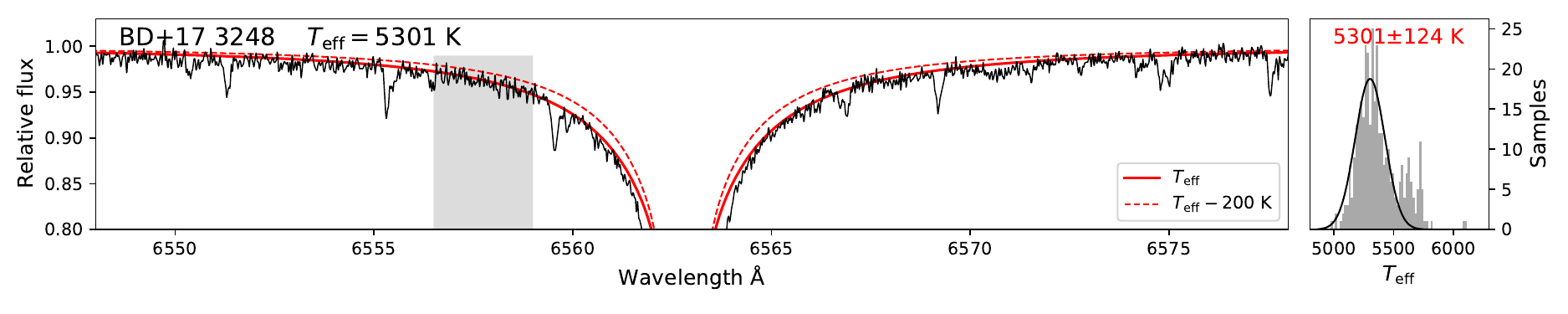}
\includegraphics[width=1\linewidth]{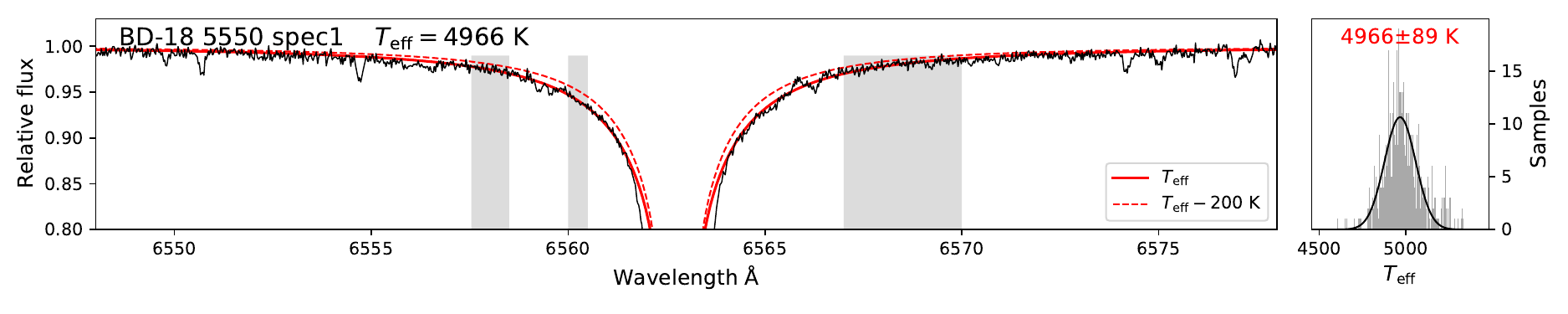}
\includegraphics[width=1\linewidth]{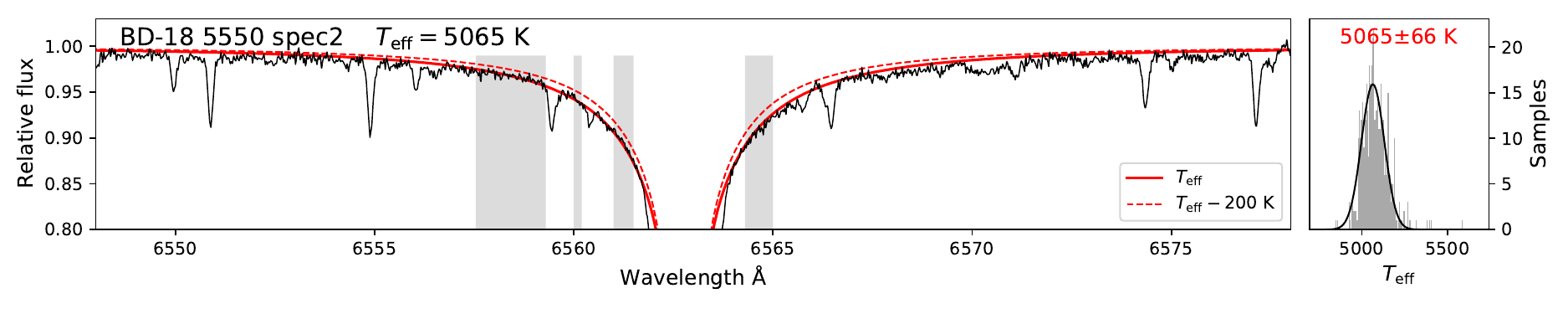}
\includegraphics[width=1\linewidth]{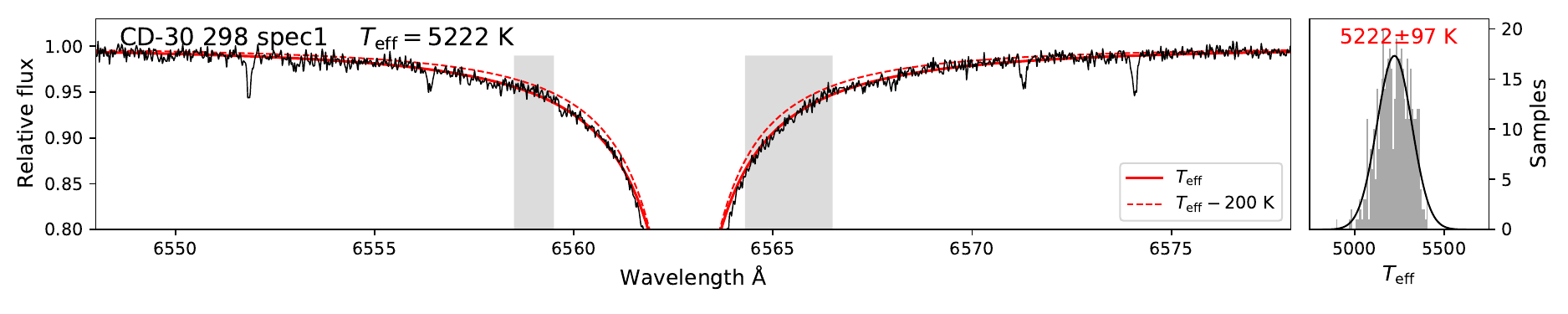}
\includegraphics[width=1\linewidth]{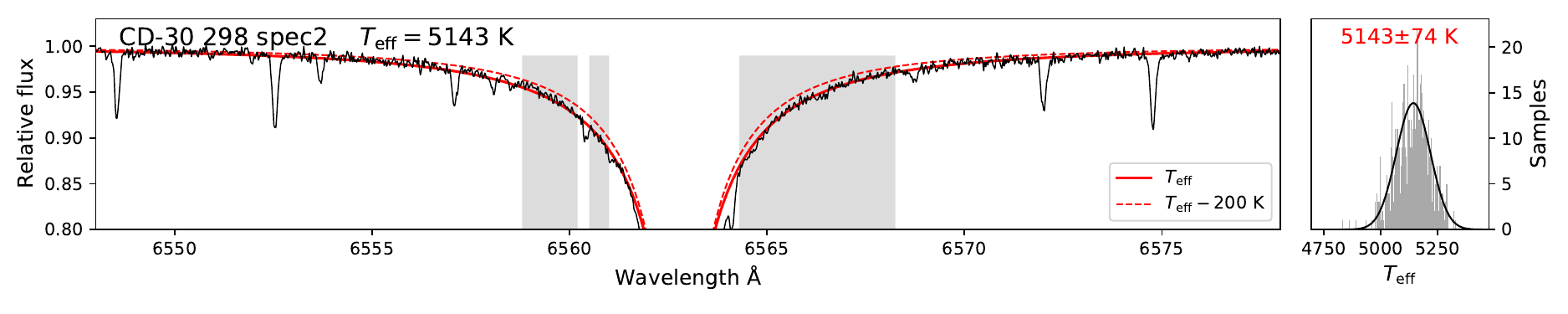}
\includegraphics[width=1\linewidth]{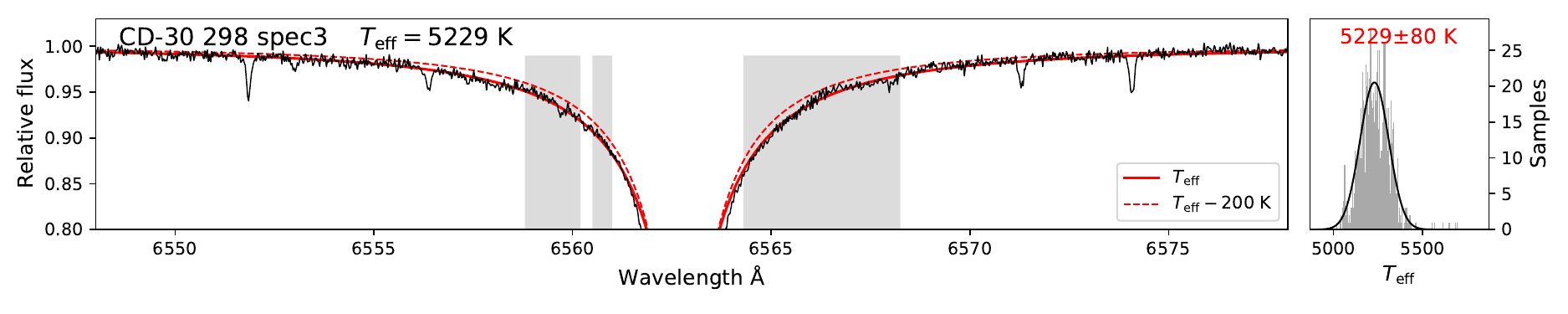}
\includegraphics[width=1\linewidth]{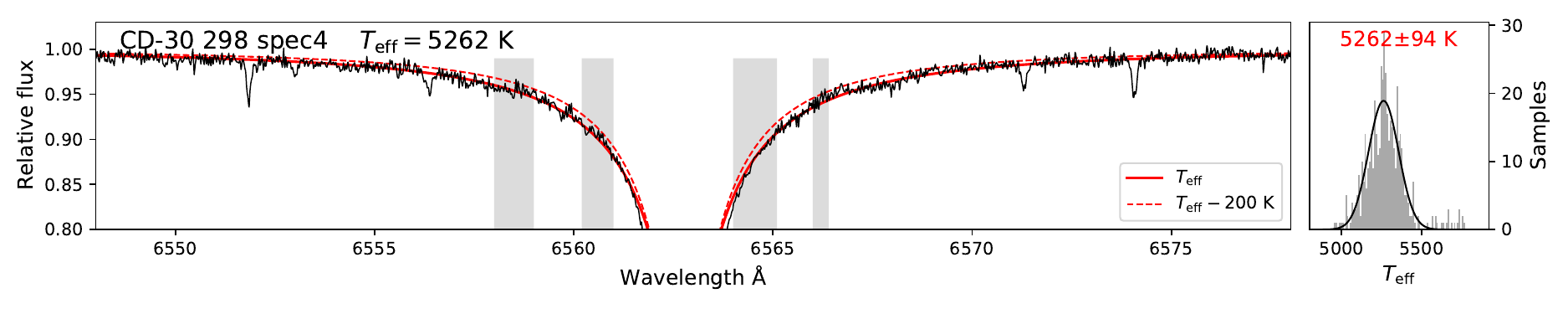}
\includegraphics[width=1\linewidth]{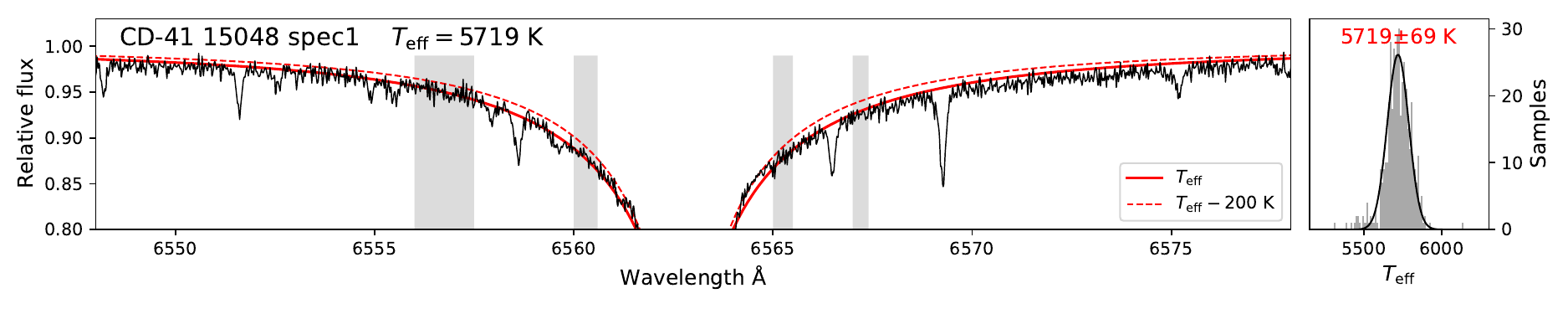}
\includegraphics[width=1\linewidth]{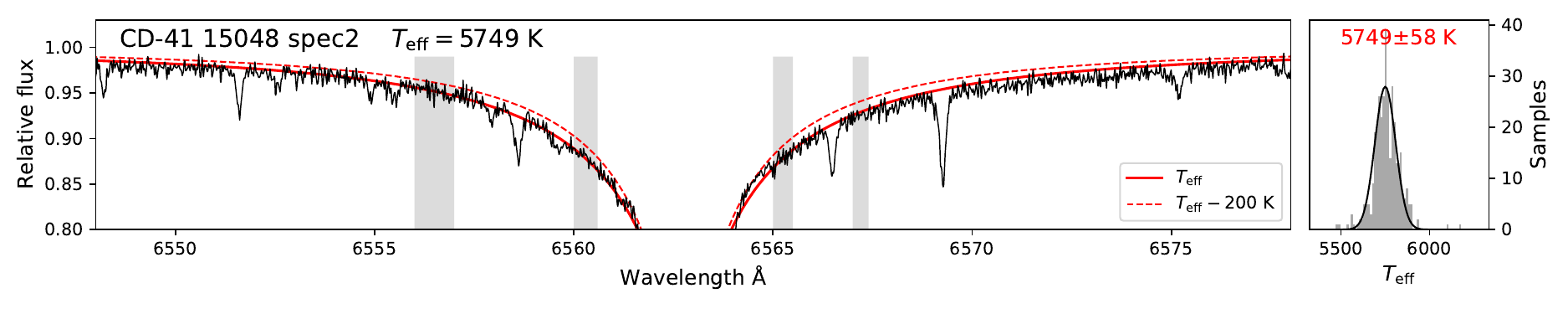}
\includegraphics[width=1\linewidth]{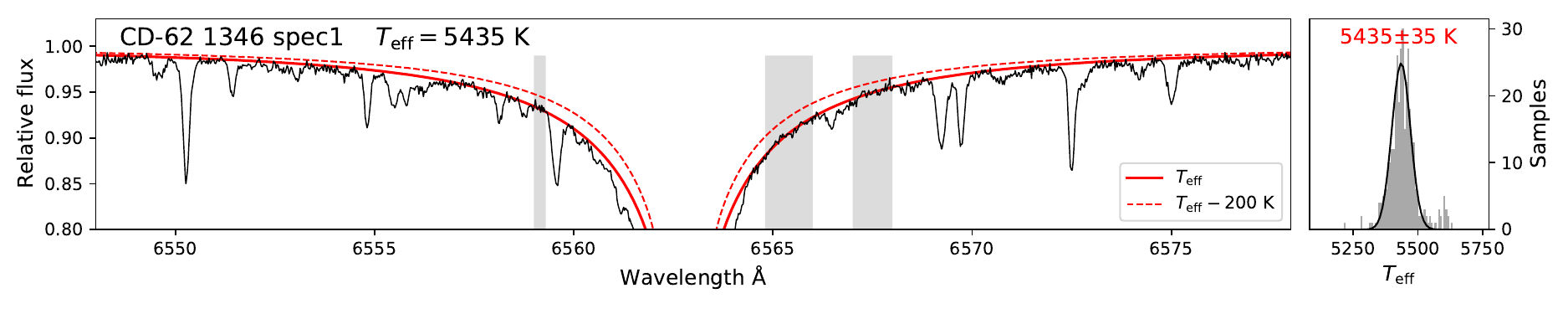}
\includegraphics[width=1\linewidth]{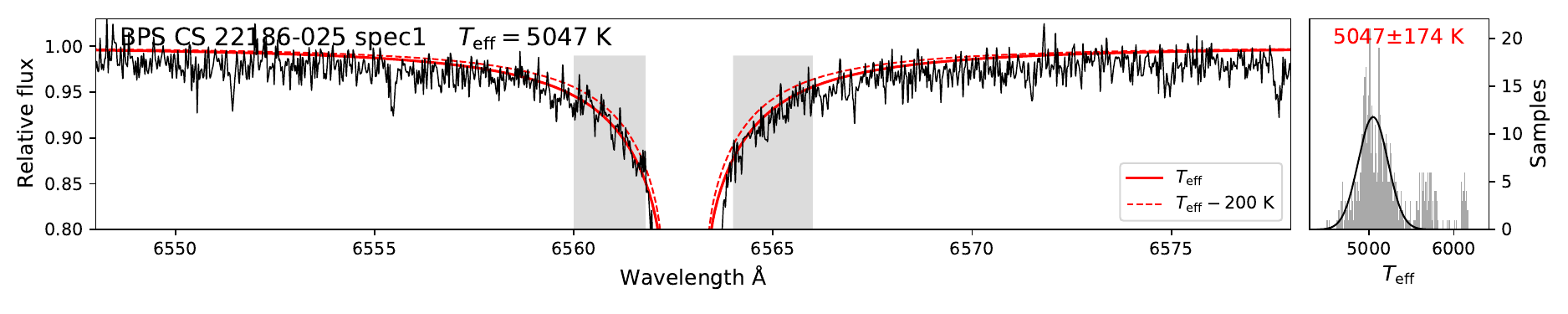}
\includegraphics[width=1\linewidth]{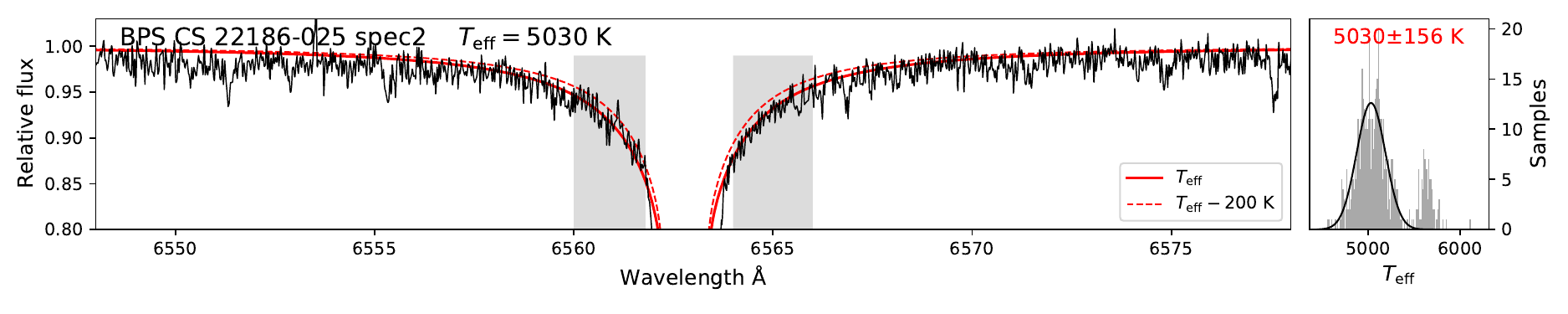}
\label{fig:Ha_other_giants2}
\end{figure}

\begin{figure}[]
    \centering
\includegraphics[width=1\linewidth]{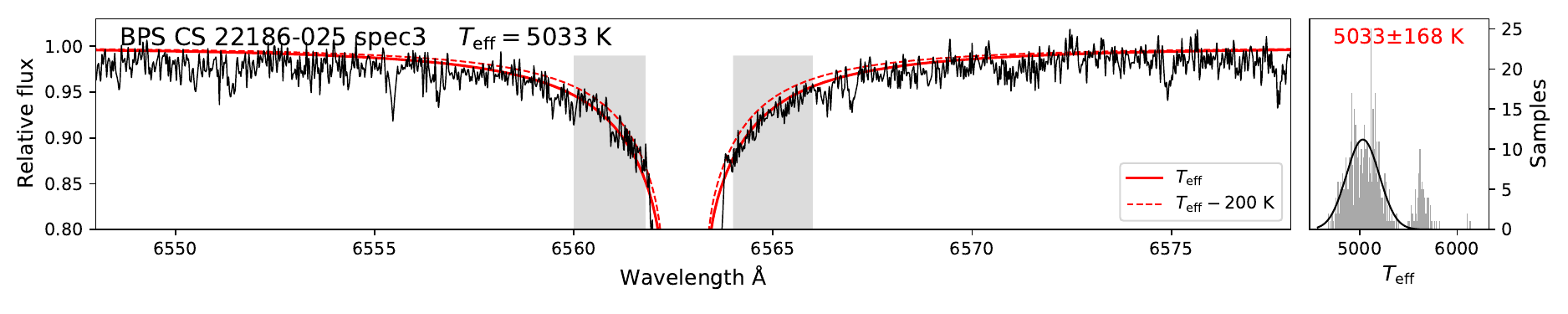}
\includegraphics[width=1\linewidth]{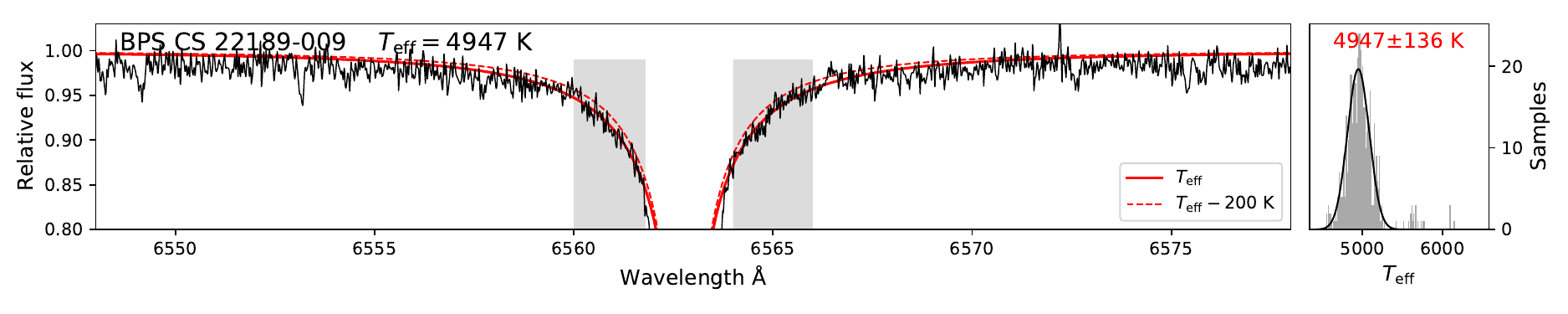}
\includegraphics[width=1\linewidth]{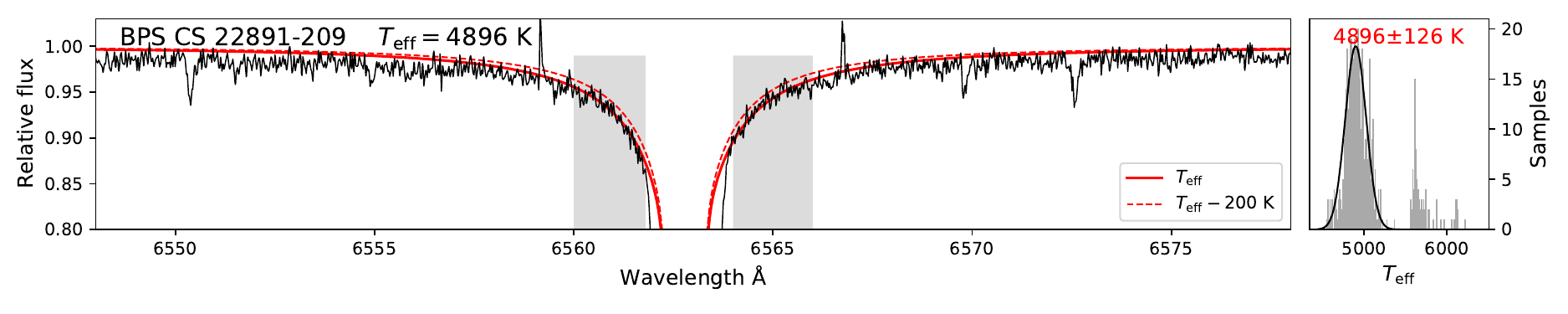}
\includegraphics[width=1\linewidth]{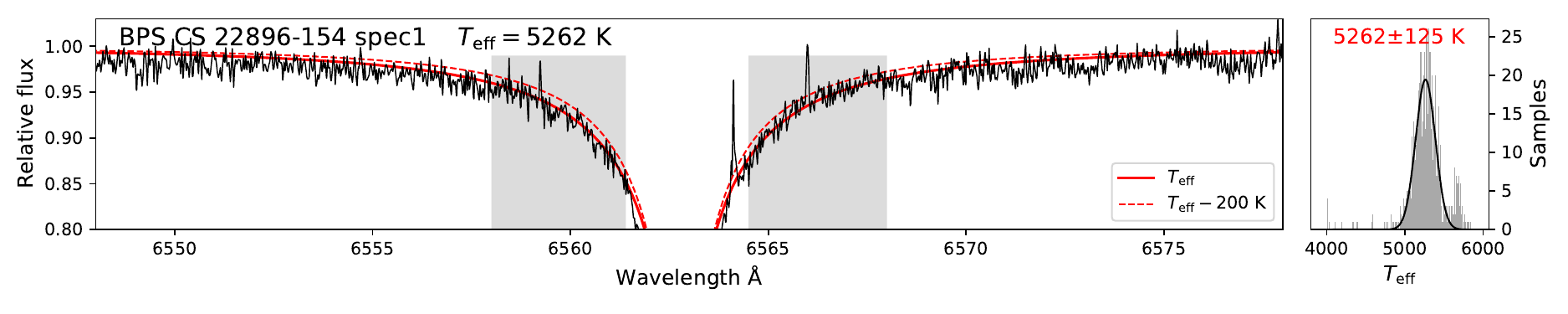}
\includegraphics[width=1\linewidth]{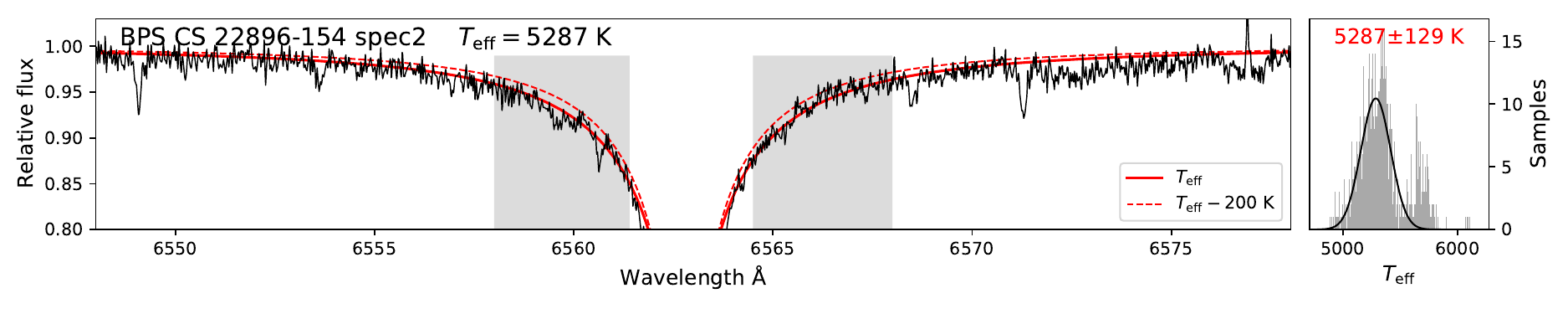}
\includegraphics[width=1\linewidth]{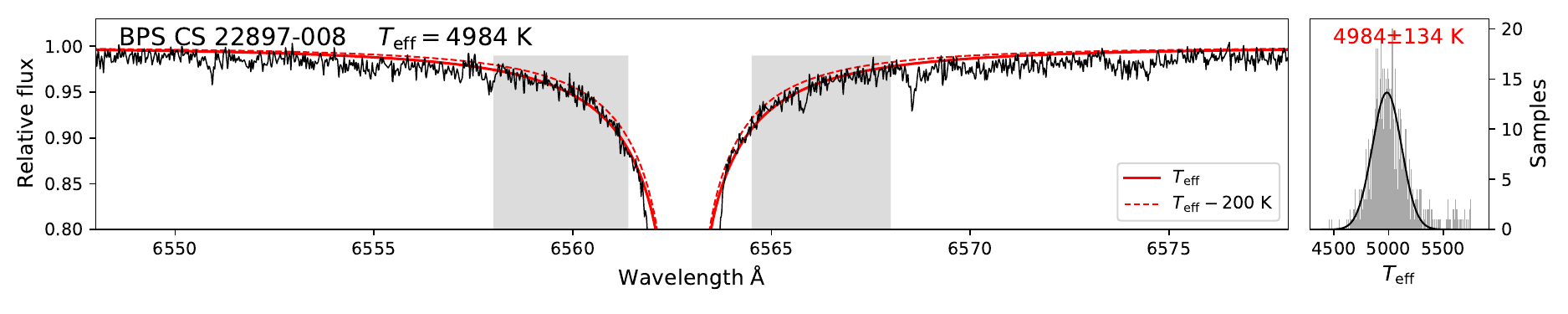}
\includegraphics[width=1\linewidth]{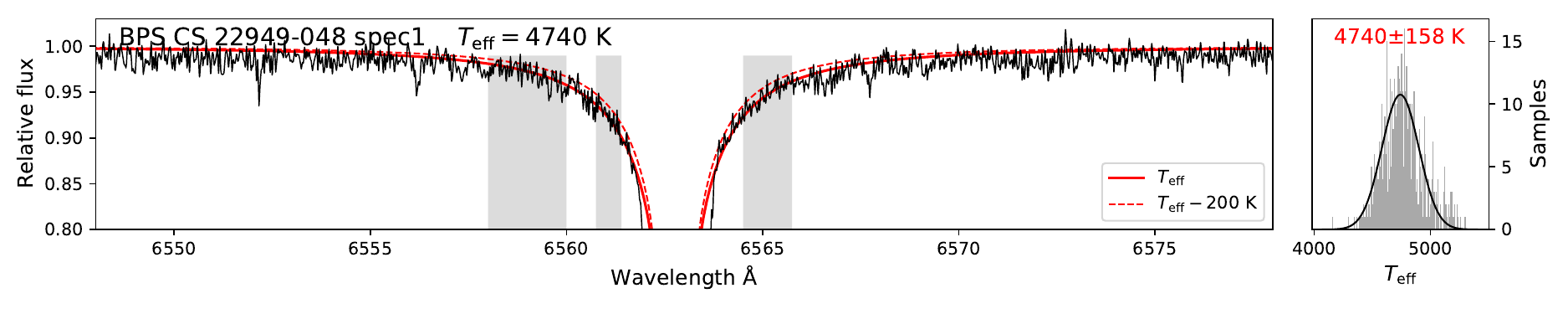}
\includegraphics[width=1\linewidth]{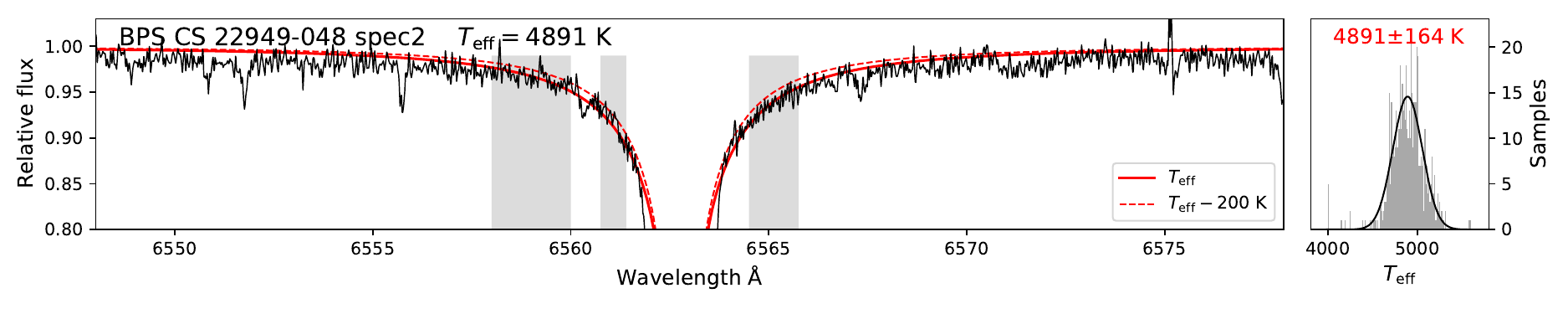}
\includegraphics[width=1\linewidth]{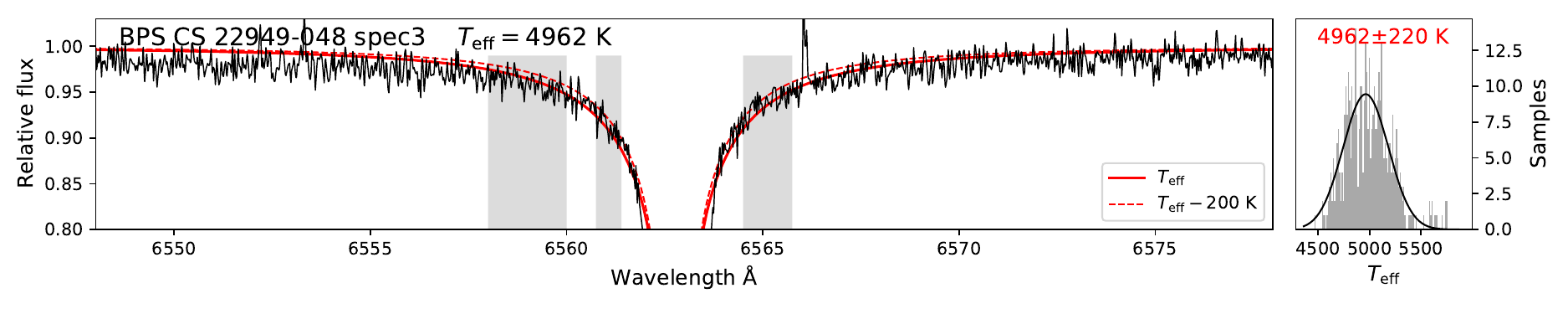}
\includegraphics[width=1\linewidth]{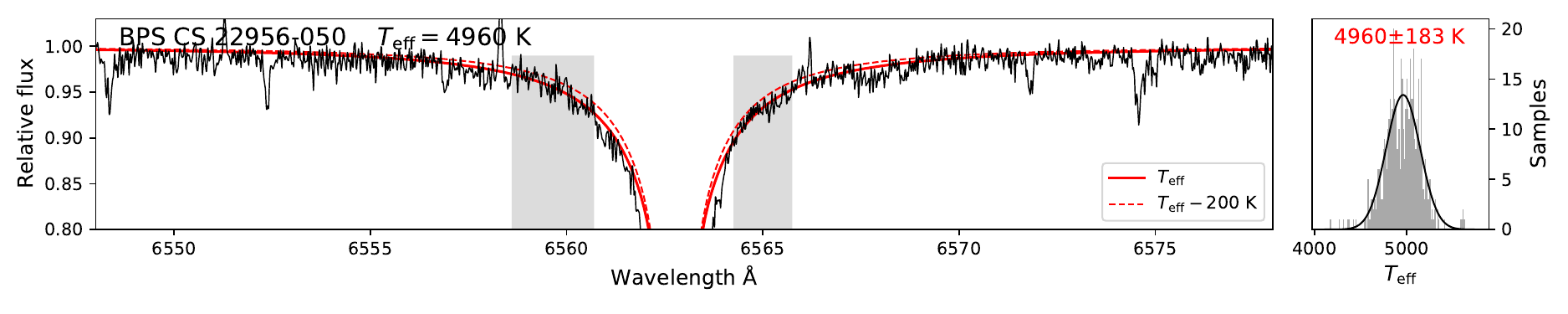}
\includegraphics[width=1\linewidth]{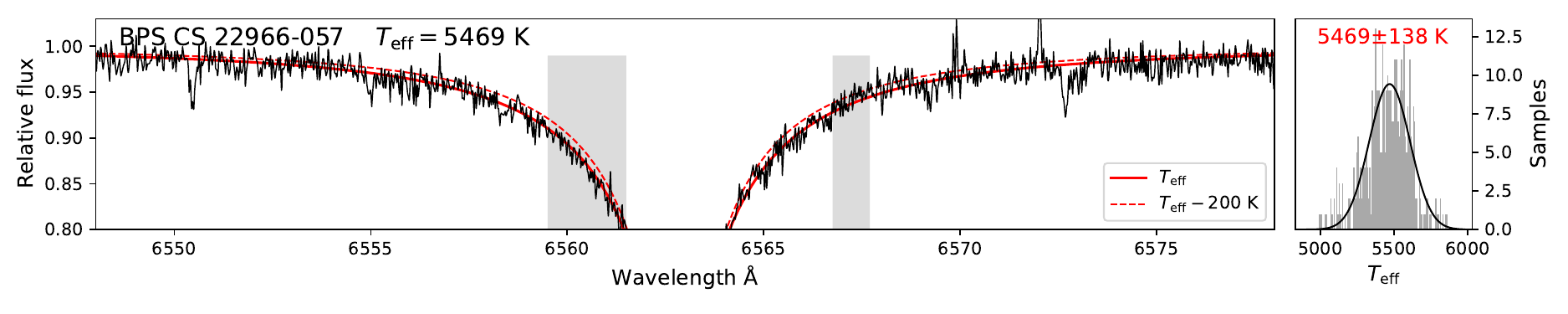}
\includegraphics[width=1\linewidth]{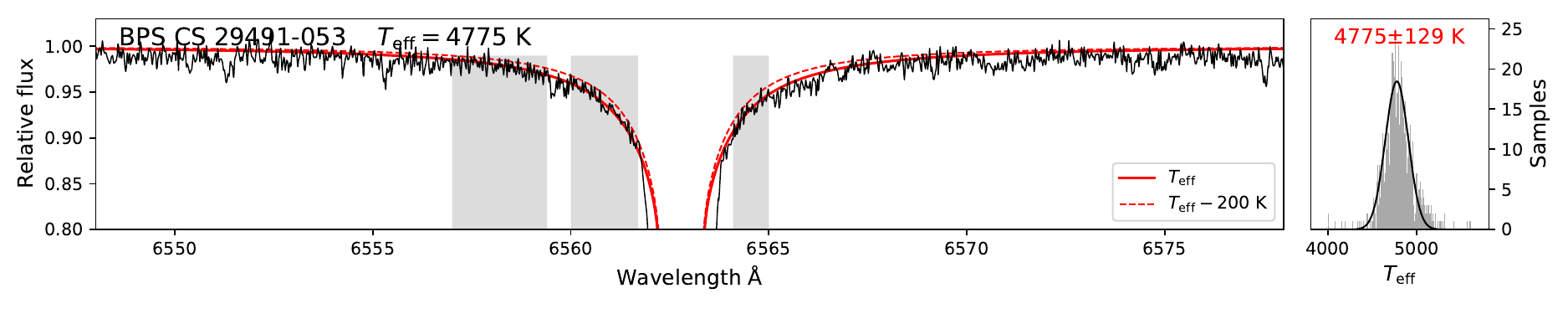}
\label{fig:Ha_other_giants3}
\end{figure}

\begin{figure}[]
    \centering
\includegraphics[width=1\linewidth]{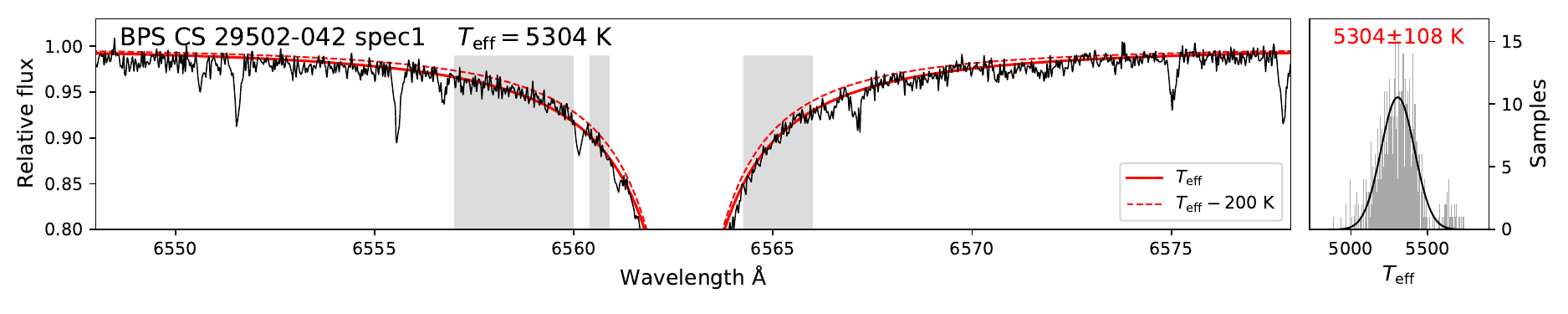}
\includegraphics[width=1\linewidth]{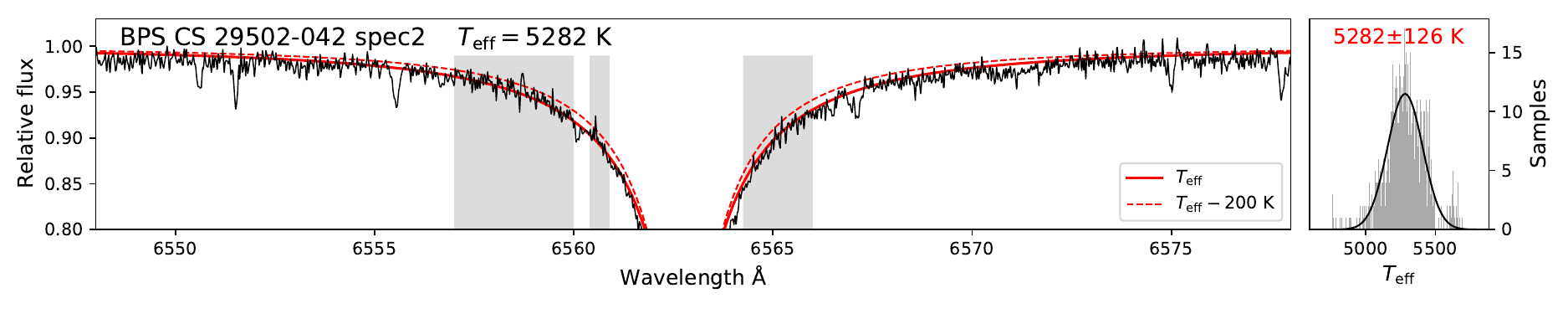}
\includegraphics[width=1\linewidth]{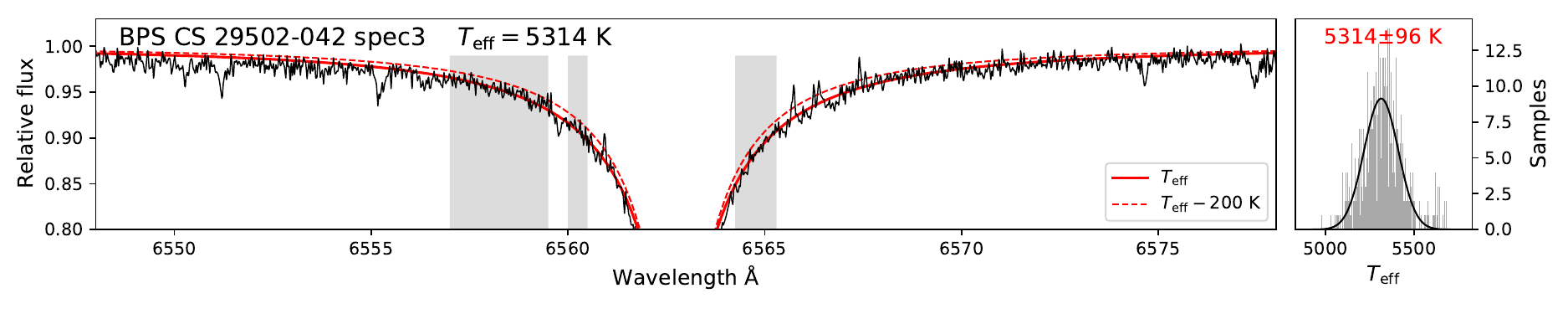}
\includegraphics[width=1\linewidth]{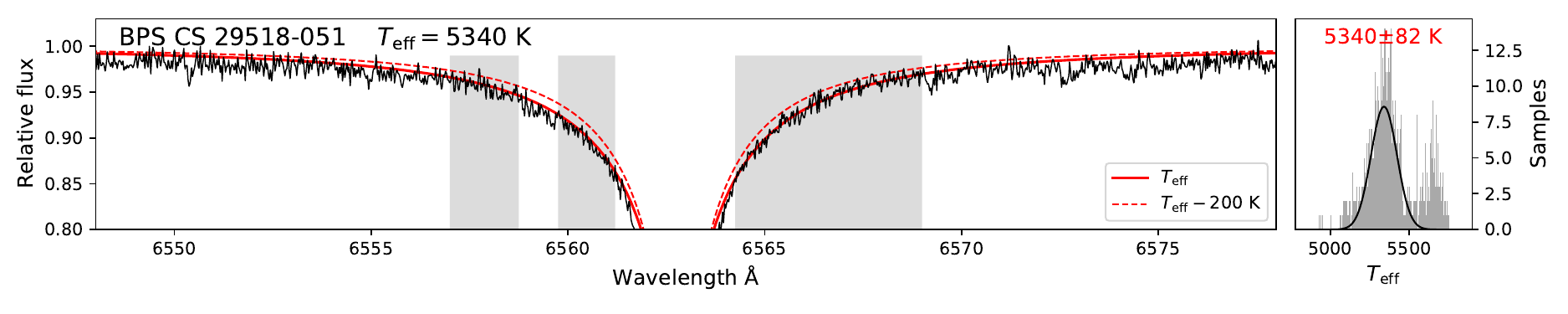}
\includegraphics[width=1\linewidth]{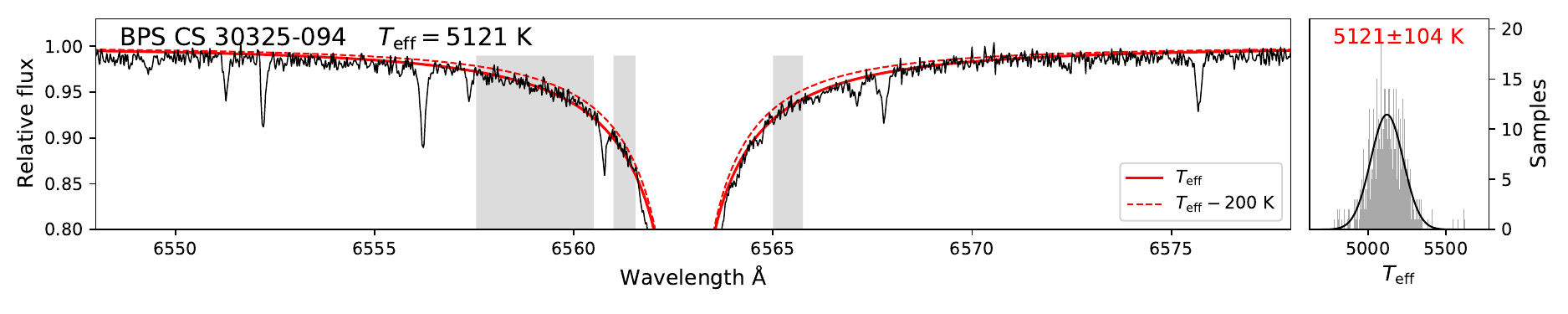}
\includegraphics[width=1\linewidth]{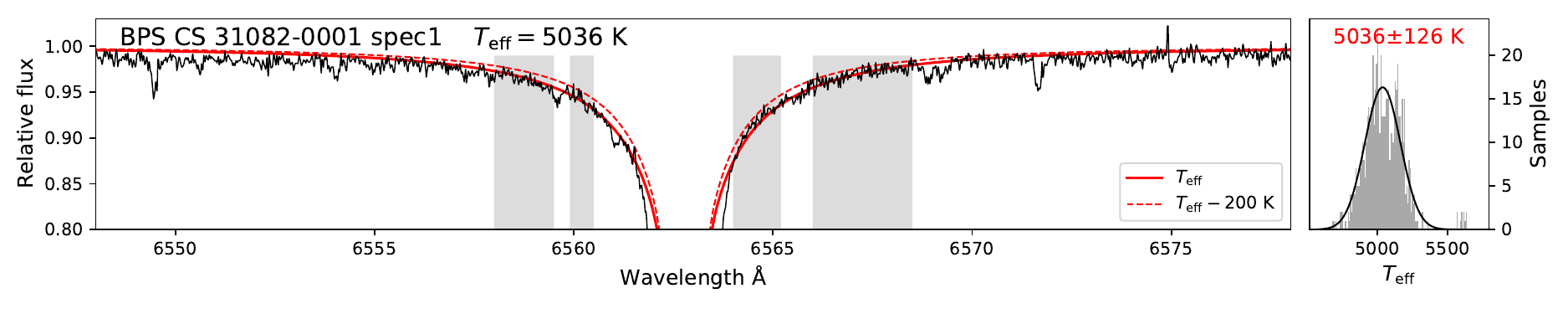}
\includegraphics[width=1\linewidth]{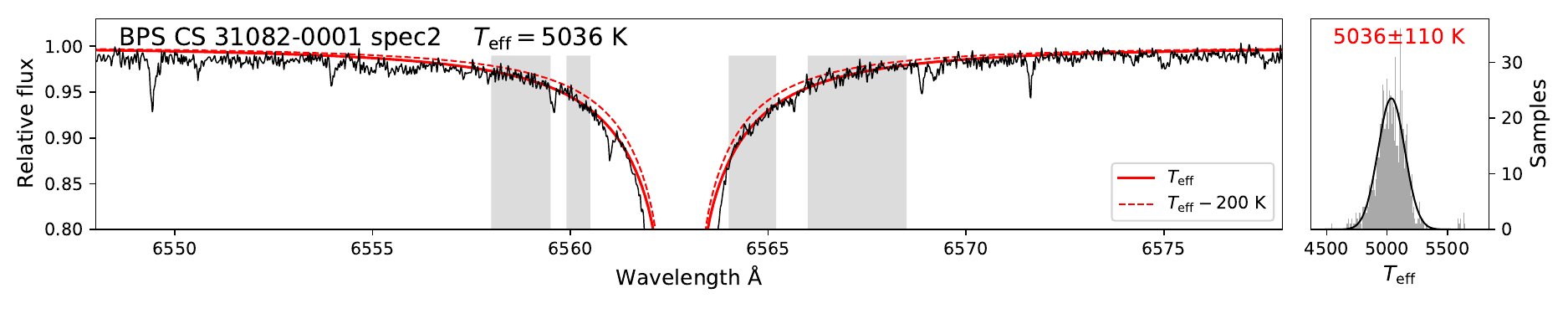}
    \caption{Similar to Fig.~\ref{fig:Ha_interferometry} for other giant stars.}
\label{fig:Ha_other_giants4}
\end{figure}

\begin{figure}[]
    \centering
    \includegraphics[width=1\linewidth]{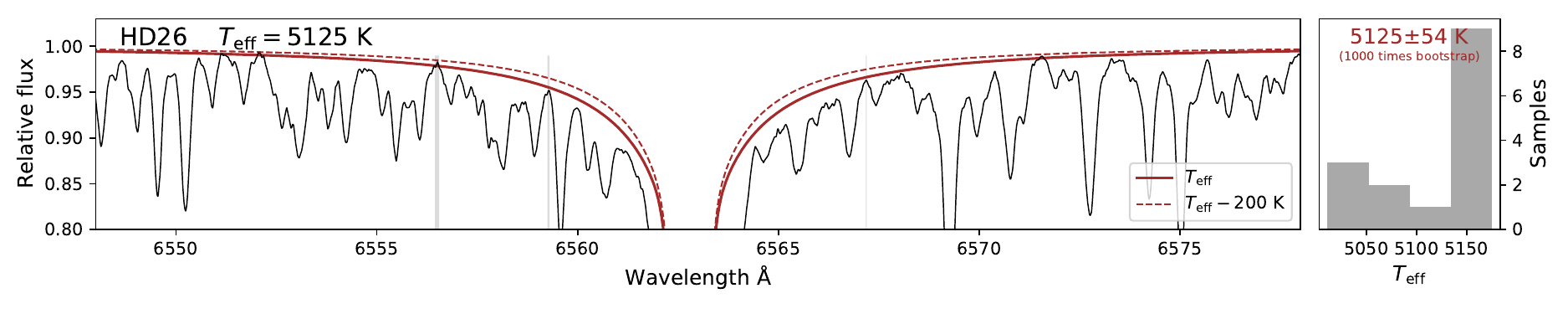}
    \includegraphics[width=1\linewidth]{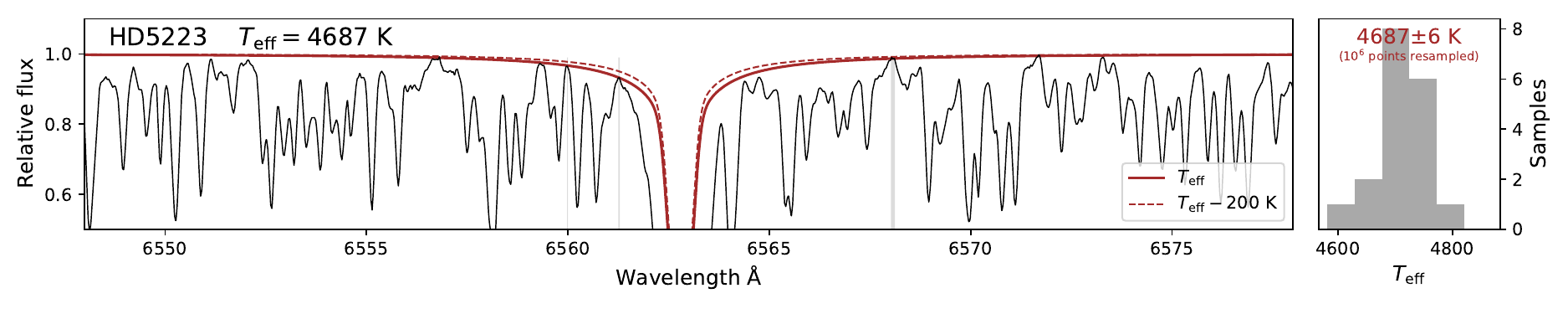}
    \includegraphics[width=1\linewidth]{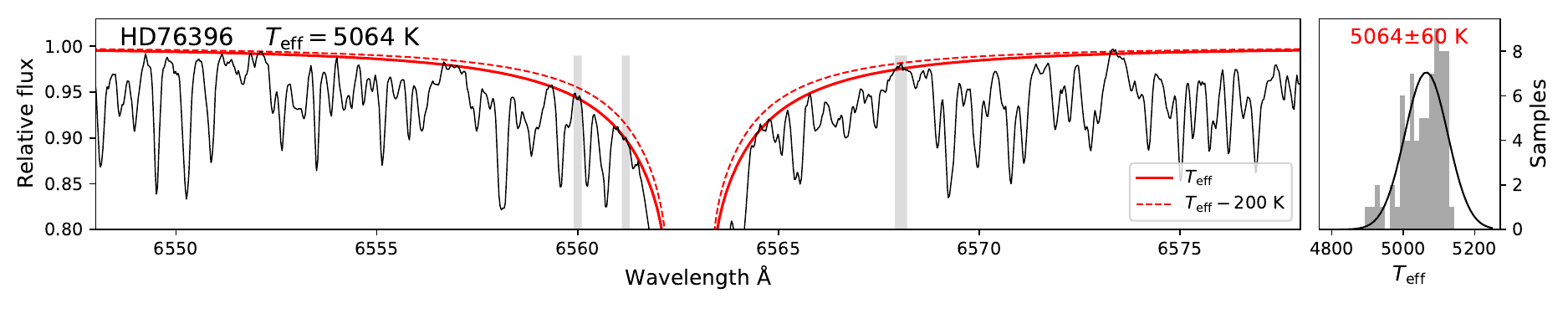}
    \includegraphics[width=1\linewidth]{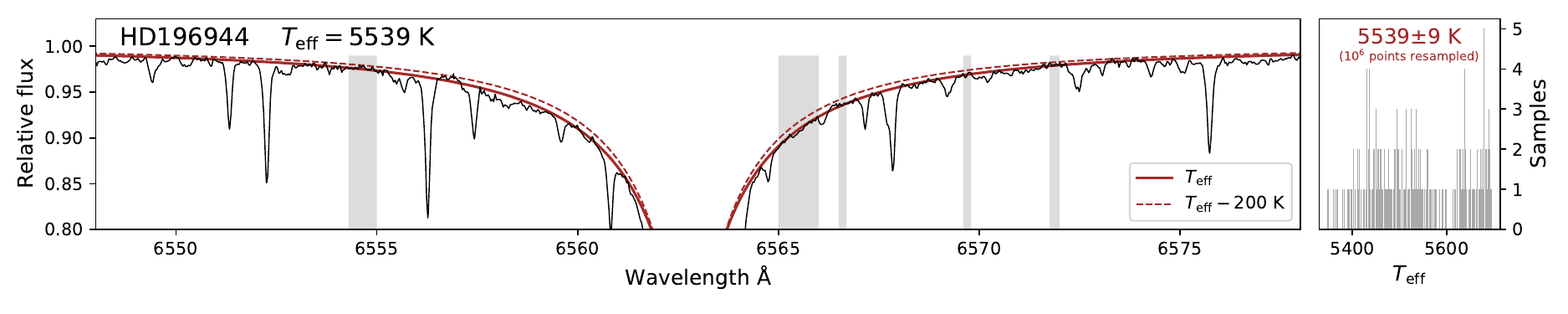}
    \includegraphics[width=1\linewidth]{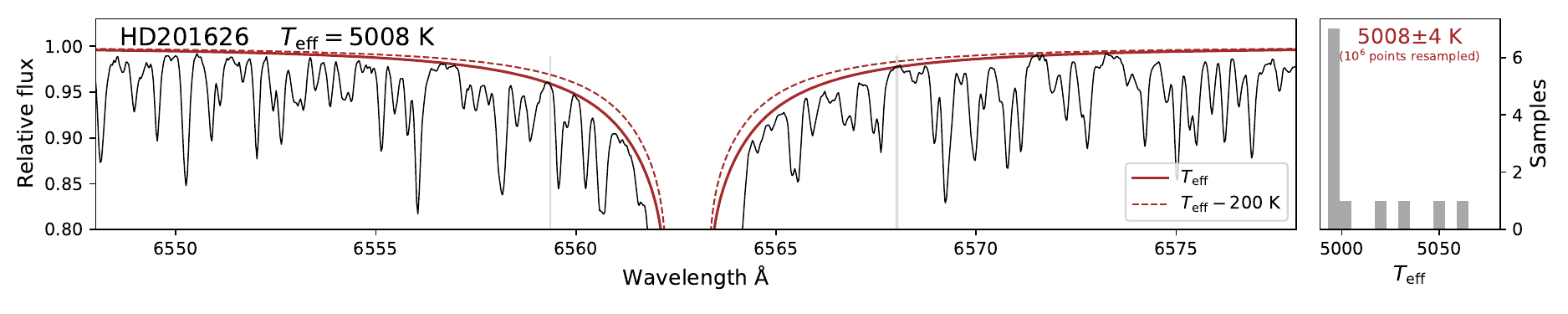}
    \includegraphics[width=1\linewidth]{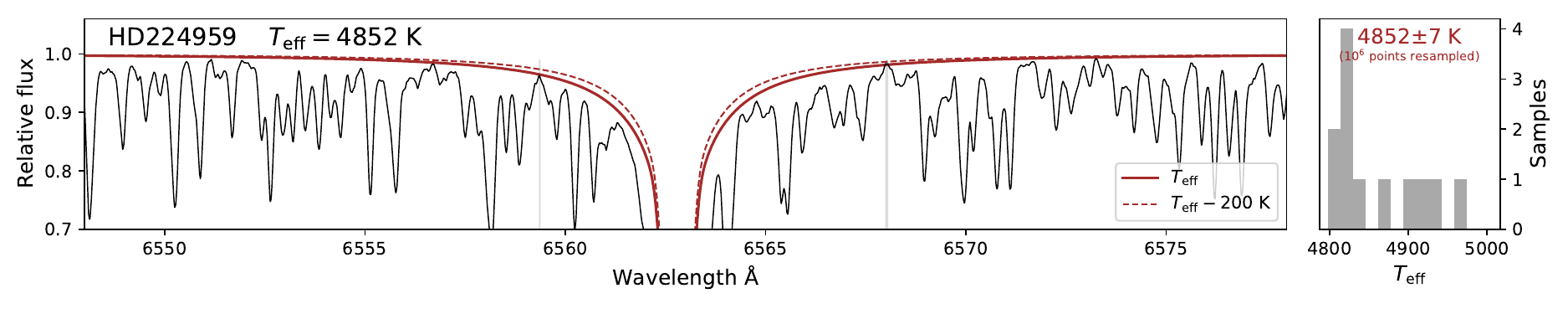}
    \includegraphics[width=1\linewidth]{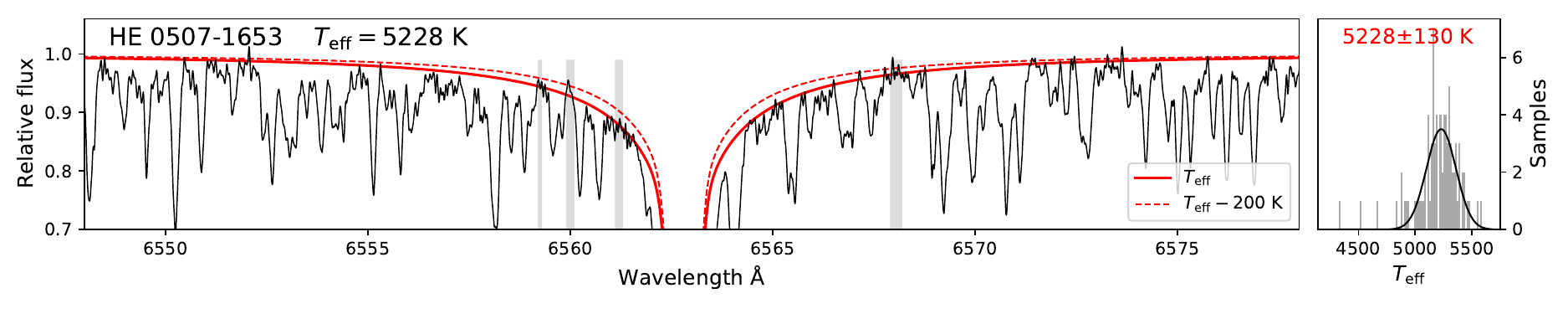}
    \caption{Similar to Fig.~\ref{fig:Ha_interferometry} for CEMP stars.}
    \label{fig:Ha_CEMP}
\end{figure}

\begin{table*}
\caption{Magnesium abundances and surface gravities.}
\label{tab:logg_Mg}
\centering
\tiny 
\begin{threeparttable}
\begin{tabular}{lccccc}
\hline\hline
Name & \logg$_{\mathrm{(Mg 5172)}}$ & \logg$_{\mathrm{(Mg 5183)}}$ & A(Mg$_{5528}$) & A(Mg$_{5711}$) & \loggiso\\
\hline
\multicolumn{6}{c}{Giants with interferometric measurements}\\
\hline
HD~2665 &  $1.96\pm0.14$ & $1.98\pm 0.20$ & 5.81 & 5.74 & $2.30 \pm 0.08$\\
HD~122563  & $1.10\pm0.04$ & $1.05\pm 0.09$& 5.27 & 5.27 & $1.39 \pm 0.06$ \\
HD~175305 &  $2.46\pm0.03$ & $2.47\pm0.24$ & 6.52 & 6.45 & $2.64 \pm 0.06$\\
HD~$221170$ & $1.08\pm0.04$ & $1.11\pm0.03$ & 6.03 & 5.80 & $1.40 \pm 0.06$\\
\hline
\multicolumn{6}{c}{Giants with direct IRFM}\\
\hline
HD 45282  & $3.04\pm0.05$ & $3.09\pm0.09$ & 6.50 & 6.48 & $3.37 \pm 0.05$\\
HD~175305 & $2.46\pm0.03$ & $2.47\pm0.24$ & 6.52 & 6.45 & $2.64 \pm 0.06$\\
HD~$218857$ & $2.54\pm0.12$ & $2.59\pm0.07$ & 5.95 & 5.93 & $2.62 \pm 0.07$\\
BPS BS 17569$-049$ & $0.99\pm0.36$ & $1.07\pm0.31$ & 5.73 & --& $2.60 \pm 0.10$\\
BPS CS 22953$-003$ & $1.88\pm0.71$ & $2.01\pm0.69$ & 5.27 & -- & $2.46 \pm 0.09$\\
BPS~CS~22892$-052$ & $1.59\pm0.43$ & $1.93\pm0.43$ & 5.10 & --& $2.00 \pm 0.15$\\
\hline
\multicolumn{6}{c}{Giants with asteroseismic measurements}\\
\hline
HD~3179 & $2.24 \pm0.06$ & $2.29\pm0.14$ & 6.93 & 6.95 & $2.93 \pm 0.07$\\ 
HD~13359 &$2.60\pm0.08$ & $2.58\pm0.04$ & 6.97 & 6.99 & $2.95 \pm 0.06$\\
HD~17072  & $2.38\pm0.04$ & $2.49\pm0.11$ & 6.91 & 6.89 & $2.96 \pm 0.06$\\
HD~221580  & $2.44\pm0.12$ & $2.51\pm0.17$ & 6.75 & 6.70 & $2.80 \pm 0.04$\\
TIC 168924748 & -- & --& --& -- & $2.95 \pm 0.04$ \\
TIC 404605506 & -- & --& -- & -- & $2.74 \pm 0.06$\\
\hline
\multicolumn{6}{c}{Other giants}\\
\hline
HD~$6229$ & $2.35\pm0.07$ & $2.35\pm0.08$ & 6.89 & 6.84 & $2.79 \pm 0.07$\\
HD~$105740$ & $2.44\pm0.12$ & $ 2.36\pm0.06$ & 7.30 & 7.46 & $2.58 \pm 0.06$\\
HD~$115444$ & $1.50\pm0.16$ & $1.55\pm0.05$  & 5.08 & 5.00 & $1.74 \pm 0.04$\\
HD~$186478$ & $1.62\pm0.02$ &  $1.58\pm0.08$ & 5.59 & -- & $1.70 \pm 0.10$ \\
BD$+09\,2860$  & $2.27\pm0.11$ & $2.25\pm0.12$ & 6.37 & 6.21 & $2.59 \pm 0.03$\\
BD$+09\,2870$ & $1.50\pm0.18$ & $1.35\pm0.09$ & 5.72 & 5.54 & $1.63 \pm 0.06$\\
BD$+11\,2998$  & $2.33\pm0.04$ & $2.32\pm0.04$ & 7.16 & 6.93 & $3.07 \pm 0.03$\\
BD$+17\,3248$ & $1.93\pm0.32$ & $2.08\pm0.18$ & 6.15 & -- & $2.40 \pm 0.13$\\
BD$-18\,5550$  & $1.30\pm0.33$ &  $1.43\pm0.33$ & 5.20 & -- & $2.22 \pm 0.07$\\
CD$-30\,298$  & $2.54\pm0.67$ & $2.45\pm0.60$ & 4.49 & -- & $2.71 \pm 0.06$\\
CD$-41\,15048$  & $2.91\pm0.29$ & $2.78\pm0.29$ & 6.03 & 5.91 & $3.07 \pm 0.03$\\
CD$-62\,1346$ & $2.25\pm0.10$ & $2.21\pm0.06$ & 6.61 & 6.39 & $2.85 \pm 0.03$\\
BPS CS 22186$-025$  & -- & -- & 5.26 & -- & $2.02 \pm 0.12$\\
BPS CS 22189$-009$   & -- & -- & 4.56 & -- & $2.20 \pm 0.11$\\
BPS CS 22891$-209$   & -- & -- & 4.91 & -- & $1.70 \pm 0.12$\\
BPS CS 22896$-154$   & $2.61\pm0.72$ & $2.61\pm0.41$ & 5.38 & -- & $2.69 \pm 0.07$\\
BPS CS 22897$-008$   & $1.98\pm0.83$ & $1.73\pm0.87$ & 4.77 & -- & $2.10 \pm 0.06$\\
BPS CS 22949$-048$  & -- & -- & 4.75 & -- & $2.26 \pm 0.09$\\
BPS CS 22956$-050$  & -- & -- & 4.92 & -- & $2.08 \pm 0.15$\\
BPS CS 22966$-057$  & $3.35\pm0.17$ & $3.36\pm0.11$ & 5.60 & -- & $3.49 \pm 0.06$\\
BPS CS 29491$-053$  & $1.55\pm0.51$ & $1.56\pm0.32$ & 5.16 & --& $1.76 \pm 0.10$\\
BPS CS 29502$-042$  & $2.45\pm0.84$ & $2.78\pm0.46$ & 5.02 & --& $2.90 \pm 0.05$\\
BPS CS 29518$-051$  & $2.37\pm0.24$ & $2.26\pm0.16$ & 5.48 & --& $2.75 \pm 0.09$\\
BPS CS 30325$-094$ & $1.93\pm0.73$ & $1.85\pm0.53$ & 4.96 & --& $2.24 \pm 0.10$\\
BPS~CS~31082$-001$ & $1.85\pm0.37$ & $2.01\pm0.26$ & 5.40 & --\\
\hline
\multicolumn{6}{c}{CEMP giants}\\
\hline
HD~$26$ & $2.16\pm0.09$ & $2.22\pm0.03$ & 7.28 & -- & --\\
HD~$5223^\S$ & $1.27\pm0.05$ & -- & -- & [5.89] & --\\
HD~$76396$ &  $2.37\pm0.08$ & $2.40 \pm0.02$  & -- & 6.34 & --\\
HD~$196944$ & $2.43 \pm 0.18$ & $2.44\pm0.18$ & 5.88 & 5.73 & -- \\
HD~$201626$ & $2.13 \pm0.02$ & $2.18 \pm 0.03$ & -- & 6.29 & --\\
HD~$224959$ & $2.03\pm0.14$ & $1.96\pm0.03$ & [5.73] & -- & -- \\
HE~0507-1653 & $2.95\pm0.11$ & $2.90\pm0.13$ & -- & 6.49 & --\\
\hline
\end{tabular}
\begin{tablenotes}
\item{} \textbf{Notes.} {Second and third columns lists surface gravities derived from the Mg triplet lines 5172 and 5183~\AA, respectively.
Fourth and fifth columns lists Mg abundances derived from the lines 5528 and 5711~\AA, respectively.
Brackets indicate uncertain values because of line blending. Last column lists \loggiso.
} 
\end{tablenotes}
\end{threeparttable}
\end{table*}

\begin{figure*}
    \centering
    \includegraphics[width=0.45\linewidth]{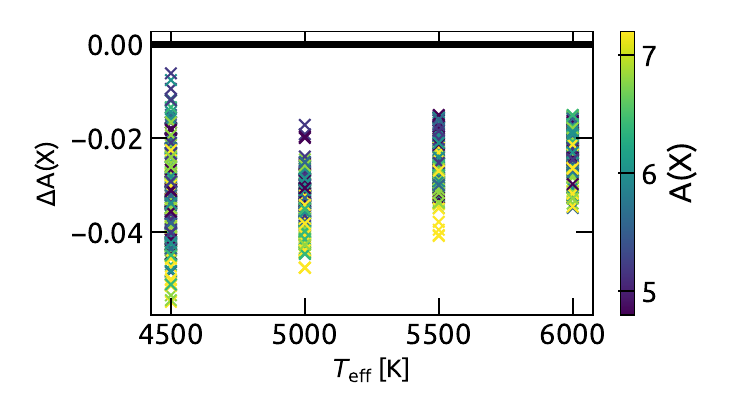}
    \includegraphics[width=0.45\linewidth]{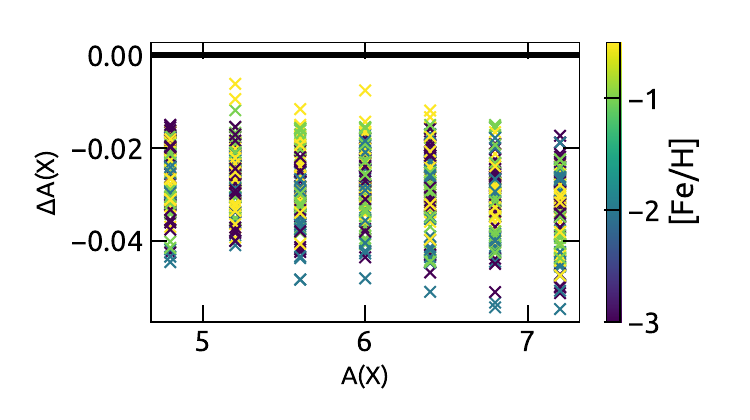}
    \includegraphics[width=0.45\linewidth]{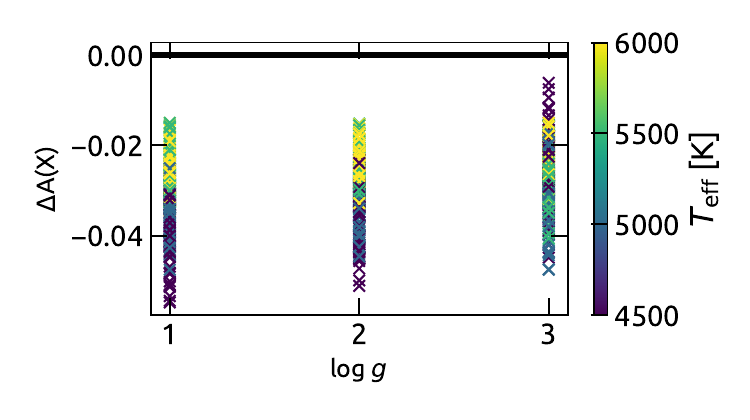}
    \includegraphics[width=0.45\linewidth]{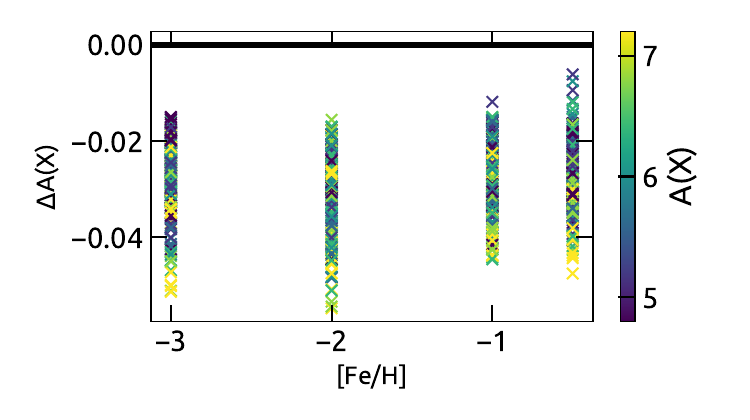}
    \caption{\tiny A(Mg) offsets induced by a $-50$~K \teff\ offset, when derived from the line at 5711~\AA. The horizontal line represents no offset. "X" represents the element, Mg in this case.}
    \label{fig:Mg_offsets}
\end{figure*}

\end{appendix}

\bibliographystyle{aa} 
\bibliography{riano}

\end{document}